\documentclass[12pt]{iopart}
\usepackage{iopams}  
\usepackage{bm}
\usepackage{graphicx}
\usepackage{slashed}
\usepackage{color,soul}
\newcommand{\hef}{$^4$He} 
\newcommand{\het}{$^3$He} 
\newcommand{\hyt}{$^3$H}

\newcommand{\fm}{fm$^{-1}$}
\newcommand{\ba}{\begin{eqnarray*}} 
\newcommand{\ea}{\end{eqnarray*}}
\newcommand{\bea}{\begin{eqnarray}} 
\newcommand{\eea}{\end{eqnarray}}
\eqnobysec
\begin{document}

\hspace*{12.5cm}{\mbox{IFUP-TH}}\\
\hspace*{13.3cm}{\mbox{JLAB-THY-15-2030}}
\topical[Electromagnetic Structure of Few-Nucleon Ground States]{Electromagnetic Structure of Few-Nucleon Ground States}

\author{L.E. Marcucci$^{1,2}$, F. Gross$^{3,4}$,  M.T. Pe\~na$^{5,6}$,  M. Piarulli$^{3,7}$, R. Schiavilla$^{3,7}$, I. Sick$^{8}$, A. Stadler$^{9,6}$, J.W. Van Orden$^{3,7}$, 
M. Viviani$^{2,1}$}

\address{$^1$ Department of Physics ``E. Fermi'', 56127 Pisa, Italy}
\address{$^2$ INFN - Sezione di Pisa, 56127 Pisa, Italy}
\address{$^3$ Jefferson Lab., Newport News, Virginia 23606, USA}
\address{$^4$ College of William and Mary, Williamsburg, Virginia 23185, USA}
\address{$^5$ Departamento de F\'isica, Instituto Superior T\'ecnico, Universidade de Lisboa, 1049-001 Lisboa, Portugal}
\address{$^6$ CFTP, Instituto Superior T\'ecnico, Universidade de Lisboa, 1049-001 Lisboa, Portugal}
\address{$^7$ Department of Physics, Old Dominion University, Norfolk, Virginia 23529, USA}
\address{$^8$ Departement f\"ur Physik, Universit\"at Basel, CH-4056 Basel, 
Switzerland}
\address{$^9$ Departamento de F\'isica, Escola de Ci\^encias e Tecnologia, Universidade de \'Evora, 7000-671 
\'Evora, Portugal}

\ead{laura.marcucci@df.unipi.it}

\begin{abstract}
Experimental form factors of the hydrogen and helium
isotopes, extracted from an up-to-date global analysis of
cross sections and polarization observables measured in
elastic electron scattering from these systems, are compared
to predictions obtained in three different theoretical approaches:
the first is based on realistic interactions and currents, including
relativistic corrections (labeled as the conventional approach);
the second relies on a chiral effective field theory description
of the strong and electromagnetic interactions in nuclei (labeled
$\chi$EFT); the third utilizes a fully relativistic treatment of nuclear
dynamics as implemented in the covariant spectator theory (labeled
CST).  For momentum transfers below $Q \lesssim 5$ fm$^{-1}$
there is satisfactory agreement between experimental data and
theoretical results in all three approaches.  However, at 
$Q \gtrsim 5$
fm$^{-1}$, particularly in the case of the deuteron, 
a relativistic treatment of the dynamics, as is done in the CST, is  necessary.
The experimental data on the deuteron $A$ structure function
extend to $Q \simeq 12$ fm$^{-1}$, and the close agreement
between these data and the CST results suggests that,
even in this extreme kinematical regime, there is no evidence for new 
effects 
coming from quark and gluon degrees of freedom at short distances.
 
\end{abstract}
\submitto{\JPG}

\maketitle

\section{Introduction}
\label{sec:intro}
Few-nucleon systems, and more generally light s- and p-shell nuclei with mass number up to $A=12$, 
offer a unique opportunity to
test our understanding of nuclear dynamics. Over the past thirty years
or so, several techniques have been developed to solve exactly the
quantum mechanical few-body problem in both non-relativistic~\cite{Nog03,Kam01} and
relativistic~\cite{Gross:1969rv,Keister:1991sb,Sta97b} regimes.
More recently, in the last decade, quantum Monte Carlo methods coupled
with improvements in algorithms and advances
in computational capabilities have made it possible to carry out also exact,
albeit non-relativistic, calculations of nuclei up to $^{12}$C~\cite{Car98,Car15}.
These technical breakthroughs have permitted first-principle studies
of the strong interaction in nuclei, as it manifests itself in terms
of two- and many-body forces among the nuclear constituents, the
protons and neutrons, and of the interactions of these constituents
with external electroweak probes in a wide range of energy and momentum
transfers. 

In the present review the focus is on the electromagnetic ground-state
structure of the hydrogen and helium isotopes.  Since the early fifties,
the associated form factors have been the subject of intense experimental
and theoretical scrutiny.   The large body of elastic electron scattering cross
section (and polarization) data from these systems---a review of these data,
and ensuing analysis, is provided in section~\ref{sec:exp}---has now led to
accurate experimental determinations of the charge and magnetic form factors
of $^2$H, $^3$H, and $^3$He, the quadrupole form factor of $^2$H, and the
charge form factor of $^4$He, up to momentum transfers $Q$ beyond
9 fm$^{-1}$, and in some instances, as for the $A$ structure function of the
deuteron, extending to $Q \simeq 12$ fm$^{-1}$.

A vast amount of theoretical work in a variety of different frameworks---purely
non-relativistic, or including relativistic corrections, or fully covariant
ones---exists for these systems, and no attempt will be made here
to summarize it.  Rather, we have chosen to focus on three representative
approaches: the first one, which we label as ``conventional'', is based
on realistic nuclear interactions and currents, including leading relativistic
corrections; the second one relies on chiral effective field theory ($\chi$EFT) for a
description of the nuclear strong and electromagnetic interactions; the third
one utilizes the fully relativistic dynamical framework 
of the covariant spectator theory (CST). 
They are reviewed
in considerable detail in section~\ref{sec:theory}, where an appraisal
of their differences and similarities is also provided.  All three approaches
have recently been used to calculate the few-nucleon form factors:
conventional and $\chi$EFT predictions are available for $A=2$--4
and CST ones for $A=2$ and 3.  These predictions are compared to
the experimental form factors extracted from the world-data analysis
of cross sections and polarizations in section~\ref{sec:res}.  
The $\chi$EFT calculation
of the $^4$He form factor are presented here for the first time.
Some
concluding remarks are  given in section~\ref{sec:conc}.  For
completeness, in the remainder of this section we recall the
basic formalism used to describe elastic electron-nucleus scattering
and the definitions of the various form factors.

\vspace{0.5in}

\subsection{Few-nucleon form factors}
\label{subsec:form}

In the one-photon-exchange approximation, the unpolarized cross section
for elastic electron-deuteron scattering can be written as
\begin{equation}
{d \sigma \over d \Omega} = \sigma_M \, f_{\rm rec}^{-1}  
\left [ A (Q) + B (Q) \tan^2 {\theta/2} \right ]  \equiv \sigma_M \, f_{\rm rec}^{-1} \,I(Q,\theta)
\label{eq:dsigma}
\end{equation}
where $\sigma_M$ is the Mott cross section, 
\begin{equation}
\sigma _M=\left[ \frac{\alpha \cos \theta/2}
{2\varepsilon\sin^{2}\theta/2 }\right]^{2}\ ,
\label{eq:smott}
\end{equation}
$f^{-1}_{\rm rec}$ is the recoil factor,
\begin{equation}
f_{\rm rec}= 1+ \frac{2\, \varepsilon}{M_d} \sin^2 \theta/2 \>\>,
\label{eq:rec}
\end{equation}
$\alpha$ is the fine structure constant,
$\varepsilon$ and $\theta$ are respectively the initial
electron energy and final electron scattering angle, and
$M_d$ is the rest mass of the deuteron.  The structure
functions $A(Q)$ and $B(Q)$ are functions only of the
four-momentum transfer 
$Q$ defined as
$Q\equiv\sqrt{Q^2}$ and 
$Q^2=4\, \varepsilon\, \varepsilon^\prime \sin^2\theta/2$,
where $\varepsilon^\prime$ is the final electron energy,
and in elastic scattering the electron energy transfer $\varepsilon-\varepsilon^\prime$ 
is related to $Q^2$ via $\varepsilon-\varepsilon^\prime=Q^2/(2\, M_d)$.
For a spin 1 nucleus like the deuteron, these structure functions can be
expressed in terms of the three form factors $G_C(Q)$,
$G_M(Q)$, and $G_Q(Q)$ (respectively
charge, magnetic, and quadrupole form factor) as
\begin{equation}
A (Q)  =  G_C^2(Q) +\frac{2}{3} \,\eta_d \,G_M^2(Q)
+\frac{8}{9} \,\eta_d^2\, G_Q^2(Q) \ ,
\label{eq:aq}
\end{equation}
and
\begin{equation}
B (Q)  = \frac{4}{3}\,  \eta_d \left( 1 +\eta_d \right) G_M^2(Q)  \ ,
\label{eq:bq}
\end{equation}
where $\eta_d=Q^2/(4\, M^2_d)$.
Note that while $G_M(Q)$ is given uniquely by $B(Q)$, $G_C(Q)$ and
$G_Q(Q)$ both appear in $A(Q)$ and, therefore, cannot be separated in
an unpolarized scattering experiment.  Separation of the charge and quadrupole
form factors requires elastic scattering involving polarization of either the
initial or final deuteron states.  For a tensor polarized initial deuteron,
the tensor analyzing power $T_{20}(Q)$ can be experimentally determined via
\begin{eqnarray}
I(Q,\theta) \,T_{20}(Q)&=& -\frac{1}{\sqrt{2}} 
 \Bigg[ \frac{8}{3}\, \eta_d \,G_C(Q)\, G_Q(Q) 
+\frac{8}{9}\, \eta_d^2\, G_Q^2(Q) \nonumber\\
&&+\frac{1}{3}\,\eta_d\, \left[ 1+2\,(1+\eta_d)\tan^2\theta/2 \right]G_M^2(Q)
\Bigg] \ ,
\label{eq:t20}
\end{eqnarray}
where $I(Q,\theta)$ is defined in equation~(\ref{eq:dsigma}).
This observable is especially sensitive to the ratio $G_Q(Q)/G_C(Q)$.
So far, the separation of the three independent form factors has been carried
out experimentally up to a momentum transfer 
$Q \simeq 8$ fm$^{-1}$.
However, data on $A(Q)$ extend up to $Q \simeq 12$ fm$^{-1}$.

The form factors defined above can be related to matrix elements
of the electromagnetic current $J^\mu$ between initial and final
deuteron states.  Introducing $J_{\lambda_\gamma}=J_\mu\, \varepsilon^\mu_{\lambda_\gamma}$, where $\varepsilon_{\lambda_\gamma}$ is the virtual photon
polarization, we define these matrix elements as 
\begin{equation}
\langle P^\prime,\lambda^\prime\!\mid J_{\lambda_\gamma}
\mid\! P,\lambda\rangle  =  g_{\mu\nu}G^{\mu}_{\lambda^\prime \lambda}(P^\prime,P)\varepsilon^\nu_{\lambda_\gamma} \equiv
G^{\lambda_\gamma}_{\lambda^\prime \lambda}(P^\prime,P) \ ,
\end{equation}
 and $\mid \! P,\lambda \rangle$ and $\mid\! P^\prime,\lambda^\prime \rangle$
are the initial and final deuteron states with four momenta $P^\mu$ and $P^{\mu\, \prime}$
and helicities $\lambda$ and $\lambda^\prime$, respectively.
Since there are only three scalar form factors, only three of the helicity
matrix elements are unique. These are chosen to be
\begin{eqnarray}
g_{-1}&\equiv&\frac{1}{2 M_d}\left< P',-1\left| J_{0}\right| P,1\right> =
\frac{1}{2 \,M_d}
G^0_{-1 1}(P',P)\ , \\
g_{0}&\equiv&\frac{1}{2 M_d}\left< P',0\left| J_{0}\right|P,0\right> =
\frac{1}{2\, M_d}
G^0_{0 0}(P',P)\ , \\
g_{+1}&\equiv&\frac{1}{2 M_d}\left< P',+1\left| J_{+1}\right|P,0\right> =
\frac{1}{2\, M_d}
G^1_{+1 0}(P',P) \ .
\end{eqnarray}
On the other hand, by invoking Lorentz invariance, parity conservation, and
current conservation, the matrix elements
$G^\mu_{\lambda^\prime  \lambda}(P^\prime,P)$ can be shown to have the
general form \cite{Gross:1964zz,Gross_proc,ACG_1}:
\begin{eqnarray}
G^\mu_{\lambda^\prime \lambda} (P',P) & = & - \Biggl [ G_1 (Q)
\,\, \xi^*_{\lambda'} \cdot \xi_{\lambda} - G_3 (Q) \frac{ \xi^*_{\lambda'}
\cdot q \,\, \xi_{\lambda} \cdot q}{2\, M_d^2} 
\Biggr ]\, (P' + P)^\mu \nonumber \\
&& -  G_2 (Q) \Bigl ( \xi^\mu_{\lambda}\,\,
\xi^*_{\lambda'} \cdot q  -
\xi^{\mu *}_{\lambda'} \,\, \xi_{\lambda} \cdot q
\Bigr ) \ , 
\label{invariantCurrent}
\end{eqnarray}
where the  four-momentum transfer is $q=P'-P$, and $\xi_{\lambda}^\mu=\xi_{\lambda}^\mu(P)$ 
[$\xi_{\lambda'}^\mu=\xi_{\lambda'}^\mu(P')$]  are  the initial (final) deuteron helicity
four-vectors.  It then follows that
\begin{eqnarray}
\!\!\!\!\!\!\!\!\!\!\!\!&&g_{-1}=\sqrt{1+\eta_d}\; G_1(Q) \ ,\\
\!\!\!\!\!\!\!\!\!\!\!\!&&g_{0}=\sqrt{1+\eta_d}\;\left[(1+2\eta_d)G_1(Q) -2\,\eta_d \,G_2(Q)
+2\,\eta_d(1+\eta_d)\,G_3(Q)\right] \ ,\\
\!\!\!\!\!\!\!\!\!\!\!\!&&g_{+1}=\sqrt{\eta_d(1+\eta_d)}\;G_2(Q)\ .
\end{eqnarray}
The charge, magnetic, and quadrupole form factors are related to the
invariant functions $G_i(Q)$'s by
\begin{eqnarray}
G_C(Q) & = & G_1(Q) + {2 \over 3} \eta_d \, G_Q(Q)\ , \\
G_M(Q) & = & G_2(Q)\ ,  \\
G_Q(Q) & = & G_1(Q) - G_2(Q) + (1 + \eta_d)\, G_3(Q)\ ,
\end{eqnarray}
or directly to the matrix elements $g_\lambda$'s by
\begin{eqnarray}
G_C(Q)&=&\frac{1}{3\,\sqrt{1+\eta_d}}\left(g_0+2\, g_{-1}\right)\ , 
\label{eq:ggcc}\\
G_M(Q)&=&\frac{1}{\sqrt{\eta_d(1+\eta_d)}}\,g_{+1} \ , 
\label{eq:ggmm}\\
G_Q(Q)&=&\frac{1}{2\,\eta_d\sqrt{1+\eta_d}}\left(g_0-g_{-1}\right)\ ,
\label{eq:ggqq}
\end{eqnarray}
and are normalized to
\begin{equation}
G_C(0)=1\ ,\>\>\>\>
G_M(0)=\frac{M_d}{m}\, \mu_d \ ,\>\>\>\>
G_Q(0)=M^2_d \, Q_d\ ,
\label{eq:gcmq_norm}
\end{equation}
where $m$ is the nucleon mass, and $\mu_d$ and $Q_d$ are respectively
the magnetic moment (in units of nuclear magnetons) and the quadrupole
moment.   In some of the calculations carried out in the conventional
approach discussed in section~\ref{sec:res}, boost corrections in the
initial and final states are ignored, and hence the factor $1/\sqrt{1+\eta_d}$
on the right-hand-side of equations~(\ref{eq:ggcc})--(\ref{eq:ggqq}) 
is not included.

Electron elastic scattering cross sections corresponding to a spin 0 target, like $^4$He,
or spin 1/2 target, like $^3$He/$^3$H, are well known~\cite{Waleckabook}, and will
not be given here.  The former can be expressed in terms of a single charge form
factor, while the latter involves a charge and a magnetic form factor.  Ignoring again boost
corrections, the $^4$He charge form factor is 
\begin{equation}
F_C(Q)=\frac{1}{Z}\, \left< 0 \left| \rho({\bf q})\right| 0 \right> \ ,
\label{eq:fc_4}
\end{equation}
while the trinucleon charge and magnetic form factors follow from
\begin{eqnarray}
F_{C}(Q)&=&\frac{1}{Z}\, \left< 1/2,{+}\left| \rho({\bf q}) \right| 1/2,{+}\right> 
= F_1(Q)-\frac{Q^2}{4M_T^2}F_2(Q)\ ,
\label{eq:cff} \\
F_{M}(Q)&=&\frac{2 \, m}{\mu}\,
 \left< 1/2,+\left|j_{x} ({\bf q}) \right| 1/2,-\right>
= \frac1{\mu}\big[F_1(Q)+F_2(Q)\big]\ ,
\label{eq:mff}
\end{eqnarray}
where $\left|0\right>$ represents the $^4$He ground state,
$\left| 1/2,\pm \right>$ represent the trinucleon ground states with
spin projections $\pm 1/2$ along the direction of the momentum
transfer ${\bf q}$,
and $F_1$ and $F_2$ are the three-body 
Dirac and Pauli form factors that appear in the relativistic current 
for a target of mass $M_T$.
The time and space parts of the four current $J^\mu$
introduced earlier are denoted respectively as $\rho$ and ${\bf j}$.
In equations~(\ref{eq:fc_4}) and~(\ref{eq:cff})--(\ref{eq:mff}) $Z$ is the
proton number, $\mu$ the trinucleon magnetic moment in
units of nuclear magnetons, and $m$ is the nucleon mass.
Hence, the $F_C(Q)$ and $F_M(Q)$ form factors are normalized to 
\begin{equation}
F_C(0)=F_M(0)=1 \ .
\end{equation}
Below we will also consider the isoscalar and isovector
combinations of the trinucleon charge and magnetic form
factors, defined as (suppressing the $Q$-dependence
for simplicity)
\begin{eqnarray}
F_C^{S/V}&=&\frac{1}{2}
\left[2\, F_C(^3{\rm He})\pm F_C(^3{\rm H})\right] \ ,
\label{eq:fcsv} \\
F_M^{S/V}&=&\frac{1}{2}
\left[\mu(^3{\rm He})F_M(^3{\rm He})\pm \mu(^3{\rm H})F_M(^3{\rm H}) \right] \ .
\label{eq:fmsv} 
\end{eqnarray}
If the $^3$H and $^3$He ground states were pure isospin $T=1/2$ states,
then $F_C^S,F_M^S$ and $F_C^V,F_M^V$ would only be affected
by, respectively, the isoscalar ($S$) and isovector ($V$) components
of the current.  However, small isospin admixtures with $T>1/2$,
induced by the electromagnetic interactions as well as charge-symmetry 
breaking terms in the strong interactions, are included in trinucleon
wave functions calculated in the conventional and $\chi$EFT approaches
discussed below.  As a consequence, isoscalar (isovector) currents 
give non-vanishing, albeit small, contributions to the isovector
(isoscalar) form factors.

\section{Overview of the world experimental data}
\label{sec:exp}
\subsection{Determination of form factors} 
\label{subsec:ff}
For the light nuclei of interest to this review, an extensive set of data
covering a large range of electron energies $\varepsilon$ 
and scattering angles $\theta$
is available. Such a set  for various reasons is not very useful for a
direct comparison to theory. Several considerations need to be 
re-emphasized:
(i) for most nuclei  the cross sections depend on two form factors, 
related to the Coulomb
monopole $C_0$ ($G_C$ for the deuteron or $F_C$ for the three- and 
four-body nuclei) and magnetic dipole $M_1$ ($G_M$ or $F_M$). 
For the case of the deuteron a third
form factor, due to the Coulomb quadrupole operator $C_2$ ($G_Q$), 
is contributing as well. As compared to the
cross sections, the individual form factors are much more sensitive to the
ingredients of the theoretical calculations. It is therefore highly desirable to
extract from the experimental cross sections the individual form factors. 
(ii) For the $A=3$ nuclei ($J_i=1/2$), the charge and magnetic form factors 
have traditionally been determined by the
authors who performed the individual experiments. For the 
deuteron case ($J_i=1$) the data
have mainly been discussed in terms of the $A(Q)$ and $B(Q)$ 
structure functions and the observable $T_{20}(Q)$, defined in 
equations~(\ref{eq:aq}),~(\ref{eq:bq}), and~(\ref{eq:t20}).

To get the most precise form factors from the {\em world} data,
it is necessary to re-analyze all  cross sections and polarization observables 
at a given momentum transfer, as only  the combination  provides an optimal
separation of the form factors. Furthermore, 
it is only with such a separation using the
{\em world} data that a reliable error bar of the form factors can be derived. 

Re-analysis of the {\em world} cross sections is required for another reason:
the
step from cross section to form factor should involve removal of the Coulomb
distortion of the electron waves, so that the form factors can be compared to
the one-photon exchange results from theory.  
This step, mainly important at low 
momentum transfer, has been
omitted in most of the past analyses determining form factors. In the re-analysis
of the {\em world} data the Coulomb corrections, which depend on both 
$\varepsilon$ and
$\theta$, can be accounted for. They are done using the approach described in
Ref.~\cite{Sick98}. 

A further complication arises from the fact that the cross sections and
analyzing powers have been measured at a variety of energies and angles. In
general even individual experiments aiming at a longitudinal/transverse (L/T)
separation have not achieved exactly the same momentum transfers at forward and
backward angles, so non-transparent interpolations/extrapolations are necessary;
this difficulty is even more serious  when combining data from different
experiments.

 In order to get the most precise information, we proceed as follows: starting
from the experimental {\em world} cross sections data, we  correct them for
Coulomb effects (and, if desired, the complete set of two-photon exchange
corrections)  and fit the resulting one-photon exchange cross sections using a very flexible
parameterization for the two (three) form factors of interest. For any interval
of momentum transfer, this  provides the most complete   information on both the
separated form factors and their uncertainty.

In this fitting procedure, the statistical uncertainties of the data can be
included in the standard way using standard error propagation and the error
matrix. For the determination of the systematic errors, we employ a rather
conservative approach: the data from individual experiments are changed by the
quoted systematic error. Then the resulting changes of the fit form factors are
quadratically added and quadratically combined with the statistical
uncertainties. 

While the results from such a fit do provide the best experimental information,
one peculiarity needs to be understood: the values of the form factors at
closely spaced values of momentum transfer are {\em not} independent. When
fitting the cross sections with parameterized form factors, the correlation
extends over an interval $\Delta Q \sim 1/R_{max}$, where $R_{max}$ is the
maximal radius at which the parameterization in radial space --- the Fourier
transform  of the parameterization  in 
momentum space --- allows for a somewhat free
variation of the density.  For the nuclei of interest here $\Delta Q \sim$ 0.2
\fm. While the numerical values of the extracted form factors no longer show
the statistical fluctuations of the data,   the error bars  do account for  both
the random and the systematic errors of the data. These error bars  are much
more quantitative  than the fluctuations of the usual  form  factors, which in
general are taken as a visual ``measure'' of the accuracy of the form factors. 

In order to determine the form factors  we have fitted the {\em world} data 
using the highly flexible sum-of-gaussians (SOG) parameterization~\cite{Sick74}.
The main restriction, introduced by the SOG in $r$-space,  is the rms-radius
of the gaussians, which is chosen to be well below the rms-radius of the
proton.  The resulting form factors and their error bars are shown in the
figures of section~\ref{sec:res}.

The approach described above may seem 'unconventional', but it is perfectly
analogous to the procedure employed since decades in nucleon-nucleon ($NN$)
scattering.  The $NN$ data (cross sections and analyzing powers) 
are no longer
used directly, but the information is condensed in  the phase-shifts
extracted from a global fit to the data. The procedure we employ corresponds to
the standard energy-dependent phase-shift analyses \cite{Stoks93}. There
are however two main
differences: 
(i) we do {\em not} prune the data, {\em i.e.} eliminate some 30\% of
the data in order to get the $\chi^2$/datum down to $\sim 1$~\cite{Stoks93};
(ii) we do take into account the absolute normalization of the data, 
rather than
floating the cross sections in order to produce a $\chi^2$/datum$\sim 1$. 

The fits to the {\em world} data also provide the most accurate values for
integral moments such as rms-radii and the  Zemach 
moments~\cite{Friar04,Friar05} of interest 
for the interpretation of transition energies in atomic nuclei.  The radii are also
quoted in section~\ref{sec:res}.
The moments in general have been obtained by constraining the {\em shape} of the
large-$r$ density to the one expected from our understanding of nuclear wave
functions; at large $r$ they must fall like a Whittaker function depending on
the nucleon separation energy. This constraint is needed as the determination of
{\em e.g.} the rms-radius involves an extrapolation from the $Q$-region
sensitive to finite size --- typically 
0.5 fm$^{-1}$$\leq Q \leq 1.2$ fm$^{-1}$ --- 
to $Q=0$
where the radii are extracted. The difficulties of this (implicit) extrapolation
are discussed in Ref.~\cite{Sick14}.

\subsection{Two-photon corrections} 
\label{subsec:tphcorr}

During the last years, it has become clear that two-photon exchange beyond the
one accounted for via the Coulomb corrections (exchange of a soft photon in
addition to hard photon responsible for the scattering) can contribute. At 
large momentum transfer the exchange of {\em two} hard photons can become
important, as was shown in particular for the proton, where the two-photon
corrections resolved the discrepancy between L/T-separation and polarization
transfer results for the proton charge form factor. For a review see
Ref.~\cite{Arrington11}. 

Two-photon exchange is difficult to calculate due to the complexities introduced
by the intermediary states, which can be the nucleus of interest in its ground
state or in any excited state.  Intermediary excitation of individual nucleons
can also contribute.  For the deuteron, calculations are available by Kobushkin
{\it et al.}~\cite{Kobushkin11}  and 
Dong {\it et al.}~\cite{Dong06,Dong09a,Dong09b}.  While
Dong {\it et al.} 
calculated the contribution where the exchange takes place with only
one nucleon, Kobushkin {\it et al.} assume that the exchange involves both nucleons. 
For \hyt , a calculation is available from Kobushkin and 
Timoshenko~\cite{Kobushkin13}. 
For the deuteron, the calculated two-photon exchange contributions are of 
the order of 10 \% of the typical differences between theory and experiment,
for $A=3$ they can, at selected places, amount to up to 30 \%.
Since the contributions of two-photon exchange at the present time
are still quite uncertain, we will not include them.

\subsection{Experiments on deuteron} 
\label{subsec:a2}

A large number of experiments on elastic electron-deuteron scattering have been 
performed since $\sim$1960, producing a total of some 500 data 
points~\cite{Abbott99}\nocite{Alexa99,Akimov79}
\nocite{Arnold75,Arnold87,Auffret85a,Benaksas66,Berard73a,Bosted90}
\nocite{Bumiller70,Buchanan65,Cramer85,Drickey62}
\nocite{Elias69,Friedman60,Galster71,Ganichot72,Goldemberg64}
\nocite{Honegger97,Grossetete66b}
\nocite{Martin77,Platchkov90,Rand67,Simon81,Stein66}-~\cite{Voitsekhovskii86}.
We discuss below only a selection which was particularly important in fixing the
deuteron form factors. For the analysis of the data in terms of form factors 
{\em all} data will be included.

The most accurate data at low momentum transfers come from the experiments of 
Simon {\it et al.}~\cite{Simon81} performed at Mainz, in the range
$Q=0.2-2$ \fm, the experiment of 
Platchkov {\it et al.}~\cite{Platchkov90} performed at the Saclay ALS, 
with $Q=1.2-4.2$ \fm, and the experiment of 
Berard {\it et al.}~\cite{Berard73a}  carried out at
the Monterey accelerator, with $Q=0.2-0.7$ \fm. These experiments
reached accuracies of the order of 1\%, employing 
liquid  deuterium targets and  high-resolution magnetic spectrometers for the
electron detection. In some 
special cases gas targets of better known thickness and
fixed-angle spectrometers of well controlled solid angle were used as a
supplement to achieve better accuracy on {\em absolute} cross sections.

Data at medium momentum transfers come, among others, from two experiments
carried out at JLab by    
Alexa {\it et al.}~\cite{Alexa99}, for $Q=4.2-12.2$ \fm, 
and Abbott {\it et al.}~\cite{Abbott99},
for $Q=4-6.6$ \fm. 
Although these experiments quote rather small errors, the data are
not entirely consistent; we suspect that, in the early days of JLab operation,
the beam energies were not very accurately known. A more recent JLab experiment
of Bosted {\it et al.}~\cite{Bosted13}, in the range 
$Q=3.7-5.7$ \fm, could not resolve the difference in
favor of one of the experiments. 

Data up to the highest momentum transfers 
were also measured at SLAC by Arnold {\it et al.}~\cite{Arnold75}, 
in the range $Q=4.5-10.1$ \fm. 
This experiment, as the ones performed at
JLab, detected the scattered electron and recoil deuteron in coincidence, such
as to cleanly identify {\em elastic} scattering despite insufficient energy
resolution. 

An important class of experiments provided the backward-angle data needed to
determine the magnetic form factor. Although these experiments provided less
accurate cross sections, they are most valuable because they are generally
totally dominated by $G_M(Q)$ and thus less dependent on the (error-enhancing)
L/T-separation.  The experiments of Benaksas {\it et al.}~\cite{Benaksas66},
with $Q=1.7-2.2$ \fm, 
and Ganichot {\it et al.}~\cite{Ganichot72}, with $Q=0.7-2.4$ \fm, were
carried out at 180$^\circ$ scattering angle at the Orsay accelerator, the data
of Rand {\it et al.}~\cite{Rand67}, with $Q=2.2-3.1$ \fm, come 
from a 180$^\circ$ experiment
done at the Stanford HEPL machine. Precise large-angle data were also provided
by the experiment of Auffret {\it et al.}~\cite{Auffret85a}, 
with $Q=2.4-4.2$ \fm, with
a scattering angle of 155$^\circ$ at the Saclay ALS.

A special class of data, only recently  accessible to experiment, involves
measurement of  tensor polarization observables, accessible with deuterons with
spin aligned in the direction of the momentum transfer. When working with a
tensor-polarized deuteron target, or when measuring the tensor polarization of
the recoiling deuteron, the quantity $T_{20}(Q)$ 
can be obtained (the other tensor
observables $T_{21}(Q)$ and $T_{22}(Q)$  are not that useful). 
With the knowledge of
$T_{20}(Q)$, 
which basically depends on the $G_Q(Q)/G_C(Q)$ ratio
[see equation~(\ref{eq:t20})],  it becomes possible to
separate the form factors $G_{C}(Q)$ and $G_{Q}(Q)$ 
which cannot be separated via
cross-section measurements alone. In the $Q$-range of interest here, 
$G_{C}(Q)$ is
of particular interest due to the presence of a diffraction feature, which is
very sensitive to the ingredients of the theoretical calculations. 

Measurements of $T_{20}(Q)$ have been performed at various 
laboratories~\cite{Schulze84} -\nocite{Dimitriev85,Gilman90,The91,Ferro96,Bouwhuis99,Abbott00a,Zhang11,Garcon94}~\cite{Nikolenko03}.  
They have been performed both at storage rings using internal, polarized
deuteron atomic beams, or with external electron beams using polarimeters to
detect the recoil tensor polarization. The most extensive set of data, measured
by Abbott {\it et al.} at JLab~\cite{Abbott00a}, comes from an experiment that used a polarimeter based
on the analyzing power of the $p(\vec{d},2p)n$ reaction which had been
calibrated using a polarized deuteron beam from the Saturne accelerator. Today,
data are available up to momentum transfers of 6.6 \fm.                                                              

For the region $Q<7$ fm$^{-1}$, where the $G_C/G_M$-separation 
can be performed, the
data base comprises some 492 data points. The fit yields a $\chi^2$ of 549
when taking into account only the statistical errors of the data.

\begin{figure}[t]
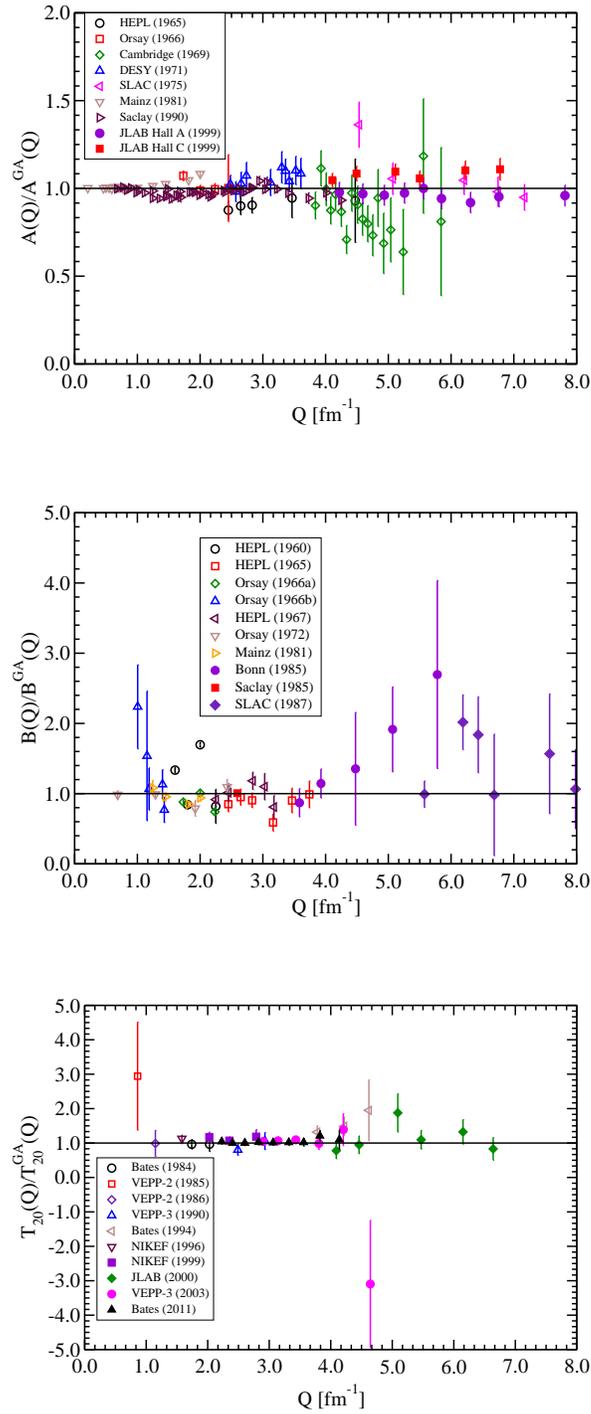

\vspace*{0.5cm}
\begin{minipage}{1\textwidth}
  \begin{center}
    \includegraphics[width=7.7cm]{A_ratio.eps}
  \end{center}
\end{minipage}
\begin{minipage}{1\textwidth}
  \begin{center}
\vspace*{1cm}
       \includegraphics[width=7.7cm]{B_ratio.eps}
  \end{center}
\end{minipage}
\begin{minipage}{1\textwidth}
  \begin{center}
\vspace*{1cm}
    \includegraphics[width=7.7cm]{T20_ratio.eps}
  \end{center}
\end{minipage}
\caption{(Color online)
Ratio of the structure functions  $A(Q)$, $B(Q)$, $T_{20}(Q)$  
as measured by the different experiments to the presented
global analysis (GA). The different labels to identify the experiments 
correspond in order, for $A(Q)$, to 
Refs.~\protect\cite{Buchanan65,Benaksas66,Elias69,Galster71,Arnold75,Simon81,Platchkov90,Alexa99,Abbott99}, for $B(Q)$ to
Refs.~\protect\cite{Friedman60,Buchanan65,Benaksas66,Grossetete66b,Rand67,Ganichot72,Simon81,Cramer85,Auffret85a,Arnold87}, and for $T_{20}(Q)$ to
Refs.~\protect\cite{Schulze84,Dimitriev85,Voitsekhovskii86,Gilman90,The91,Garcon94,Ferro96,Bouwhuis99,Abbott00a,Nikolenko03,Zhang11}. Note that
Refs.~\protect\cite{The91} and~\protect\cite{Garcon94} refer both to
the label ``Bates (1994)''.}
\label{fig:ratios} 
\end{figure}

To make a connection between the present global analysis and
the previous measurements discussed above, 
we show in  figure~\ref{fig:ratios} 
the ratio of the experimental values for the usual
structure functions $A(Q)$, $B(Q)$, $T_{20}(Q)$ existing in the literature
to the ones obtained in the present global analysis. 
Note, however, that in our global analysis
we use roughly twice as many data points, since there are many data available
that have not led to a  determination of the deuteron structure functions. 
It also should be
noted that the structure functions from the literature have been derived {\em
without} accounting for Coulomb distortion. 

\subsection{Experiments for $^3$H} 
\label{subsec:h3}

The \hyt~ nucleus represents a particular challenge due to its
radioactive nature (with a half-life of 12 years); experiments
involve up to $10^5$ Curies of material. This requires strict safety measures,
particularly since  high-intensity  beams as  available at modern accelerators can
easily melt a hole into the windows of the  target container.

 The earliest experiment was performed at the Stanford HEPL accelerator by
Collard {\it et al.}~\cite{Collard65} using a high-pressure (100 bar) gas-target. This
experiment reached $Q=2.8$ \fm. 
Another experiment with a gas target was performed
about 20 years later at the MIT/Bates accelerator. This experiment, 
by Beck {\it et al.}~\cite{Beck87}, used tritium stored in an Uranium-oven; 
in the target the
tritium was cooled down to 45 K, 
thus allowing to run at a more modest pressure
of 15 bar. This experiment reached a similar maximal momentum transfer
as the previous one, but 
better statistical accuracy due to the achieved higher luminosity. 

The most extensive set of data comes from the experiment performed at the Saclay
ALS accelerator by Amroun {\it et al.}~\cite{Amroun94,Juster85}. In order to reach a
much higher luminosity, this experiment used {\em liquid} tritium, cooled to a
temperature of 20 K. With a novel target system --- cooled 5 cm long 
cylindrical
target plus warm storage vessel of 0.2 liter volume ---   
a permanently sealed
system could be employed with most of the tritium (98\%) in the target at a
pressure as low as  3 bar when the target was in operation.  This led to the
highest luminosity with only 10$^4$ Curies of tritium. 
This experiment reached a maximum
momentum transfer of $Q=5$ \fm~ 
and covered the region of the diffraction minimum
and maximum expected from the then already known data on \het. 

All three experiments provided data at both forward and backward angles,
collecting about 190 data points, so that 
a Rosenbluth separation into charge and magnetic
form factors is possible. This separation has been performed as described above.
The $\chi^2$ of the SOG-fit is $\sim$340, mainly due to a difference in
normalization of the Saclay/Bates data sets. 
 
\subsection{Experiments on $^3$He}
\label{subsec:he3}

A good target for \het~ represents a lesser challenge than an \hyt -target,
although still a major effort is needed  to reach an adequate target thickness
and luminosity.  

Data at low momentum transfers have been measured by 
Szalata {\it et al.}~\cite{Szalata77} using the NBS accelerator ($Q<0.6$ \fm ), 
vonGunten~\cite{vonGunten82} at the Darmstadt machine, 
and Ottermann {\it et al.}~\cite{Ottermann85}
at Mainz ($Q<1.9$ \fm ).  
The experiment of Dunn {\it et al.}~\cite{Dunn83}, performed at
the Bates laboratory, provided data up to $Q\sim$3.3 \fm .

The region of the predicted diffraction minimum and maximum of the charge form
factor was reached by the experiment of 
McCarthy {\it et al.}~\cite{McCarthy70,McCarthy77}. Contrary to all other experiments, 
McCarthy {\it et al.}
used a {\em liquid} \het -target, cooled by superfluid \hef. This produced a 
much higher target thickness and luminosity, and allowed to measure very small
cross sections, leading to a maximum momentum transfer of 4.5 \fm.  Use of
liquid \het~ came at the expense of a rather complicated target setup.




\begin{figure}
\begin{center}
\includegraphics[scale=0.55,clip]{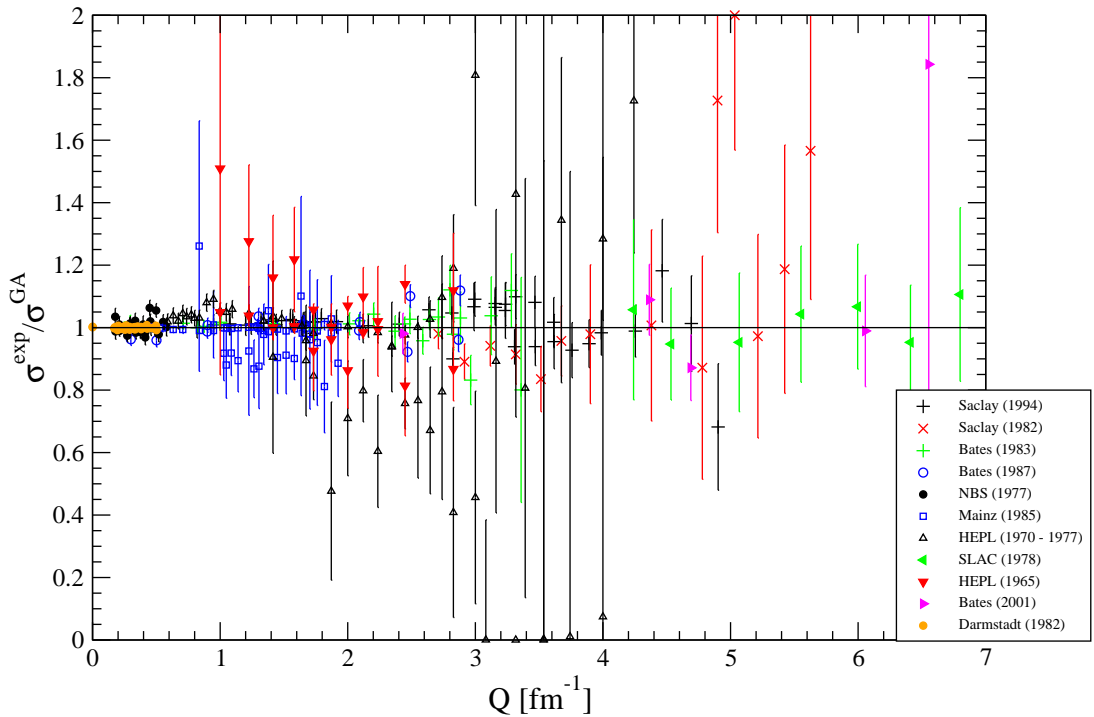}
\caption{(Color online) 
Ratio of experimental cross section over the one obtained
by the present global analysis for $^3$He. 
The different labels to identify the experiments 
correspond in order to 
Refs.~\protect\cite{Amroun94,Cavedon82,Dunn83,Beck87,Szalata77,Ottermann85,McCarthy70,McCarthy77,Arnold78,Collard65,Nakagawa01,vonGunten82}.}
\label{he3012.rat.ps}
\end{center}  
\end{figure}

The diffraction feature in the magnetic form factor was
observed in the experiment of Cavedon {\it et al.}~\cite{Cavedon82} performed at the
Saclay ALS accelerator. Additional backward-angle points were measured at
160$^\circ$ at Bates by Nakagawa {\it et al.}~\cite{Nakagawa01}. With this experiment,
the data on the magnetic form factor now reach a maximum transfer of  
6.5 \fm .

Data up to the highest momentum transfers, $Q \sim$10 \fm , were provided by
Arnold {\it et al.}~\cite{Arnold78} with an experiment carried out at SLAC at
forward scattering angles. With a high-pressure gas target cooled by liquid
hydrogen, this experiment could reach the extremely small form factors near the
second diffraction minimum. Due to the limited energy resolution of the
spectrometers, scattered electron and recoil \het\ had to be detected in
coincidence in order to cleanly identify elastic scattering. 

The data set for
\het\ comprises some 310 data points; the fit has a $\chi^2$ of 347 when ignoring
the systematic errors of the data. 
The ratio of experimental cross section over the one obtained by
the present global analysis for $Q<7$ fm$^{-1}$ is
shown in figure~\ref{he3012.rat.ps}.  Note that an experiment going to
similarly large $Q$ values has been performed at JLab, but the data have not
yet been published.

For the $A=3$ system, it is of interest to study, 
besides the form factors of the
individual nuclei, the isoscalar and isovector 
combinations as well. In particular, the isoscalar charge
form factor, as defined in equation~(\ref{eq:fcsv}), 
is more directly comparable to the charge
form factors of deuteron and \hef. These isoscalar  and isovector
combinations can be easily calculated, together with their uncertainties,  from
the form factors fitted to the experimental data. 

\subsection{Experiments on $^4$He}
\label{subsec:he4}

This nucleus is particularly tightly bound, resulting in the largest density in
the inner region. This property makes the form factors of 
$^4$He especially
interesting. Due to its $J_i=0$ nature, the determination of the form factor is
easier than for all other light nuclei, and a total of 190 data points are now
available.

Experiments at low momentum transfers have been carried out   by 
vonGunten~\cite{vonGunten82} and Erich {\it et al.}~\cite{Erich68} at Darmstadt, 
and by Ottermann {\it et al.}~\cite{Ottermann85} at Mainz, the latter reaching 
$Q\sim$2 \fm. The data at
medium momentum transfers come from two experiments performed at the Stanford
HEPL machine by Frosch {\it et al.}~\cite{Frosch67} and McCarthy {\it et al.}~\cite{McCarthy77}.
These experiments, performed using liquid targets,  covered the region up to   
$Q=$4.5 \fm\ and thus the region of the first diffraction minimum and maximum. 

The highest-$Q$ data were measured  by Arnold {\it et al.}~\cite{Arnold78} at
SLAC, reaching $Q=7.9$ \fm . The recent publication by  Camsonne {\it et
al.}~\cite{Camsonne13} provided data up to $Q=8.8$ \fm .  Profiting from the
high luminosity achievable at JLab,  this experiment covered the region of the
second diffraction feature.   In the region of overlap with the data of
Ref.~\cite{Arnold78} some disagreement is visible, whose origin, at the present
time,  has not been cleared up. 

\section{Theoretical approaches}
\label{sec:theory}

In this section we discuss the different theoretical frameworks
adopted to study $A=2-4$ electromagnetic form factors. Section~\ref{subsec:conv} reviews
the conventional approach, which uses phenomenological
realistic models for the nuclear interactions
and currents, section~\ref{subsec:chieft} reviews  the chiral effective field
theory approach, in which both currents and interactions are 
consistently derived from chiral Lagrangians, and section~\ref{subsec:cst} reviews results from the relativistic 
covariant spectator theory, 
where (using the language of covariant field theory) currents  consistent with the relativistic dynamics are constructed.  Finally, section~\ref{subsec:comp} presents
an appraisal and comparison of these various methods.

\subsection{The conventional approach}
\label{subsec:conv}
The conventional approach views the nucleus as made up of nucleons
interacting among themselves via two- and many-body potentials, and
with external electroweak fields via one- and many-body currents.  It
assumes that all other sub-nucleonic degrees of freedom, involving, for
example, the excitation of nucleon resonances such as $\Delta$ isobars,
can be eliminated in favor of these effective potentials and currents, acting
on nucleon coordinates.  
(For the size of effects of single $\Delta$-isobar excitations in two- and three-nucleon form factors see, for instance, the pioneering work by Sauer and collaborators \cite{Strueve:1987ty}\nocite{Pena:1990jk,Henning:1992qv}
-\cite{Henning:1995fk}.)
The validity of this greatly simplified description,
in which color-carrying quarks and gluons (the degrees of freedom of
quantum chromodynamics, QCD) are assembled into colorless clusters
(the nucleons), and these clusters are taken as effective constituents
of the nucleus,  
ultimately rests on the success it has achieved in the
quantitative description of many nuclear properties.

\subsubsection{Nuclear potentials}

In the current version of the conventional approach, the nuclear Hamiltonian
is taken to consist of a non-relativistic kinetic energy term and of two- and three-body
potential energy terms, 
\begin{equation}
  H=\sum_i {p_i^2\over 2m}+\sum_{i<j} v_{ij} + \sum_{i<j<k}
     V_{ijk}\ .\label{eq:nucH}
\end{equation}
The kinetic energy term is predominantly charge-independent, though it
has a small charge-symmetry-breaking (CSB) component due to the
difference between proton and neutron masses which leads to a
very small correction to nuclear energies~\cite{Nog03}.  The two-body
potential consists of a long-range part, for inter-nucleon separation
$r \gtrsim 2$ fm, due to one-pion exchange (OPE), and intermediate-
and short-range parts, for, respectively, 1 fm $\lesssim r \lesssim 2$ fm
and $r\lesssim 1$ fm, which are derived from theoretical arguments
and are constrained by fits to nucleon-nucleon ($NN$) elastic scattering
data typically up to lab kinetic energies of 350 MeV, slightly
above the pion production threshold.  These potentials are
customarily referred to as ``realistic potentials'' in the literature.
In the case of the CD-Bonn (CDB) momentum-space (and strongly non-local)
potential~\cite{CDB}, the short- and intermediate-range parts are
parametrized in terms of $\rho$ and $\omega$ vector-meson exchanges
as well as the exchange of two effective scalar mesons, whose
masses are about 450 (350) MeV and 1220 (800) MeV, as determined
by fits to $pp$ data ($np$ data in isospin channel $T=0$).
In the case of the Argonne $v_{18}$ (AV18) configuration-space
potential~\cite{Wir95}, the intermediate range part is parametrized
in terms of two-pion exchange (TPE), based on, but not consistently
derived from, a field-theory analysis of (direct and crossed) box diagrams
with intermediate nucleons and $\Delta$ isobars~\cite{Pandha76},
while its short-range part is represented by spin-isospin (and momentum-dependent)
operators multiplied by Woods-Saxon radial functions.  This potential
is used in most of the results obtained in the conventional approach
and presented in later sections of this review; for details
see Ref.~\cite{Wir95}.

Modern realistic potentials contain isospin-breaking (IB) terms.  At the level
of accuracy required, electromagnetic interactions, along with strong interactions,
have to be specified in order to fit the data precisely.  These electromagnetic
interactions consist of one- and two-photon Coulomb terms as well as 
Darwin-Foldy, vacuum polarization, 
and magnetic moment contributions~\cite{Stoks90}.
The full potential $v_{ij}$ is then the sum of isospin-conserving strong-interaction terms $v^{\rm IC}_{ij}$, specified electromagnetic-interaction terms $v^\gamma_{ij}$ up to order
$\alpha^2$, where $\alpha$ is the fine structure constant, and finally additional
isospin-breaking strong-interaction terms $v^{\rm IB}_{ij}$.

In the context of the conventional framework described here,
it is an established fact that accurate calculations based on
realistic two-nucleon potentials find that the observed energy
spectra of light nuclei with mass number $A$ in the range
$3 \leq A \leq 12$ are under-predicted~\cite{Kie08,Wir01} and that the empirical
saturation density of nuclear matter is over-predicted 
by roughly a factor of 2~\cite{Akmal,DMC}.  In the specific case of the few-nucleon
systems of interest in this review, (essentially exact) hyperspherical-harmonics
calculations lead to triton binding energies ranging 
from about 7.6 MeV for a local potential
like the AV18 to about 8.0 MeV for a strongly non-local potential
like the CDB, compared with the experimental value of 8.48 MeV.  For $^4$He,
the binding energy ranges from 24.0 MeV for the AV18 to 26.3 MeV for
the CDB, compared to the experimental value of 28.3 MeV~\cite{Kie08}.

Several effects could be important in reproducing the binding energies
of nuclei. Two of them are immediately apparent: relativistic corrections
and three-nucleon interactions.  It has long been known that these effects
cannot be completely separated, {\it i.e.} that they are theoretically related
(for a brief discussion see Section~\ref{subsec:comp} below).
Furthermore, their contributions are comparable. In the $^4$He ground state,
for example the (non-relativistic) kinetic energy is about 100 MeV, and one
would expect 1--2 \% relativistic corrections of this value, amounting to 1--2 MeV. 
Three-nucleon potentials lead to corrections of similar size.  At long range,
the three-nucleon potential is of the well known Fujita-Miyazawa type~\cite{Fuj57},
corresponding to single-pion exchanges between three nucleons with intermediate
excitation of a $\Delta$-isobar resonance. 
In coupled-channel calculations of the three-nucleon bound state in which the 
$\Delta$-isobar is treated as an active degree of freedom, this three-nucleon 
force yields almost 1 MeV of additional binding energy, but it is to a large 
part cancelled by the dispersive contribution of intermediate 
$\Delta$-excitations to two-nucleon 
scattering~\cite{Hajduk:1983rm,Stadler:1992fk,Pena:1990jk}. 
In $^4$He,  the presence of this relatively low-lying
resonance produces  a three-nucleon potential whose contribution
is of roughly a few
MeV.

The 
 Fujita-Miyazawa three-nucleon potential has the
following structure
\begin{equation}
\fl V^{2\pi}_{ijk}=A_{2\pi} \left[
\left\{X_{ij}\, , \, X_{ik} \right\} \left\{ {\bm \tau}_i\cdot {\bm \tau}_j \, , \, {\bm \tau}_i \cdot {\bm \tau}_k \right\}
 +\frac{1}{4} \left[ X_{ij}\, , \, X_{ik} \right] 
\left[ {\bm \tau}_i\cdot {\bm \tau}_j \, , \, {\bm \tau}_i \cdot {\bm \tau}_k\right] \right]  \ ,
\end{equation}
where $X_{ij}=Y_\pi (r_{ij})\, {\bm \sigma}_i\cdot{\bm \sigma}_j+T_\pi(r_{ij})\, S_{ij}$, $\{X_{ij},X_{ik}\}$ ($[X_{ij},X_{ik}]$) is the anticommutator (commutator),
and the $Y_\pi(r)$ and $T_\pi(r)$ radial functions are those entering the OPE
two-nucleon potential.  In the approach developed in Ref.~\cite{Pud95} and
adopted in the calculations of few-nucleon form factors reported in this
review, the attractive three-nucleon potential above is supplemented by a purely
central short-range term, {\it i.e.},
\begin{equation}
\label{eq:tni9}
V_{ijk}=V^{2\pi}_{ijk}+V^R_{ijk}\ ,
 \qquad V^R_{ijk}=U_0 \sum_{\rm cyc} T^2_\pi(r_{ij}) \, T^2_\pi(r_{ik}) \ .
\end{equation}
The $V^R_{ijk}$ term is of two-pion-exchange range on each of the two
legs.  It is meant to simulate the dispersive effects that are
required when integrating out $\Delta$ degrees of freedom.  These
terms are repulsive and are taken to be independent of spin and
isospin.  The constant $A_{2\pi}$ in $V^{2\pi}_{ijk}$ and $U_0$ in
$V^R_{ijk}$, in combination with the AV18 two-nucleon potential,
are adjusted to reproduce the triton binding energy and to provide
additional repulsion in hypernetted-chain variational calculations
of nuclear matter near the equilibrium density.  The three-nucleon potential
in equation~(\ref{eq:tni9}) is denoted as the Urbana IX (UIX) model in what
is to follow, and AV18/UIX is used to denote the Hamiltonian including
the AV18 two-nucleon and UIX three-nucleon potentials.

Before moving on to a discussion of the nuclear electromagnetic current
operator, it should be pointed out that recent developments in quantum
Monte Carlo methods and, in particular, Green's function Monte Carlo
methods have made it possible to carry out essentially exact calculations of
the energy spectra of low-lying states of light (s- and p-shell) nuclei in the
mass range up to $A=12$ ($^{12}$C).  These calculations have exposed
the inadequacy of the AV18/UIX model to satisfactorily reproduce the observed
spectra of $A$=6--12 nuclei, and have led to the development of a new model
for the three-nucleon potential---the Illinois-7 (IL7) model~\cite{Pieper}.
The latter incorporates $V^{2\pi}_{ijk}$, but also includes terms involving
multi-pion exchanges and intermediate $\Delta$'s as well as a representation
of the short-range term $V^R_{ijk}$ which now retains isospin dependence.
It is characterized by three parameters, which have been determined, in
combination with the AV18, by fitting the low-lying states of $A$=3--10
nuclei.  The resulting AV18/IL7 Hamiltonian then leads to predictions of
several ground- and excited-state energies, including the $p-^3$He 
elastic scattering observables~\cite{Viv13} and
the $^{12}$C
ground- and Hoyle-state energies~\cite{Pieper}, 
in very good agreement with the
corresponding empirical values.  However, in the few-nucleon systems of
interest here, bound-state energies obtained with either the AV18/IL7 or
AV18/UIX Hamiltonian are not significantly different.

\subsubsection{Nuclear electromagnetic charge and current operators} 
\label{subsec:2bcurrent}
A fundamental aspect in the description of electromagnetic (and weak) processes
in nuclei is the derivation of a consistent set of nuclear electromagnetic (and weak)
currents.  The leading terms are expected to be those associated with the charges
and convection currents of the individual protons, and the spin-magnetization currents of the individual protons and neutrons, which
in configuration space are
\begin{eqnarray}
 \label{eq:chargenr}
\fl \rho_{i}({\bf q})&=&\left[ \frac{1}{\sqrt{1+Q^2/4m^2}} \, \epsilon_i(Q)-
 \frac{i}{4m^2} \left[2\, \mu_i(Q)-\epsilon_i(Q)\right]{\bf q}
 \cdot\left( {\bm \sigma}_i\times {\bf p}_i\right)\right]
  {\rm e}^{{\rm i} {\bf q}\cdot{\bf r}_i}\ , \\
\fl   {\bf j}_i({\bf q})&=&\frac{\epsilon_i(Q)}{2m}\,
\{{\bf p}_i\, ,\, {\rm e}^{{\rm i}{\bf q}\cdot{\bf r}_i}\}
-\frac{\rm i}{2m}\, \mu_i(Q)\, {\bf q}\times{\bm \sigma}_i\,
{\rm e}^{{\rm i}{\bf q}\cdot{\bf r}_i}\ ,
\label{eq:1bcurr}
\end{eqnarray}
and follow from a non-relativistic expansion of the covariant single-nucleon
current, including corrections to the operators up to order $Q^2/m^2$. Here
${\bf q}$ and $\omega$ are as 
defined in section~\ref{sec:intro}, 
${\bf p}_i$ is the momentum
operator of nucleon $i$ with its charge and magnetization
distributions described by the form factors $\epsilon_i(Q)$ and $\mu_i(Q)$,
 \begin{eqnarray}
  \epsilon_i(Q)&=&\frac{1}{2}\left[G_E^S(Q) 
 +G_E^V(Q)\, \tau_{i,z}\right]\ ,\label{eq:ei} \\
  \mu_i(Q)&=&\frac{1}{2}\left[G_M^S(Q) 
 +G_M^V(Q)\, \tau_{i,z}\right]\ ,
 \label{eq:mui}
\end{eqnarray}
where $G_E^S(Q)$ and $G_M^S(Q)$, and $G_E^V(Q)$ and $G_M^V(Q)$,
are, respectively, the  isoscalar electric and magnetic, and isovector electric
and magnetic, combinations of the proton and neutron form factors,
normalized as $G_E^S(0)=G_E^V(0)=1$,
$G_M^S(0)=\mu^S$, and $G_M^V(0)=\mu^V$, with $\mu^S$ and $\mu^V$
denoting the isoscalar and isovector combinations of the proton and neutron
magnetic moments, $\mu^S=0.88$ and $\mu^V=4.706$ in units of
nuclear magnetons.  These form factors are obtained from fits
to elastic electron scattering data off the proton and deuteron~\cite{Hyde}.

The current and charge operators given in equations~(\ref{eq:chargenr})
and~(\ref{eq:1bcurr}) are usually referred to as 
impulse approximation (IA).
There is ample evidence that these IA
operators are inadequate, especially
for the description of isovector currents.
This evidence comes from studies of a variety of photo- and
electron-nuclear observables at low and intermediate values of energy and
momentum transfers, especially in light nuclei ($A \leq 12$) for which essentially
exact calculations can be carried out.  Experimental data are poorly reproduced
in IA.  Well known examples are the, classic by now, 10\% underestimate
of the $n-p$ radiative capture cross section at thermal neutron energies, which
in fact provided the initial impetus to consider two-body terms in the nuclear current
operator~\cite{Ris85a}\nocite{Ris85b,Ris85c}-~\cite{Buc85}, the 15\% underestimate of the isovector magnetic moment
of the trinucleons, the large discrepancies between the experimental and calculated
charge and magnetic form factors of the hydrogen and helium isotopes,
particularly in the first diffraction region at momentum transfers in the range
of 
3--3.5 fm$^{-1}$, the large underestimate
of the 
$n-d$ and $n-^3$He radiative captures~\cite{Mar05,Gir10a},
and, finally, the significant underestimate, in some cases even of
40\%, of magnetic moments and $M_1$ radiative transition rates
in $A$=7--9 nuclei~\cite{Pastore}.

Many-body terms in the nuclear electromagnetic charge and current operators
arise quite naturally in a meson-exchange picture or when the excitations of nucleon
resonances, such as the $\Delta$ isobar, are taken into account. There is a very large
body of work dealing with the problem of constructing these electromagnetic many-body
operators from meson-exchange theory, and we defer to a number of 
reviews~\cite{Ris89,Che71} for
a summary of efforts along those lines.  Here we will describe an approach, originally
proposed by Riska~\cite{Ris85a,Ris85b,Ris85c}, that leads to conserved currents, even in the
presence of two- and three-nucleon potentials, not necessarily derived from
meson-exchange mechanisms (as is the case for the AV18 and UIX models).
This approach has been consistently used by the ANL/JLab/LANL/Pisa group 
to study many electromagnetic processes
in light nuclei (up to $^{12}$C),
and has proved to be quite successful in providing predictions 
systematically in close agreement
with experiment.

The dominant part of any realistic $NN$ potential consists of
static (momentum-independent) terms, and leading electromagnetic
two-body charge and current operators are derived from these terms,
specifically the isospin-dependent central, spin, and tensor components.
The latter are assumed to be due to exchanges of effective pseudo-scalar
(PS or $\pi$-like) and vector (V or $\rho$-like) mesons, and the corresponding charge
and current operators are constructed from non-relativistic reductions
of Feynman amplitudes with the $\pi$-like and $\rho$-like effective propagators.
For the $\pi$-like case (we defer to Ref.~\cite{Car98} for a complete listing),
they read 
\begin{eqnarray}
\fl{\bf j}_{ij}^{PS}({\bf k}_i,{\bf k}_j)
&=&i\,G_{E}^{V}(Q)
   ({\bm\tau}_i \times {\bm\tau}_j)_z \, v_{PS}(k_j)\, 
   \bigg[  {\bm\sigma}_i 
 -{ {\bf k}_i - {\bf k}_j \over k_i^2 -k_j^2 }\,
  {\bm\sigma}_i \cdot {\bf k}_i \bigg] {\bm\sigma}_j \cdot {\bf k}_j + i \rightleftharpoons j \ ,
  \label{eq:jps} \\
\fl\rho_{ij}^{PS}({\bf k}_i,{\bf k}_j)&=& \left[ F_1^S(Q)   \, {\bm \tau}_i \cdot {\bm \tau}_j 
+F_1^V(Q) \,\tau_{z,j} \right] \frac{v_{PS}(k_j)}{2\,m}
 {\bm \sigma}_i \cdot {\bf q} \,\,   {\bm \sigma}_j \cdot {\bf k}_j  + i \rightleftharpoons j \ ,
\label{eq68a}  
\end{eqnarray}
where ${\bf k}_i$ and ${\bf k}_j$ are the fractional momenta delivered to 
nucleons $i$ and $j$, with ${\bf q}={\bf k}_i+{\bf k}_j$, and
$v_{PS}(k)$ is projected out of the (isospin-dependent) spin
and tensor components of the potential (see
equation (2.26) of Ref.~\cite{Mar05}). The
Dirac nucleon electromagnetic form factors $F^{S/V}_1$ are related
to those introduced previously via $F_1^{S/V}=\left(G_E^{S/V} +\eta \, G_M^{S/V}\right)/(1+\eta)$ with $\eta=Q^2/(4\,m^2)$, and therefore differ from $G^{S/V}_E$ by relativistic
corrections proportional to $\eta$. By construction, the longitudinal components
of the resulting ${\bf j}_{ij}^{PS}$ and ${\bf j}_{ij}^V$ currents satisfy current
conservation with the static part of the AV18, denoted $v_{ij}({\rm static})$ below,
\begin{equation}
{\bf q}\cdot \left[ {\, \bf j}_{PS}({\bf q})+{\bf j}_V({\bf q})\right]= [\, v_{ij}({\rm static})\, ,\, \rho_{i}({\bf q})
   +\rho_{j}({\bf q})\, ]
 \ . \label{eq:ccr2} 
 \end{equation}
Hence the use in equation~(\ref{eq:jps}) of the form factor $G_E^V(Q)$
entering $\rho_i({\bf q})$.  Of course, the continuity equation poses
no restrictions on transverse components of the current, in particular
on electromagnetic hadronic form factors that may be used
in these components.  Ignoring this ambiguity, the choice $G_E^V$
is made here for both longitudinal {\it and} transverse components.

Additional conserved currents follow from minimal substitution in the
momentum dependent part of $v_{ij}$, denoted as
$v_{ij}({\rm non}$-static). This momentum dependence
enters explicitly via the spin-orbit, quadratic orbital angular momentum,
and quadratic spin-orbit operators, and implicitly via ${\bm \tau}_i\cdot {\bm \tau}_j$,
which for two nucleons can be expressed in terms of space- and
spin-exchange operators as in
\begin{equation}
  {\bm \tau}_i\cdot{\bm \tau}_j = -1-
  (1+{\bm \sigma}_i\cdot{\bm \sigma}_j) \, {\rm e}^{-i\, {\bf r}_{ij}\cdot\left( {\bf p}_i
  - {\bf p}_j\right)} \ .
  \label{eq:tt}
\end{equation}
Both the explicit and implicit (via ${\bm \tau}_i\cdot {\bm \tau}_j$) momentum-dependent 
terms need to be gauged in order to construct exactly conserved currents with 
$v_{ij}({\rm non}$-static).  The procedure, including the ambiguities inherent
in its implementation, is described in Ref.~\cite{Mar05}.   In contrast to the purely
isovector ${\bf j}^{PS}_{ij}$ and ${\bf j}^V_{ij}$, the currents from
$v_{ij}({\rm non}$-static) have both isoscalar and isovector terms, which,
however, due to their short-range nature, lead to contributions that are typically
much smaller (in magnitude) than those generated by ${\bf j}^{PS}_{ij}$ and
${\bf j}^V_{ij}$.

Finally, conserved three-body currents associated with the $V^{2\pi}_{ijk}$ term of
the $V_{ijk}$ have also been derived by assuming that this term originates
from the exchange of effective PS and V mesons with excitation of
an intermediate $\Delta$ isobar~\cite{Mar05}.  However, their
contributions have been found to be generally negligible,
except for some of the polarization observables, like $T_{20}$ and $T_{21}$,
measured in the proton-deuteron radiative capture at low energy~\cite{Mar05}.

It is important to stress that the two- and three-body charge and current operators
discussed so far have no free parameters, and that their short-range behavior
is consistent with that of the potential, which is ultimately constrained by
$NN$ scattering data.
It is also worthwhile noting that the two-body charge operators vanish
at vanishing momentum transfer, as they must in order to conserve the
overall charge of the nucleus.  They also vanish in the static limit
($m \rightarrow \infty$).  It was pointed out by Friar~\cite{Friar} long ago that
a proper derivation of the leading two-body charge operator of pion range,
the $\rho^{PS}_{ij}$  in equation~(\ref{eq68a}), necessarily entails the study of non-static
corrections to the OPE potential, and that in particular its form depends on the
specific, but arbitrary, off-the-energy shell extension---that is, on the corrections
beyond the static limit, such as those induced by retardation effects---adopted
for it. Furthermore, he showed that these different operators
and corresponding (non-static) OPE potentials are related to each other
by a unitary transformation, which implies that their intrinsic lack
of uniqueness has no consequence for physical observables.
However, non-static corrections to the OPE term are not considered in the
AV18.  It is also reassuring to note that a pion-range charge operator
of the form given in equation~(\ref{eq68a}) has been derived in the
context of chiral effective field theory; see below.

Additional short-range isoscalar (isovector) two-body charge and current
operators follow from the $\rho\pi\gamma$ ($\omega\pi\gamma$)
transition mechanism.  The  $\rho\pi\gamma$ and $\omega\pi\gamma$
currents are purely transverse and therefore unconstrained by current
conservation.  The coupling constants, and hadronic and electromagnetic
form factors at the $\rho NN$, $\omega NN$, $\rho\pi\gamma$, and
$\omega\pi\gamma$ vertices are poorly known~\cite{Car98}.  However,
with the exception of the $\rho\pi\gamma$ current which gives
a significant contribution to the deuteron magnetic form factor,
generally these operators lead to very small corrections
to the charge and magnetic form factors of the few-nucleon
systems of interest in this review.

Finally, there are purely transverse many-body currents arising from
$M_1$-excitation of $\Delta$ resonances.  They have been derived in a
number of different approaches, the most sophisticated of which is
based on the explicit inclusion of $\Delta$ isobar degrees of freedom
in nuclear wave functions. In this approach, known as the
transition-correlation-operator (TCO) method and
originally developed in Ref.~\cite{Sch92}, the nuclear wave function
is written as
\begin{equation}
\Psi_{N+\Delta}=\left[ {\cal S} \prod_{i <j}\left( 1+U^{\rm TR}_{ij} \right) \right] \Psi
\end{equation}
where $\Psi$ is the purely nucleonic component and ${\cal S}$ is the symmetrizer.
The transition operators $U_{ij}^{\rm TR}$ convert $NN$ into $N\Delta$ and $\Delta\Delta$
pairs and are obtained from two-body bound and low-energy
scattering solutions of the full $N+\Delta$ coupled-channel problem~\cite{Sch92},
including transition potentials  $v^{\rm TR}_{ij}(NN \rightarrow N\Delta)$
and $v^{\rm TR}_{ij}(NN \rightarrow \Delta\Delta)$.  Indeed, the simpler
perturbative treatment of $\Delta$-isobar degrees of freedom, commonly
used in estimating the $\Delta$-current contributions, uses the approximation
\begin{equation}
U^{\rm TR}_{ij}=\frac{1}{m-m_\Delta}\left[ v^{\rm TR}_{ij}(NN \rightarrow N\Delta)
+i\rightleftharpoons j\right] + \frac{ v^{\rm TR}_{ij}(NN 
\rightarrow \Delta\Delta) }{2\,(m-m_\Delta)} \ ,
\end{equation}
and $m_\Delta$ (1232 MeV) is the $\Delta$ mass.
This perturbative treatment has been found to be inappropriate,
since it overestimates $\Delta$ contributions.
In the presence of an electromagnetic field, $N \rightleftharpoons \Delta$
and $\Delta \rightarrow \Delta$ couplings need to be accounted for, and
these couplings (and associated electromagnetic form factors) are taken
from $N(e,e^\prime)$ data in the resonance region.  In practice, the
currents arising from $\Delta$ resonance excitation
can be reduced to effective two- and many-body operators
depending on $U^{\rm TR}_{ij}$, but acting only  on the nucleonic
component $\Psi$ of the full wave function.  The TCO method
is used in some of the calculations reported in the present review.

In the conventional framework based on instant-form Hamiltonian dynamics,
it is possible to perform calculations within a $v/c$ expansion scheme,
in which the Poincar\`e covariance of the theory is satisfied to order
$(v/c)^2$~\cite{Foldy61,Krajcik74,Friar75}. In this approach the many-body Hamiltonian is written as
\begin{equation}
\label{eq:hhrr}
\fl H=\sum_i \left( \sqrt{p_i^2+m^2} -m\right)+ \sum_{i<j}\left[ \, \overline{v}_{ij}+
\delta v_{ij}({\bf P}_{ij})\right] + \sum_{i<j<k} \left[\, \overline{V}_{ijk}+\delta V_{ijk}({\bf P}_{ijk}) \right] \ ,
\end{equation} 
where the relativistic expression for the kinetic energy is used, and  $\overline{v}_{ij}$
and $\overline{V}_{ijk}$ are, respectively, the two- and three-nucleon potentials in
the corresponding rest frames, while the so-called boost corrections $\delta {v}_{ij}({\bf P}_{ij})$
and $\delta {V}_{ijk}({\bf P}_{ijk})$ depend on the total momenta ${\bf P}_{ij}$ and ${\bf P}_{ijk}$
of, respectively, the two- and three-body subsystems, and vanish when ${\bf P}_{ij}=0$ and
${\bf P}_{ijk}=0$.  These boost corrections are related to the rest-frame potentials
by the requirement that the commutation relations of the Poincar\`e group be satisfied to order
$(v/c)^2$~\cite{Friar75,Car93}.  The effects of these boost corrections on the binding
energies of light nuclei have been studied in Refs.~\cite{Car93,Wir01}.  In these calculations,
the rest-frame two-body potential $\overline{v}_{ij}$ must be refitted to $NN$ data, and obviously
the parameters present in $\overline{V}_{ijk}$ must also be recalibrated.  The results of a comparison
with a phase-equivalent nonrelativistic Hamiltonian show that the relativistic corrections to the binding energies
are repulsive: in $^4$He, for example, they amount to about 2 MeV.

The approach above has been used to carry out a deuteron form factor calculation~\cite{Sch02}, by 
consistently including also the $(v/c)^2$ corrections arising from the boosting of the deuteron wave
function,
\begin{equation}
\psi_{\bf v}({\bf p}) = \frac{1}{\sqrt{\gamma}}\left[ 1-\frac{i}{4\, m}
{\bf v}\cdot\left({\bm \sigma}_1 -{\bm \sigma}_2\right) \times{\bf p}\right]
\psi_0({\bf p}_\parallel/\gamma,{\bf p}_\perp)
\end{equation}
where ${\bf v}$ is the velocity of the moving frame, $\gamma=1/\sqrt{1-v^2}$,
${\bf p}_\parallel$ and ${\bf p}_\perp$ are the components of the
momentum ${\bf p}$ parallel and perpendicular to the velocity ${\bf v}$,
and $\psi_0$ is the rest frame wave function.  
The calculation of Ref.~\cite{Sch02} also
retained the full covariant structure of the one-body current
and two-body $\rho\pi\gamma$ current, 
and the pion-like and $\rho$-like two-body terms (correct to order $(v/c)^2$) that contribute to the isoscalar charge operator.
Some of the results of this approach
are presented below.

\subsection{The chiral effective field theory approach}
\label{subsec:chieft}

The last two decades have witnessed significant developments in nuclear chiral
effective field theory ($\chi$EFT), originally proposed by Weinberg in a series
of papers in the early nineties~\cite{Weinberg90_a,Weinberg90_b,Weinberg90_c}. 
The (approximate) chiral symmetry exhibited
by quantum chromodynamics (QCD) severely restricts the form of the interactions
of pions among themselves and with other particles.  In particular, the pion couples
to baryons, such as nucleons and $\Delta$-isobars, by powers of its momentum
${\cal Q}$, and the Lagrangian describing these interactions can be expanded in powers
of ${\cal Q}/\Lambda_\chi$, where $\Lambda_\chi \sim 1$ GeV specifies the chiral-symmetry
breaking scale.  In what follows, ${\cal Q}$ will represent not only the momentum of the pion, but may also the generic value of the 
momentum of other particles. It is assumed to be less or equal to the pion mass.  As a result, classes of Lagrangians emerge, each characterized
by a given power of ${\cal Q}/\Lambda_\chi$ and each involving a certain number of
unknown coefficients, the so called low-energy constants (LEC's).  These LEC's are then
determined by fits to experimental data (see, for example, the review papers~\cite{Bedaque02}
and~\cite{Epelbaum09}, and references therein).  Thus, $\chi$EFT provides, on the one hand,
a direct connection between QCD and its symmetries, in particular chiral symmetry, and the
strong and electroweak interactions in nuclei, and, on the other hand, a practical calculational
scheme, which can,  at least in principle, be improved systematically.  In this sense, it can be
justifiably argued to have put low-energy few-nucleon physics on a more fundamental basis.

Within the nuclear $\chi$EFT approach, a variety of studies have been carried out
in the strong-interaction sector dealing with the derivation of two- and three-nucleon
potentials~\cite{Ord95}
\nocite{Epelbaum98,Ent03,Machleidt11,Nav07,Epe02,Bira94,Bern11}-~\cite{Gir11}
and accompanying isospin-symmetry-breaking corrections~\cite{Fri99}
\nocite{Epe99,Fri04}-~\cite{Fri05},
and in the electroweak sector dealing with the derivation of parity-violating
two-nucleon potentials induced by hadronic weak 
interactions~\cite{Hax13}\nocite{Zhu05,Gir08}-~\cite{Viv14}
and the construction of nuclear electroweak currents~\cite{Park93}.  In this review, the focus
is on nuclear electromagnetic charge and current operators.
These were originally derived up to one loop level  in
the heavy-baryon formulation of covariant perturbation theory  by Park {\it et al.}~\cite{Park96}.
More recently, however, two independent derivations, based
on time-ordered perturbation theory (TOPT), have appeared in the literature, one
by some of the present authors~\cite{Pastore09,Pastore11,Piarulli13} and the other by K\"olling
{\it et al.}~\cite{Koelling09,Koelling11}.  In the following, we outline the derivation of these
operators, deferring the discussion of some of the more technical aspects to the
original papers~\cite{Pastore09,Pastore11,Piarulli13}.  

\subsubsection{Interaction Hamiltonians}
\label{sec:hint}

In the simplest implementation, $\chi$EFT Lagrangians are constructed
in terms of nucleon and pion degrees of freedom.  By now, the procedure
by which this is accomplished has been codified in a number of papers~\cite{Fettes00}, and
$\pi N$ and $\pi \pi$ Lagrangians, denoted respectively as ${\cal L}^{(n)}_{\pi N}$ and
${\cal L}^{(m)}_{\pi \pi}$, have been derived up to high order in the
chiral expansion.  Contributions arising from the inclusion of additional
degrees of freedom, such as $\Delta$-resonances and heavier mesons, are
effectively subsumed in the LEC's entering these ${\cal L}^{(n)}_{\pi N}$ and
${\cal L}_{\pi\pi}^{(m)}$ Lagrangians.  In principle, they contain an infinite number
of interactions compatible with the QCD symmetries, but as mentioned above, 
the transition
amplitudes obtained from them can be expanded in powers of ${\cal Q}/\Lambda_\chi$,    
and at each given order of the expansion, the number of terms contributing to the
amplitude is finite~\cite{Weinberg90_a,Weinberg90_b,Weinberg90_c}.
The canonical formalism is used to construct the Hamiltonians from the chiral
Lagrangians. 
Only terms entering the
two-nucleon potential and electromagnetic charge and current operators up to one loop are listed below. 
As it will become apparent later, this requires the potential up to order 
$({\cal Q}/\Lambda_\chi)^2$
and the electromagnetic operators up to order $({\cal Q}/\Lambda_\chi)^1$ for a two-nucleon
system. However, it is worthwhile pointing out that at the present time two-nucleon potentials
have been derived and widely used
up to order $({\cal Q}/\Lambda_\chi)^4$ (requiring two-loop
contributions). Very recently, a new derivation up to order
$({\cal Q}/\Lambda_\chi)^5$ has appeared~\cite{Epelbaum:2014sza}.
Indeed, some of these high-order potentials have been used,
in conjunction with the one-loop electromagnetic operators, in the calculations of
static properties and form factors of $A$=2 and 3 nuclei reported below.
From a chiral counting perspective, these calculations are not
consistent, since strict adherence to consistency would require going
up to order $({\cal Q}/\Lambda_\chi)^3$ in the derivation of the electromagnetic
operators, a rather daunting task.  It is also unclear at this point
how many new LEC's would enter, in addition to the five at order $({\cal Q}/\Lambda_\chi)^1$
(see below); 
if there were to be too
many of them, this would obviously reduce substantially the predictive power
of the theory, since there are only a limited number of electromagnetic
observables in the few-nucleon systems (including single nucleons)
to constrain these LEC's. 

Setting aside these considerations, the subset of interaction terms in
${\cal L}^{(1)}_{\pi N}$, ${\cal L}^{(2)}_{\pi N}$, and ${\cal L}^{(3)}_{\pi N}$ in the $\pi N$
sector, and ${\cal L}^{(2)}_{\pi \pi}$  in the $\pi\pi$ sector, relevant to the
derivation of potentials and electromagnetic operators at one loop level leads
to the following Hamiltonians
\begin{eqnarray}
\!\!\!\!\!\!\!\!\!\!\!\!\!\!\!\!H_{\pi N} \!&=&\!\int\!\! {\rm d}{\bf x}\,
 N^\dagger\Big[ \frac{g_A}{2 f_\pi}\,\tau_a\,
 {\bm \sigma} \cdot {\bm \nabla} \pi_a \!+\!\frac{1}{4f_\pi^2}
{\bm \tau}\cdot
({\bm \pi}\times \partial^{\, 0} {\bm \pi})+\dots \Big]N  \ ,
\label{eq:hpiN} \\
\!\!\!\!\!\!\!\!\!\!\!\!\!\!\!H_{\gamma N}\!&=&\! e\int\!\!{\rm d}{\bf x}\, N^\dagger \Big[ e_N \, A^0 +
i \,\frac{e_N}{2\,m} 
\left( -\overleftarrow{\bm \nabla}\cdot {\bf A}
+ {\bf A} \cdot \overrightarrow{\bm \nabla} \right)
-\frac{\mu_N}{2\, m} \, 
{\bm \sigma}\cdot  {\bm \nabla} \times {\bf A} \nonumber \\
 \!\!\!\!\!\!\!\!\!\!\!\!\!\!\!&&- \frac{2\, \mu_N - e_N }{8\, m^2} 
\left( {\bm \nabla }^2 A^0
 + {\bm \sigma} \times  {\bm \nabla} A^0 \cdot \overrightarrow{\bm \nabla}
- \overleftarrow{\bm \nabla} \cdot {\bm \sigma} \times  {\bm \nabla} A^0 \right) +\dots \Big] N \ ,
\label{eq:hgn} \\
\!\!\!\!\!\!\!\!\!\!\!\!\!\!\!H_{\gamma \pi} \!\!&=&\!e\int\!\!{\rm d}{\bf x}\,\left[
A^0\left( {\bm \pi} \times  \partial^{\,0} {\bm \pi} \right)_z +\epsilon_{zab}\,
 \pi_a\, \left( {\bm \nabla} \pi_b\right) \cdot  {\bf A}+
\dots \right] \ ,
\label{eq:hgpi}\\
\!\!\!\!\!\!\!\!\!\!\!\!\!\!\! H_{ \gamma \pi N}\!&=&\! \frac{e}{2f_\pi} \int\!\!{\rm d}{\bf x}\, N^\dagger\Bigg[ 
  \frac{ g_A}{2\, m} \left( {\bm \tau}\cdot {\bm \pi}+\pi_z \right)\,
    {\bm \sigma}\cdot {\bm \nabla} A^0\nonumber \\
\!\!\!\!\!\!\!\!\!\!\!\!\!\!\!    &&+\left(  d^\prime_8 {\bm \nabla} \pi_z +
d^\prime_9  \tau_a  {\bm \nabla} \pi_a
+d^\prime_{21}\, \epsilon_{zab} \tau_b\, {\bm \sigma}\times
 {\bm \nabla}\pi_a  \right) \cdot {\bm \nabla} \times {\bm A} \,
+\dots  \Bigg] N\  ,
\label{eq:hgpin}
\end{eqnarray}
where $g_A $, $f_\pi$, $e$, and $m$ are, respectively, the
nucleon axial coupling constant, pion decay amplitude, proton electric charge, and
nucleon mass, and the parameters $d^\prime_i$ are (yet to be determined)
LEC's.  The isospin doublet of (non-relativistic) nucleon
fields, isospin triplet of pion fields, and electromagnetic vector field are
denoted by $N$, ${\bm \pi}$, and $A^\mu$, respectively, and
${\bm \sigma}$ and ${\bm \tau}$ are spin and isospin Pauli
matrices.  The arrow over the gradient specifies whether it acts on
the left or right nucleon field.  The isospin operators $e_N$ and $\mu_N$
are defined as
\begin{equation}
e_N = (1+\tau_z)/2 \ , \,\,\,
\kappa_N =  (\kappa_S+ \kappa_V \, \tau_z)/2 \ , \,\,\, \mu_N = e_N+\kappa_N  \ ,
\label{eq:ekm}
\end{equation}
and $\kappa_S$ and $\kappa_V$ are the isoscalar and isovector combinations
of the anomalous magnetic moments of the proton and neutron.
The power counting of the resulting vertices follows by noting that each
gradient brings in a factor of ${\cal Q}$, so, for example, the two terms in $H_{\pi N}$
are each of order ${\cal Q}$, while (ignoring the counting ${\cal Q}$ assumed for the
external field $A^\mu$) the first term in $H_{\gamma \pi N}$ is of
order ${\cal Q}$, and the remaining ones  in the second line of equation~(\ref{eq:hgpin})
are of order ${\cal Q}^2$.

In addition to the chiral Hamiltonians above, up to and including order
$({\cal Q}/\Lambda_\chi)^2$ there are fourteen contact interaction terms allowed by the
symmetries of the strong interactions, each multiplied by a LEC. Two of these contact
terms (proportional to the LEC's $C_S$ and $C_T$ in standard notation) are of a
non-derivative type, and therefore are of order $({\cal Q}/\Lambda_\chi)^0$,
while the remaining twelve (proportional to the LEC's $C_i^\prime$) of order $({\cal Q}/\Lambda_\chi)^2$
involve two gradients acting on the nucleon fields---they are listed in Ref.~\cite{Gir10}.
The contact potential at order $({\cal Q}/\Lambda_\chi)^2$ derived from them
in the two-nucleon center-of-mass system in fact depends on $C_S$ and $C_T$,
and seven linear combinations of the $C_i^\prime$, customarily denoted
as $C_1,\dots,C_7$.  The remaining five linear combinations of $C_i^\prime$
have been shown to be related to $C_S$ and $C_T$ by requiring
that the Poincar\'e covariance of the theory be satisfied to order
$({\cal Q}/\Lambda_\chi)^2$~\cite{Gir10}.  So the $({\cal Q}/\Lambda_\chi)^2$
potential involves nine independent LEC's. (As a side remark, we note that the contact
potential at order $({\cal Q}/\Lambda_\chi)^4$ requires an additional fifteen independent LEC's.)
These LEC's are determined by fitting two-nucleon
elastic scattering data. Minimal substitution in the gradient terms leads to a
(two-nucleon) contact current~\cite{Pastore09,Piarulli13}.

Lastly, non-minimal couplings through the electromagnetic field tensor
$F_{\mu \nu}$ are also allowed.  It can be shown~\cite{Pastore09} that the only
two independent operator structures at order $({\cal Q}/\Lambda_\chi)^1$ are
\begin{eqnarray}
H_{CT}^{\gamma{\rm nm}}&=&e \int {\rm d}{\bf x}\,
\Big[ C_{15}^\prime\,  N^\dagger {\bm \sigma} N \,\, N^\dagger N + 
C_{16}^\prime\,\Big( N^\dagger {\bm \sigma}\, \tau_z N \,\, 
N^\dagger N \nonumber \\
&&- N^\dagger {\bm \sigma} N \,\,
N^\dagger \tau_z N \Big) \Big] \cdot {\bm \nabla}\times{\bf A} \ ,
\label{eq:gnm}
\end{eqnarray}
where the isoscalar $C_{15}^\prime$ and isovector $C_{16}^\prime$
LEC's (as well as the $d^\prime_i$'s multiplying the higher order terms
in the $\gamma\pi N$ Hamiltonian) can be determined by fitting photo-nuclear data in the
few-nucleon systems.  We will return to this issue below. 

\subsubsection{From amplitudes to currents}
\label{sec:road}

We present the expansion of the amplitudes $T$ and $T_\gamma$ for the processes
$NN \longrightarrow NN$ and $NN\gamma \longrightarrow NN$ based on
TOPT.  Terms in this
expansion are conveniently represented by diagrams.  Before discussing
them in some detail, it is worthwhile to make some preliminary considerations.
We distinguish between reducible diagrams (diagrams which involve
a pure nucleonic intermediate state) and irreducible diagrams (diagrams
which include both pionic and nucleonic intermediate states).  The former
are enhanced with respect to the corresponding irreducible contributions
by a factor of ${\cal Q}$ for each pure nucleonic intermediate state.  In the static limit---that is,
in the limit in which $m \rightarrow \infty$ or, equivalently, neglecting nucleon kinetic
energies---reducible contributions are infrared-divergent.  The prescription
proposed by Weinberg~\cite{Weinberg90_a,Weinberg90_b,Weinberg90_c} to
treat the latter is to define the nuclear potential (and currents) as given by the
irreducible contributions only.  Reducible contributions, instead, are generated
by solving the Lippmann-Schwinger (or Schr\"odinger) equation iteratively with
the nuclear potential (and currents) arising from the irreducible amplitudes.

The formalism developed by some of the present authors is based on this prescription.
However, the omission of reducible contributions from the definition of nuclear operators
needs to be dealt with care when the irreducible amplitudes are evaluated 
in the static limit approximation, which is usually the case.
The iterative process will then generate only part of the reducible
amplitude.  The reducible part of the amplitude beyond the static limit approximation needs to be incorporated
order by order---along with the irreducible amplitude---in the definition of nuclear operators.
This scheme in combination with TOPT, which is best suited to separate the reducible
content from the irreducible one, has been implemented in Refs.~\cite{Pastore09,Pastore11,Piarulli13}
and is described below.  The method leads to nuclear operators which are not uniquely
defined due to the non-uniqueness of the transition amplitude off-the-energy shell.  The
operators, while non unique, are unitarily equivalent, and therefore the description of
physical systems is not affected by this ambiguity.  We defer to Ref.~\cite{Pastore11} 
for a discussion of these technical aspects. 

Another approach, implemented to face the difficulties posed by the reducible amplitudes,
has been introduced by Epelbaum and collaborators~\cite{Epelbaum98}.  The method,
referred to as the unitary transformation method, is based on TOPT and exploits the Okubo
(unitary) transformation~\cite{Okubo54} to decouple the Fock space of pions and nucleons
into two subspaces, one containing only pure nucleonic states and the other involving
states which retain at least one pion.  In this decoupled space, the amplitude
does not involve enhanced contributions associated with the reducible diagrams.
The subspaces are not-uniquely defined, since it is always possible to perform
additional unitary transformations onto them, with a consequent change in the formal definition
of the resulting nuclear operators.  This, of course, does not affect the physical results.

The two TOPT-based methods outlined above lead to formally equivalent operator structures
for the nuclear potential and electromagnetic currents up to loop-corrections included~\cite{Piarulli13},
which makes it plausible to conjecture that the two methods are closely related.  However, this
topic has not been investigated further. 

In what follows, we outline briefly the methods developed in Refs.~\cite{Pastore09,Pastore11,Piarulli13}
and sketch how nuclear operators are derived from transition amplitudes.  We start from the conventional
perturbative expansion of the $NN$ scattering amplitude $T$, which reads
\begin{equation}
 \langle f \!\mid T\mid\! i \rangle= 
 \langle f\! \mid H_1 \sum_{n=1}^\infty \left( 
 \frac{1}{E_i -H_0 +i\, \eta } H_1 \right)^{n-1} \mid\! i \rangle \ ,
\label{eq:pt}
\end{equation}
where $\mid\! i \rangle$ and $\mid\!\! f \rangle$ represent the initial
and final $NN$ states of energy $E_i=E_f$, $H_0$ is the Hamiltonian
describing free pions and nucleons, and $H_1$ is the Hamiltonian
describing interactions among these particles (see section~\ref{sec:hint}).
The evaluation of this amplitude is carried out in practice by inserting
complete sets of $H_0$ eigenstates between successive terms of $H_1$.
Power counting is then used to organize the expansion.

In the perturbative series, equation~(\ref{eq:pt}), a generic (reducible or
irreducible) contribution is characterized by a certain number, say
$M$, of vertices, each scaling as ${\cal Q}^{\alpha_i}\times {\cal Q}^{-\beta_i/2}$
($i$=$1,\dots,M$), where $\alpha_i$ is the power counting implied by the
relevant interaction Hamiltonian and $\beta_i$ is the number of
pions in and/or out of the vertex, a corresponding $M-1$ number of energy
denominators, and possibly $L$ loops.  Out of these $M-1$ energy
denominators, $M_K$ of them will involve only nucleon kinetic
energies, which scale as ${\cal Q}^2$, and the remaining $M-M_K-1$ will involve,
in addition, pion energies, which are of order ${\cal Q}$.  Loops, on the other hand,
contribute a factor ${\cal Q}^3$ each, since they imply integrations over intermediate
three momenta.  Hence the power counting associated with such a contribution is
\begin{equation}
\left(\prod_{i=1}^M  {\cal Q}^{\alpha_i-\beta_i/2}
\right)\times \left[ {\cal Q}^{-(M-M_K-1)}\, {\cal Q}^{-2M_K} \right ]
\times {\cal Q}^{3L} \ .
\label{eq:count}
\end{equation}
Clearly, each of the $M-M_K-1$ energy denominators can be further expanded as
\begin{equation}
\frac{1}{E_i-E_I-\omega_\pi}= -\frac{1}{\omega_\pi}
\bigg[ 1 + \frac{E_i-E_I}{\omega_\pi}+
\frac{(E_i-E_I)^2}{\omega^2_\pi} + \dots\bigg] \ ,
\label{eq:deno}
\end{equation}
where $E_I$ denotes the kinetic energy of the intermediate two-nucleon state,
$\omega_\pi$ the pion energy (or energies, as the case may be), and the
ratio $(E_i-E_I)/\omega_\pi$ is of order ${\cal Q}$.  The terms proportional to
powers of $(E_i-E_I)/\omega_\pi$ lead to non-static corrections.

The ${\cal Q}$-scaling of the interaction vertices and the considerations 
above show that $T$
admits the following expansion:
\begin{equation}
 T=T^{(\nu_{min})} + T^{(\nu_{min}+1)} + T^{(\nu_{min}+2)} + \dots \ ,
\label{eq:tmae}
\end{equation}
where $T^{(n)} \sim {\cal Q}^n$, and chiral symmetry ensures that
$\nu_{min}$ is finite. In the case of the two-nucleon potential 
$\nu_{min}=0$. A two-nucleon potential $v$ can then be derived
which, when iterated in the Lippmann-Schwinger (LS) equation,
\begin{equation}
v+v\, G_0\, v+v\, G_0 \, v\, G_0 \, v +\dots \ ,
\label{eq:lse}
\end{equation}
leads to the on-the-energy-shell ($E_i=E_f$) $T$-matrix in equation~(\ref{eq:tmae}),
order by order in the power counting.  In practice, this requirement can only be
satisfied up to a given order $n^*$, and the resulting potential, when inserted into
the LS equation, will generate contributions of order $n > n^*$, which do not match $T^{(n)}$.
In equation~(\ref{eq:lse}), $G_0$ denotes the free two-nucleon propagator, $G_0=1/(E_i-E_I+i\, \eta)$,
and we assume that
\begin{equation}
v=v^{(0)}+v^{(1)}+v^{(2)}+\dots\ ,
\end{equation}
where the yet to be determined $v^{(n)}$ is of order ${\cal Q}^n$.  We also note that, generally, a term
like $v^{(m)}\, G_0 \, v^{(n)}$ is of order ${\cal Q}^{m+n+1}$, since $G_0$ is of order ${\cal Q}^{-2}$
and the implicit loop integration brings in a factor ${\cal Q}^3$.  Having established the above power counting,
we obtain
\begin{eqnarray}
v^{(0)} &=& T^{(0)} \ , \label{eq:v0}\\
v^{(1)} &=& T^{(1)}-\left[ v^{(0)}\, G_0\, v^{(0)}\right] \ , \\
v^{(2)} &=& T^{(2)}-\left[ v^{(0)}\, G_0\, v^{(0)}\, G_0\, v^{(0)}\right] \nonumber\\
&&\qquad-\left[ v^{(1)}\, G_0 \, v^{(0)}
+v^{(0)}\, G_0\, v^{(1)}\right] \ . \label{eq:v2}
\end{eqnarray}
The leading-order (LO) ${\cal Q}^0$ term, $v^{(0)}$, consists of (static) 
OPE and two (non-derivative) contact interactions, 
while the next-to-leading (NLO)
${\cal Q}^1$ term, $v^{(1)}$, is easily seen to vanish~\cite{Pastore11}, since the leading non-static corrections
of order ${\cal Q}$ in $T^{(1)}$ to the (static) OPE amplitude add up to zero on the energy
shell, while the remaining diagrams in $T^{(1)}$ represent iterations of $v^{(0)}$,
whose contributions are exactly canceled by $\left[ v^{(0)}\, G_0\, v^{(0)}\right]$
(complete or partial cancellations of this type persist at higher $n\ge 2$ orders).
The next-to-next-to-leading (N2LO) ${\cal Q}^2$ term, which follows from equation~(\ref{eq:v2}),
contains two-pion-exchange (TPE) and contact (involving two gradients of the nucleon fields)
interactions. 

The inclusion (in first order) of electromagnetic interactions in the perturbative
expansion of equation~(\ref{eq:pt}) is in principle straightforward.
The transition operator can be expanded as~\cite{Pastore11}
\begin{equation}
T_\gamma=T_\gamma^{(-3)}+T_\gamma^{(-2)}+T_\gamma^{(-1)} +\dots \ ,
\end{equation}
where $T_\gamma^{(n)}$ is of order $e\, {\cal Q}^n$ ($e$ is the electric charge).
The nuclear charge, $\rho$, and current, ${\bf j}$, operators follow
from $v_\gamma= A^0\, \rho-{\bf A}\cdot {\bf j}$, where
$A^\mu=(A^0,{\bf A})$ is the electromagnetic vector
field, and it is assumed that $v_\gamma$ has a similar expansion
as $T_\gamma$.  The requirement that, in the context of the LS equation,
$v_\gamma$ matches $T_\gamma$ order by order in the power counting
implies relations for the $v^{(n)}_\gamma=A^0\, \rho^{(n)}-{\bf A}\cdot {\bf j}^{(n)}$,
which can be found in Ref.~\cite{Pastore11},  similar to those derived above
for $v^{(n)}$, the strong-interaction potential.  

The lowest order contributing to the charge operator has $n=-3$,
\begin{equation}
\rho^{(-3)}=e\, \frac{ 1+\tau_{1,z}}{2} + (1 \rightleftharpoons 2) .
\label{eq:r-3}
\end{equation}
There is no $n=-3$ contribution to ${\bf j}$, and the lowest order ($n=-2$) consists
of the single-nucleon convection and spin-magnetization currents,
\begin{equation}
{\bf j}^{(-2)}=\frac{e}{2\, m}
\left(2\,  {\bf K}_1 \, \frac{ 1+\tau_{1,z}}{2} +i\,{\bm \sigma}_1\times {\bf q }\, 
 \frac{ \mu^S+\mu^V \tau_{1,z}}{2} \right)+
 (1 \rightleftharpoons 2)\ ,
 \label{eq:jlo}
\end{equation}
where ${\bf q}$ is the momentum carried by the external field,
${\bf k}_i$ and ${\bf K}_i$ denote hereafter the combinations of
initial and final nucleon momenta
\begin{equation}
{\bf k}_i={\bf p}_i^\prime-{\bf p}_i \ ,\qquad {\bf K}_i=({\bf p}_i^\prime+{\bf p}_i)/2\ .
\end{equation}
Note that $\rho^{(-3)}$ and ${\bf j}^{(-2)}$ are the same operators as
the one-body terms used in the conventional approach,
see equations~(\ref{eq:chargenr}) and~(\ref{eq:1bcurr}).
The counting $e\, {\cal Q}^{-3}$ ($e\, {\cal Q}^{-2}$) 
in the charge (current) operator follows from the product of a factor $e\, {\cal Q}^0$ ($e\, {\cal Q}^1$)
associated with the $\gamma NN$ vertex, and a factor ${\cal Q}^{-3}$ due to the momentum-conserving
$\delta$-function implicit in a disconnected term of this type. 

\begin{figure}[t]
\vspace*{0.6cm}
\centerline{\includegraphics[width=8cm]{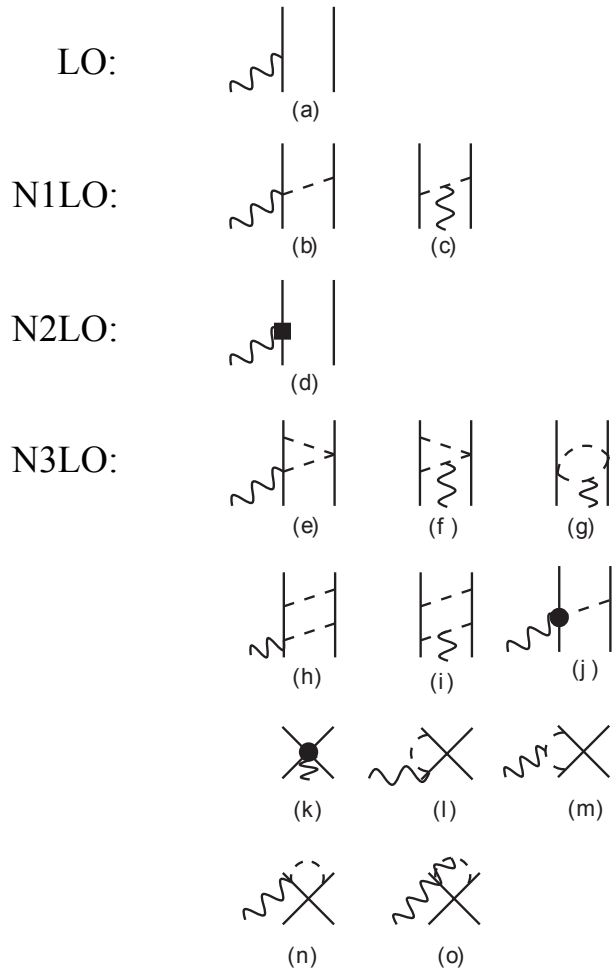}}
\vspace{8pt}
\caption{Diagrams illustrating one- and two-body currents entering at $e\, {\cal Q}^{-2}$ (LO),
$e\, {\cal Q}^{-1}$ (N1LO), $e\, {\cal Q}^{\,0}$ (N2LO), and $e\, {\cal Q}^{\,1}$ (N3LO).  Nucleons, pions,
and photons are denoted by solid, dashed, and wavy lines, respectively.  The square
in panel (d) represents the $({\cal Q}/m)^2$ relativistic correction to the LO one-body current;
the solid circle in panel (j) is associated with the $\gamma \pi N$ current coupling
of order $e\, {\cal Q}$, involving the LEC's $d_8^\prime$, $d_9^\prime$, and $d_{21}^\prime$;
the solid circle in panel (k) denotes two-body contact terms of minimal and non-minimal
nature, the latter involving the LEC's $C_{15}^\prime$ and $C_{16}^\prime$.
Only one among all possible time orderings is shown for the NLO and N3LO currents, so that all box contributions also include crossed box contributions.}
\label{fig:f2}
\end{figure}

\begin{figure}[t]
\vspace*{0.6cm}
\centerline{\includegraphics[width=8cm]{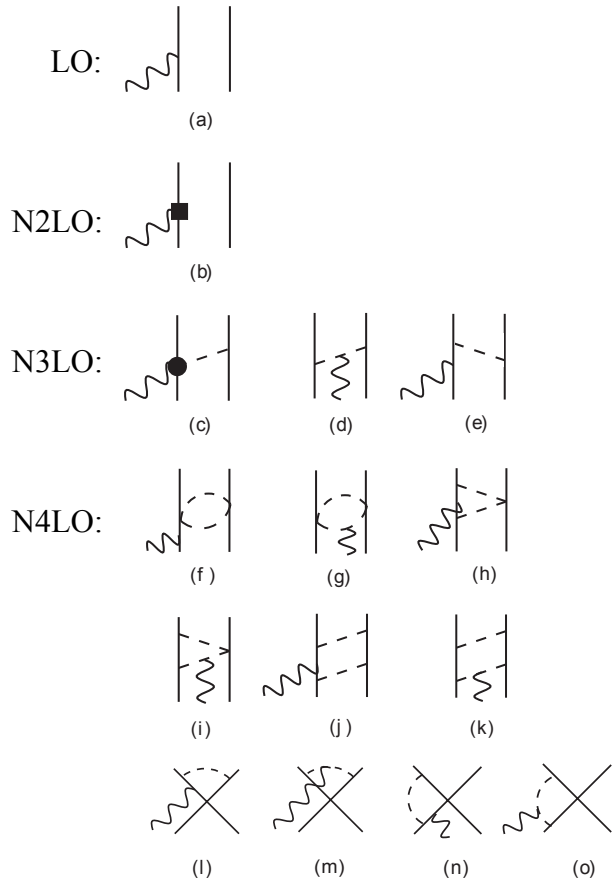}}
\vspace{8pt}
\caption{Diagrams illustrating one- and two-body charge operators entering at $e\, {\cal Q}^{-3}$ (LO),
$e\, {\cal Q}^{-1}$ (N2LO), $e\, {\cal Q}^{0}$ (N3LO), $e\, {\cal Q}^{1}$ (N4LO).  
The square in panel (b) represents
the $({\cal Q}/m)^2$ relativistic correction to the LO one-body charge operator, whereas the solid circle
in panel (c) is associated with a $\gamma \pi N$ charge coupling of order $e\, {\cal Q}$.
As in figure~\ref{fig:f2}, only one among the possible time orderings is shown for the N3LO and N4LO charge operators.}
\label{fig:f5}
\end{figure}

The contributions to the electromagnetic current and charge operators up to one loop
are illustrated diagrammatically in figures~\ref{fig:f2} and~\ref{fig:f5}.
As already noted, the LO starts at $n=-2$ for the current and at $n=-3$ for the
charge operator; N$n$LO corrections to both of them are labelled as ${\cal Q}^n \times \,{\rm LO}$.  

The currents from LO, N1LO, and N2LO terms and from N3LO loop corrections depend
only on the known parameters $g_A$ and $f_\pi$ (N1LO and N3LO), and the nucleon
magnetic moments (LO and N2LO).  Unknown LEC's enter the N3LO OPE contribution
involving the $\gamma \pi N$ vertex of order $e\, {\cal Q}^2$ from $H_{\gamma \pi N}$, second
line of equation~(\ref{eq:hgpin}), as well as the contact currents implied by non-minimal
couplings, equation~(\ref{eq:gnm}),  discussed in the next
subsection.  On the other hand, in the charge operator there are no unknown LEC's  up
to one loop level, and OPE contributions, illustrated in panels
(c)-(e) of figure~\ref{fig:f5}, only appear at N3LO.
The contributions in panels (d) and (e) involve non-static corrections~\cite{Pastore11}, while the contribution
in panel (c) is associated with the $\gamma \pi N$ coupling of order $e\, {\cal Q}$ originating
from the first term in equation~(\ref{eq:hgpin}).  It leads to a two-body charge operator
given by
\begin{equation}
\rho^{(0)}_\pi =\frac{e}{2\, m} \frac{g_A^2}{F_\pi^2} \left( {\bm \tau}_1 \cdot {\bm \tau_2}
+ \tau_{2z}\right)\, \frac{{\bm \sigma}_1 \cdot {\bf q} \,\, {\bm \sigma}_2 \cdot {\bf k}_2}{k^2_2+m_\pi^2}
+ (1 \rightleftharpoons 2) \ .
\label{eq:pich}
\end{equation}
In the present $\chi$EFT context, $\rho^{(0)}_\pi$ was derived first by
Phillips in 2003~\cite{Phillips03}.  However, 
the operator of equation~(\ref{eq:pich}) 
is the same as the $\pi$-exchange contribution derived within the
conventional approach described in section~\ref{subsec:2bcurrent}
(see Ref.~\cite{Car98} and references therein).  This operator plays an important role
in yielding predictions for the $A$=2--4 charge form factors
that are in very good agreement with the experimental data at low 
and moderate values of the momentum transfer 
($Q \lesssim 5$ fm$^{-1}$)~\cite{Car98,Piarulli13}.   The
calculations in Ref.~\cite{Piarulli13} also showed that
the OPE contributions from panels (d) and (e) of figure~\ref{fig:f5} are typically an order of
magnitude smaller than those generated by that in panel (c).

The loop integrals in the N3LO and N4LO diagrams of figures~\ref{fig:f2} and~\ref{fig:f5}
are ultraviolet divergent and are regularized in dimensional regularization~\cite{Pastore09,Pastore11}.
In the current the divergent parts of these loop integrals are reabsorbed by the LEC's
$C_i^\prime$~\cite{Pastore09}.  In the charge, however, they cancel out, in line with fact
that there are no counter-terms at N4LO~\cite{Pastore11}.  Finally, the resulting renormalized
operators have power-law behavior for large momenta, and must be further regularized
before they can be sandwiched between nuclear wave functions.  This is accomplished by
the inclusion of a momentum-space cutoff of the type $C_\Lambda(k)={\rm exp}(-k^4/\Lambda^4)$
with $\Lambda$ in the range $\simeq (500$--700) MeV/c.  The expectation is that observables,
like the few-nucleon form factors at low momentum transfer of interest here, are fairly
insensitive to variations of $\Lambda$ in this range.  

\subsubsection{Determining the LEC's}
\label{lecs}

We now turn our attention to the determination of the LEC's.
The $C_i$ in the minimal contact current, corresponding to the
$\Lambda$ cutoffs of 500 and 600 MeV/c in $C_\Lambda(k)$, are
taken from fits to the two-nucleon scattering data~\cite{Machleidt11}.  The
$d_i^{\, \prime}$, entering the OPE N3LO current, could be
fitted to pion photo-production data on a single nucleon, or related
to hadronic coupling constants by resonance saturation arguments.
Both procedures have drawbacks.  While the former achieves consistency
with the single-nucleon sector, it nevertheless relies on single-nucleon
data involving photon energies much higher than those relevant to the
threshold processes under consideration and real (in contrast to virtual)
pions.  The second procedure is questionable because of poor knowledge
of some of the hadronic couplings, such as $g_{\rho NN}$.  Alternative strategies
have been investigated for determining the LEC's $d_i^\prime$ as well as $C_{15}^\prime$
and $C^\prime_{16}$~\cite{Piarulli13}.  In this respect, it is convenient
to define the adimensional LEC's $d_i^{S,V}$ (in units
of the cutoff $\Lambda$) related to the original set via
\begin{eqnarray}
C_{15}^\prime&=&d_1^S/\Lambda^4  \ , \qquad d_9^\prime=d_2^S/\Lambda^2 , \nonumber \\
C_{16}^\prime&=&d_1^V/\Lambda^4 \ , \qquad d_8^\prime=d_2^V/\Lambda^2\ ,
\qquad d_{21}^\prime=d_3^V/\Lambda^2 \ ,
\end{eqnarray}
where the superscript $S$ or $V$ on the $d^{S,V}_i$ labels 
the isospin of the
associated operator.  

The isoscalar $d_1^S$ and $d_2^S$ have been fixed by reproducing the experimental
deuteron magnetic moment $\mu_d$ and isoscalar combination $\mu^S$ of the
trinucleon magnetic moments.  Their values are listed in table~\ref{tb:tds}.
The LEC $d_1^S$ multiplying the contact
current is rather large, but not unreasonably large, while the LEC $d_2^S$
is quite small~\cite{Piarulli13}.
\begin{table}[bth]
\caption[Table]
{\label{tb:tds}Adimensional values of the isoscalar LEC's corresponding
to cutoff parameters $\Lambda$ in the range 500--600 MeV/c obtained
for the N3LO/N2LO 
Hamiltonian.}
\begin{center}
\begin{tabular}{ccc}
\br
$\Lambda$  & $d_1^S$  & $d_2^S\times 10 $ \\
\mr
500 &  4.072 
& 2.190 
  \\
\mr
600  & 11.38 
& 3.231 
 \\
\br
\end{tabular}
\end{center}
\end{table}

The isovector LEC's $d_1^V$, $d_2^V$, and $d_3^V$
have been determined in the following three different ways (denoted as
I, II, and III in table~\ref{tb:tdv}).  In all cases I-III,
we have assumed $d_3^V/d_2^V=1/4$ as suggested by $\Delta$ dominance
in a resonance saturation picture of the N3LO OPE current of panel (j) in 
figure~\ref{fig:f2}.
In set I, $d_1^V$ and $d_2^V$ have been constrained to reproduce the
experimental values of the $np$ radiative capture cross section $\sigma_{np}$
at thermal neutron energies and the isovector combination $\mu^V$ of the
trinucleon magnetic moments.  This, however, leads to unreasonably
large values for both LEC's, and is clearly unacceptable~\cite{Piarulli13}.
In sets II and III, the LEC $d_2^V$ is fixed by assuming $\Delta$ dominance
(see equation (4.2) of Ref.~\cite{Piarulli13}),
while the LEC $d_1^V$
multiplying the contact current is fitted to reproduce either $\sigma_{np}$
in set II or $\mu^V$ in set III.  Both alternatives still lead to somewhat
large values for this LEC, but the degree of unnaturalness is tolerable
in this case.  There are no three-body
currents at N3LO~\cite{Pastore09}, and therefore it is reasonable to fix the strength of the
two-nucleon contact operators by fitting a three-nucleon observable
such as $\mu^S$ and $\mu^V$.
Note that the values of the LEC's in table~\ref{tb:tds} 
and~\ref{tb:tdv}, as well as the results presented in section~\ref{sec:res},
have been obtained using the chiral $NN$ potential derived 
up to N3LO by Entem and Machleidt in Ref.~\cite{Ent03}, augmented,
in the $A=3,4$ cases, by the three-nucleon interaction 
derived up to N2LO, in the local version of Ref.~\cite{Nav07}. 
\begin{table}[t]
\caption[Table]
{\label{tb:tdv} Adimensional values of the isovector LEC's corresponding
to cutoff parameters $\Lambda$ in the range 500--600 MeV/c obtained
for the N3LO/N2LO 
Hamiltonian.  In sets II and III the values
of $d_2^V$ have been fixed assuming a $\Delta$ dominance model.
See text for further explanation.}
\begin{center}
\begin{tabular}{c|cc|cc|cc} 
\br
$\Lambda$  & $d_1^V$(I) & $d_2^V$(I) & $d_1^V$(II) & $d_2^V$(II) & $d_1^V$(III) & $d_2^V$(III)  \\
\mr
500 &  10.36 
& 17.42 
& --13.30 
& 3.458  & --7.981 
&  3.458  \\
\mr
600 &  41.84 
&  33.14 
& --22.31 
& 4.980  &   --11.69 
& 4.980  \\
\br
\end{tabular}
\end{center}
\end{table}

Finally, we conclude by noting that hadronic electromagnetic form factors
need to be included in the nuclear $\chi$EFT charge and current operators. The
latter could be consistently calculated in chiral perturbation theory ($\chi$PT)~\cite{Kubis01},
but the convergence of these calculations in powers of the momentum transfer appears
to be rather poor.  For this reason, in the results reported below for the few-nucleon form factors,
they are taken from fits to available electron scattering data~\cite{Piarulli13}.

\subsection{The covariant spectator theory (CST) }
\label{subsec:cst}

The covariant spectator theory (CST)~\cite{Gross:1969rv,Gross:1972ye,Gro82a}, 
when applied to few-nucleon form factors, starts from the premise that they  
can be calculated from a covariant field theory in which nucleons and the 
lightest mesons are treated as the effective degrees of freedom.   
A covariant equation is constructed that provides an approximate solution to 
the field theory, and from this the 
currents are determined and the form factors calculated.

\subsubsection{CST two-body equations}

\begin{figure}
\begin{center}
\includegraphics[width=0.7\textwidth]{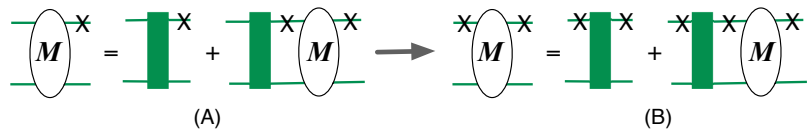}
\caption{
(Color online) Panel A shows the two-body CST-BS equation where the amplitude 
for both nucleons off-shell in the final state can be calculated from the CST 
amplitude.  
Panel B gives a diagrammatic representation of the two-body CST (or Gross) 
equation for the $NN$ scattering amplitude. In both panels,
crosses indicate on-shell particles and the (green) solid box is the kernel.
}
\label{fig:Meqoffa}
\end{center}
\end{figure}

Figure~\ref{fig:Meqoffa} shows a diagrammatic representations of two forms 
(discussed below) of the CST equation for  two-nucleon scattering.  
All diagrams are Feynman diagrams, so iterating the CST equation gives the 
sum $\sum_{n=1}^\infty M^n$, where  $M^n$ is the Feynman diagram with $n$ 
kernels connected by a two-nucleon propagator with one nucleon on mass shell.
The CST equation differs from the more familiar Bethe-Salpeter (BS) 
equation~\cite{Salpeter:1951sz}  in that one of the intermediate nucleons 
(labeled by an $\times$ in the diagrams) is always on-shell, while in the BS 
equation both are off-shell.  It is important to realize that both the BS 
and the CST equations are exact, provided the kernel is exact, so that the 
approximation to the dynamics lies in the choice of kernel, not in the 
choice of equation.  In this way the CST is similar to the conventional 
approach described above; the kernel plays the role of a generalized 
potential, with a theoretical structure based on physical insight and 
parameters adjusted to fit the data.  However, a significant difference 
here is that the structure of the kernel corresponds to a selection of 
Feynman diagrams from which the electromagnetic current can be determined 
consistently (as in $\chi$EFT).

\begin{figure}
\begin{minipage}{0.5\textwidth}
\includegraphics[width=1.2\textwidth]{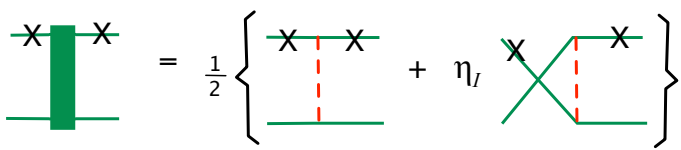}
\end{minipage}
 \begin{minipage}{0.5\textwidth}
\caption{
(Color online) Diagrammatic representation of the antisymmetrized two-body 
one boson exchange kernel.  
}
\label{fig:CSTkernel}
\end{minipage}
\end{figure}  

In the following, we use the notation 
\bea
\int_{k} \equiv \int\frac{d^3k}{(2\pi)^3}\frac{m}{E_k}\qquad  \int_{k_1k_2} 
\equiv \int\frac{d^3k_1}{(2\pi)^3}\frac{m}{E_{k_1}} \int\frac{d^3k_2}{(2\pi)^3}
\frac{m}{E_{k_2}} \, ,
\eea
where $m$ is the nucleon mass and $E_k=\sqrt{m^2+{\bf k}^2}$ the on-shell 
energy.
The specific form of the CST equation for the $NN$ scattering amplitude $M$, 
with particle 1 on-shell in the initial state and {\it both\/} particles 
off-shell in the final state, as illustrated in panel (A) of 
figure~\ref{fig:Meqoffa}, 
is 
\bea
\fl M_{12}(p_1^*,  p_1'; P)=
 \overline {V}_{12}(p^*_1, p'_1; P)
-\int_{k_1} 
\overline {V}_{12}(p^*_1, k_1;P)S_2(k_2)
{M}_{12}(k_1,p'_1;P)\, ,
\label{eq:CSTBS-NN}
\eea
where the momentum of the internal particle 1 is on-shell 
(so that $k_1=\{E_k,{\bf k}_1\}$), $p_1^*$ is off-shell 
(so that $p_1^*=\{p_0,{\bf p}_1\}$), and $\overline V$ is the 
antisymmetrized kernel.   In the rest frame 
$P \equiv p_1+p_2=\{W,{\bf 0}\}$, so the 
momentum of the off-shell particle 2 is $k_2=P-k_1=\{W-E_k,-{\bf k}_1\}$, 
and its propagator is $S_2(k_2)=(m-\slashed{k}_2-i\epsilon)^{-1}$.   
[In many references, the relative momenta $p=\frac12(p_1-p_2)$ are used in 
place of $p_1$ (and similarly for $k$), and sometimes $G$ is used in place of 
$S$; use care in reading the literature.]   Since $p_1^*$ is off-shell in 
equation~(\ref{eq:CSTBS-NN}), this is referred to as the CST-BS equation; 
it gives 
an expression for the fully off-shell scattering amplitude once the CST 
amplitude is known.  

To find the CST amplitude, one merely sets $p^*_1\to p_1$, giving a 
closed equation for $M$~\cite{Gro92,Gro08} 
\bea
\fl M_{12}(p_1,  p'_1; P)=
\overline {V}_{12}(p_1, p'_1; P)
-\int_{k_1} 
\overline {V}_{12}(p_1, k_1;P)S_2(k_2)
{M}_{12}(k_1,p'_1;P)\, .
\label{eq:CST-NN}
\eea
This is the equation illustrated in panel (B) of figure~\ref{fig:Meqoffa}.  
Since particle 1 is on shell throughout,  $M_{12}$ may be defined to be the 
matrix element of the Feynman scattering amplitude ${\cal M}$ between 
positive-energy Dirac spinors of particle 1
\bea
\fl M_{12}(p_1, p'_1; P)\equiv M_{\lambda\lambda',\alpha\alpha'}(p_1, p'_1; P)
=\bar u_\beta({\bf p}_1,\lambda){\cal M}_{\beta\beta',\alpha\alpha'}(p_1, p'_1; P)
u_{\beta'}({\bf p'}_1,\lambda') \, .
\label{eq:M12_contracted}
\eea
The index 2 still refers collectively to the Dirac index of particle 2, 
$\{\alpha\alpha'\}$, but index 1 may refer either to the helicities of 
particle 1  $\{\lambda\lambda'\}$ or the Dirac indices  $\{\beta\beta'\}$ 
[the replacement of a Dirac index ($\alpha_i$, $\beta_i$, $\dots$) by a 
helicity index ($\lambda_i$) always indicates a corresponding contraction 
with a positive-energy helicity spinor].   
For applications, it is sufficient to also place particle 2 on mass shell 
in the initial state.  The two-body interaction kernel $\overline V$ that 
enters equations~(\ref{eq:CSTBS-NN}) or~(\ref{eq:CST-NN}), sometimes 
called the ``potential'' because of its close connection with the 
nonrelativistic potential, is constructed by explicitly antisymmetrizing the 
kernel $V$, as illustrated in figure~\ref{fig:CSTkernel}.   In its Dirac form 
it is
%
\bea
{\overline V}_{\beta\beta',\alpha\alpha'}(p_1,k_1;P)
=\frac12
\left[ V_{\beta\beta',\alpha\alpha'}(p_1,k_1;P)+\eta_I V_{\alpha\beta',\beta\alpha'}
(p_2,k_1;P)
\right] \, ,\qquad \label{eq:kernel}
\eea
%
%
where  the factor $\eta_I=\zeta(-)^{I+1}$ (with $I$=0 or 1 the isospin of 
the $NN$ state) and $\zeta=1$ for bosons and $-1$ for fermions.

\begin{figure}
\begin{center}
\includegraphics[width=0.9\textwidth]{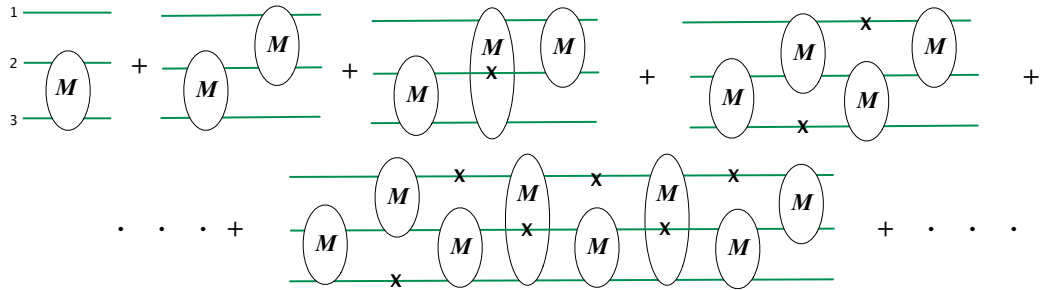}
\caption{Three-body diagrams (with internal spectators on-shell) that 
contribute to the infinite class of successive two-body
scatterings (represented by the ovals).   
These are not time-ordered diagrams, but as in all other diagrams in this
section, the order of the interactions develops from right to left.
In this example, all of the diagrams but the first contribute to the
subamplitude  
${T}^{13}$ where particle 1 (on the top) is the last spectator and
particle 3 (on the bottom) is the first spectator.  The first diagram 
contributes 
${T}^{11}$.}
\label{fig:3Bscattering}
\end{center}
\end{figure}

For the calculation of the deuteron form factors, a relativistic description 
of the deuteron bound state is needed.  The deuteron appears as a pole in the 
scattering amplitude $M$ at $P^2=M_d^2$. At the pole, the scattering amplitude 
can be written
\begin{equation}
\fl M_{\lambda\lambda',\alpha\alpha'}(p_1,p_1';P)=-\sum_{\lambda_d}
\frac{{\cal G}^{\lambda_d}_{\alpha\lambda}(p_1,P) 
\overline{{\cal G}}^{\lambda_d}_{\lambda'\alpha'}(p_1',P) }{M_d^2-P^2} + 
R_{\lambda\lambda',\alpha\alpha'}(p_1,p_1';P) ,\label{eq:deuteronpole}
\end{equation}
where ${\cal G}^{\lambda_d}(p_1,P)={\cal G}^\mu(p_1,P)\xi_\mu^{\lambda_d}$ is 
the vertex function for an incoming deuteron with four-momentum $P$ and 
helicity $\lambda_d$ (including the charge conjugation matrix; for details 
see Ref.~\cite{Gross:2014zra}), and  $R$ is a non-singular remainder.
At the pole, these vertex functions  satisfy a homogeneous integral equation 
of the  type~(\ref{eq:CSTBS-NN}) or~(\ref{eq:CST-NN}) depending on whether 
or not both particles are off-shell. For the discussion of the form factors, 
it is useful to {\it define\/} a relativistic wave function, $\Psi$, equal to 
\begin{eqnarray}
\Psi^{\lambda_d}_{\alpha\lambda_1}(k_1,P)=
S_{\alpha\alpha'}(k_2)
{\cal G}^{\lambda_d}_{\alpha'\lambda_1}(k_1,P)
\, .
\label{wavefunction2}
\end{eqnarray}
%

\subsubsection{CST three-body equation}
 
In the absence of irreducible three-body interactions (which is the case for 
CST one-boson-exchange models), any three-body scattering amplitude $T$ can 
be viewed as a sum of successive two-body scattering processes, with one 
particle uniquely identified as the spectator 
(see figure~\ref{fig:3Bscattering}). It can then be written 
$T=\sum_{i,j=1}^3 T^{ij}$, where each component $T^{ij}$ is the total amplitude 
of all processes in which particles $i$ and $j$ are the spectators in the 
final and initial states, respectively.  Summing these contributions leads 
to a coupled set of integral equations with a Faddeev-like structure, but 
because these equations  sum Feynman diagrams and not the  time-ordered 
diagrams used in Hamiltonian theories, their physical content is quite 
different~\cite{Gro14}.

As in the two-body case, a three-body bound state with mass  $M_t$ appears 
as a pole in the scattering amplitude $T$ at $P^2= M_t^2$, where the conserved 
total four-momentum  is $P=k_1+k_2+k_3$, with  $k_i$  the individual particle 
momenta.  As in equation~(\ref{eq:deuteronpole}),
near the pole the subamplitudes can be written
\begin{equation}
T^{ij}=-\frac{|\Gamma^i \rangle \langle \Gamma^j |}{M_t^2-P^2} + R^{ij} \, ,
\end{equation}
where $\Gamma^i$ is a vertex function with particle $i$ as spectator, and 
$R^{ij}$ is a non-singular remainder.
These vertex functions $\Gamma^i$ satisfy a homogeneous integral equation 
of the Faddeev type,  whose kernel contains the total two-body scattering 
amplitude~\cite{Sta97b}.   

In the CST, the spectator is always on mass shell. 
Because the structure of the Faddeev equation dictates that after each 
two-body interaction a different particle becomes spectator, a second 
particle $j$ must be placed on-shell in order to obtain a closed set of 
equations for the CST vertex function.  We will use the notation  
$\Gamma^i_j$ to denote the subamplitude where particle $i$ is the spectator 
and $j\ne i$ is the second on-shell particle (which must be one of the 
particles in the ``last'' interacting pair).  
When the three particles are identical, the two-body amplitudes in the kernel 
are symmetrised, and the various vertex functions $\Gamma^i_j$ are related 
by symmetry relations. As a consequence, it is sufficient to determine 
just one of them, by convention $\Gamma^1_2$ (the indices 1 and 2 on $\Gamma$ 
will be suppressed from here on). This vertex function describes the 
structure of a three-body bound state in the CST, and it also determines 
the three-body form factors.

For three identical nucleons with mass $m$,  the bound-state Faddeev vertex 
function to be calculated is 
$\Gamma_{\lambda_1\lambda_2\alpha}(k_1,k_2,k_3) = \bar u_{\alpha_1}(k_1,\lambda_1) 
\bar u_{\alpha_2}(k_2,\lambda_2) \Gamma_{\alpha_1\alpha_2\alpha}(k_1,k_2,k_3)$, 
where nucleons 1 and 2 are on mass shell ($k_1^2=k_2^2=m^2$), with helicities 
$\lambda_1$ and $\lambda_2$, and particle 3 is off mass shell with the 
associated Dirac index $\alpha$.    
In the following the alternative notation $\Gamma(k_1,k_2;P)$ is used for the 
vertex function; it is more convenient because $k_3=P-k_1-k_2$ is a dependent 
variable.

\begin{figure}
\begin{minipage}{0.4\textwidth}
\includegraphics[width=1.3\textwidth]{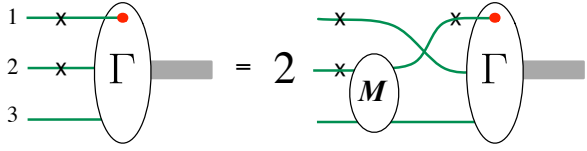}
\end{minipage}
 \begin{minipage}{0.6\textwidth}
\caption{
(Color online) Diagrammatic representation of the Faddeev 
equation~(\ref{eq:3NFad}). The interior dot labels the spectator to 
the ``last'' interaction.  
}
\label{fig:faddeev}
\end{minipage}
\end{figure}  


The vertex function satisfies the Faddeev equation (shown diagrammatically 
in figure~\ref{fig:faddeev})
\begin{eqnarray}
\fl \Gamma_{\lambda_1\lambda_2\alpha}(k_1,k_2;P)=-\sum_{\lambda'_2}\int_{k'_2}
M_{\lambda_2\lambda'_2,\alpha\alpha'}(k_2,k_2';P_{23})\,
2\,\zeta\,{\cal P}_{12}\,
\Psi_{\lambda_1\lambda'_2\alpha'}(k_1,k'_2;P)
\, ,
\label{eq:3NFad}
\end{eqnarray}
%
%
where $P_{23}=P-k_1$ is the total four-momentum of the pair, $M$ is the 
two-body scattering amplitude satisfying equation~(\ref{eq:CST-NN}),  
$\zeta=+1$ for bosons and $\zeta=-1$ for fermions,   
%
%
%
and ${\cal P}_{12}$ is the permutation operator that interchanges particles 
1 and 2 (for details, see Ref.~\cite{Sta97b}).  As in the two-body case, 
it is useful to {\it define\/} a three-body relativistic wave function $\Psi$ 
as
\begin{eqnarray}
\Psi_{\lambda_1\lambda_2\alpha}(k_1,k_2;P)=
S_{\alpha\alpha'}(k_3)
\Gamma_{\lambda_1\lambda_2\alpha'}(k_1,k_2;P)
\, .
\label{wavefunction}
\end{eqnarray}

\subsubsection{Properties of OBE models used with the CST equations}
\label{sec:OBE}

\begin{table}[b]
\caption[Table]{\label{tab:Ls}
Mathematical forms of the $bNN$ vertex functions, with 
$\Theta(p)\equiv(m-\slashed{p})/2m$.
The vector propagator is $\Delta_{\mu\nu}=g_{\mu\nu}-q_\mu q_\nu/m_v^2$ 
with the boson momentum $q=p_1-k_1=k_2-p_2$.}
\begin{tabular}{lccl}
\br
$J^P (b)\quad$ & $\epsilon_b$& $\quad\Lambda_1\otimes\Lambda_2\quad$  & $\Lambda(p,k)$ or $\Lambda^\mu(p,k)$ \cr
\mr
\noalign{\smallskip}
$0^+ (s)$ & $-$& $ \Lambda_1 \Lambda_2$ &
$g_s-\nu_s\left[\Theta(p)+\Theta(k)\right] $\cr
$0^- (p)$ & + &$ \Lambda_1 \Lambda_2$ &
$g_p\gamma^5$
$-g_p(1-\lambda_p)\left[\Theta(p)\gamma^5+\gamma^5\Theta(k)\right]$\cr
$1^- (v)$ & $+$ & $\Lambda_1^\mu \Lambda_2^\nu \Delta_{\mu\nu}$
& $g_v\left[\gamma^\mu +  \frac{\kappa_v}{2M}i\sigma^{\mu\nu}(p-k)_\nu\right]$
$+g_v\nu_v \left[\Theta(p)\gamma^\mu  +  \gamma^\mu \Theta(k) \right]$\cr
$1^+ (a)$ & $+$  & $\Lambda_1^\mu \Lambda_2^\nu g_{\mu\nu}$ & $g_a\gamma^5\gamma^\nu$ \\
\br
\end{tabular}
\end{table}

It has been found that a one-boson-exchange (OBE) kernel, illustrated in 
figure~\ref{fig:CSTkernel}, can provide a precision fit to the data.  The 
bosons required include the familiar six bosons $\pi, \eta, \sigma_0, 
\sigma_1,  \rho, \omega$  plus, in some cases, the heavier axial vector 
isoscalar $h_1$ and isovector $a_1$ mesons.    There is a long and continuing 
debate as to whether the scalar mesons $\sigma_I$ (where $I$ is the isospin) 
are ``real'' mesons, but there is certainly a strong two-pion interaction 
in the isoscalar channel and it is well known that the nuclear force 
requires a significant two-pion exchange mechanism beyond that obtained from 
the iteration of two noninteracting pions.  In the CST these mesons are 
treated as important phenomenological degrees of freedom with an effective 
mass chosen to fit the data (which comes out in the vicinity of 400-500 MeV).
The simplicity of OBE models makes them very attractive, and they have been 
perused for over 40 years, but only with the recent fits using the CST 
equation has it been found that such models can give precision fits to the 
data.  This success may be due in part to the 
{\it cancellation theorem\/}~\cite{Gro93}, 
discussed briefly in section~\ref{subsec:comp} below.  

\begin{table}
\caption[Table]{\label{tab:cstOBEa}
Values of the 13 parameters for the 6 bosons of  Model IIB and the 16 
parameters for the 6 bosons of Model W16.  All masses and energies are in 
MeV; other couplings are dimensionless;  $G_b=g_b^2/4\pi$.  Parameters 
in {\bf bold} were varied during the fit; those labeled with an $^*$ were  
constrained to equal the one above.  The triton binding energy is $E_t$ 
(with its experimental  value in parentheses).  
For model W16, $\nu_s=g_s\nu_s^*$. 
No $E_t$ was calculated for
model IIB.
}
\resizebox{1\textwidth}{!}{%
\begin{tabular}{lc|rccc|ccrcl}
\br
& & \multicolumn{4}{c}{Model IIB (1992)}&\multicolumn{5}{c}{Model W16 (1997)}\\ 
\mr
$b$ & I & $\quad G_b\quad$ & $m_b$ & $\lambda_b$  & $\quad\kappa_v\quad$  & 
$\quad G_b\quad$ & $m_b$ & $\lambda_b/\nu_b^*$  & $\quad\kappa_v\quad$ & 
$\Lambda_b$
\cr
\mr
$\pi^0,\pi^\pm$ & $\quad1\quad$ & {\bf 13.37758}&138.0 &--- & --- 
& {\bf 13.34}&138.0 & 0.0& --- &{\bf 2075}\cr
$\eta$ & $0$ & {\bf 5.30321} & 548.8 & --- & --- & {\bf 2.969} & 548.8 
& $0.0$& --- & {\bf 1206}\cr
$\sigma_0$ & $0$ & {\bf 4.86870} & {\bf 522.0} & --- & ---  & {\bf 4.99887} 
& {\bf 506} & ${\bf -1.2}$& --- &1206$^*$\cr
$\sigma_1$ & $1$ & {\bf 0.24372} & {\bf 482.0} &  --- & ---  & {\bf 0.62818} 
& {\bf 512} & ${\bf 4.16}$& --- &1206$^*$ \cr
$\omega$ & $0$ & {\bf 8.86086} & 782.8 & 1.0 & ${\bf 0.22069}$  & {\bf 14.879} 
& 782.8 & $0.0$& ${\bf 0.195}$ &1206$^*$ \cr
$\rho$ & $1$ & {\bf 0.60318} & 760.0 & ${\bf 0.82989}$\,\,$$& ${\bf 5.66983}$  
& {\bf 0.899} & 760.0 & ${\bf  1.556}$& ${\bf 6.267}$ &  1206$^*$  \cr
\mr
& & \multicolumn{4}{c}{$\Lambda_N={\bf 1675};\; \Lambda_b={\bf 2185}$} 
& \multicolumn{5}{c}{$\Lambda_N={\bf 1822};\; E_t=-8.49\, (-8.48)$ }\\
\br
 \end{tabular}
}
\end{table}

\begin{table}
\caption[Table]{Values of the 27 parameters  for WJC-1 with 7 bosons and 
2 axial vector contact interactions, and the 15 parameters for WJC-2 with 
no axial vectors, obtained by fitting $np$ data.  For additional explanation, 
see table~\ref{tab:cstOBEa}.  
}
\bigskip
\label{tab:cstOBE}
\resizebox{1\textwidth}{!}{%
\begin{tabular}{lc|rcrcl|ccrcl}
\br
& & \multicolumn{5}{c}{Model WJC-1 (2008)}&
\multicolumn{5}{c}{Model WJC-2 (2008)}\\ 
\mr
\noalign{\smallskip}
$b$ & I & $\quad G_b\quad$ & $m_b$ & $\lambda_b/\nu_b$  
& $\quad\kappa_v\quad$ & $\Lambda_b$ & $\quad G_b\quad$ & $m_b$ 
& $\lambda_b/\nu_b$  & $\quad\kappa_v\quad$ & $\Lambda_b$\cr
\mr
\noalign{\smallskip}
$\pi^0$ & $\quad1\quad$ & {\bf 14.608}&134.9766 &{\bf 0.153}$\,\,$& --- &
{\bf 4400}& {\bf 14.038}&134.9766 & 0.0& --- &{\bf 3661}\cr
$\pi^\pm$ & $1$ & {\bf 13.703} &139.5702 & ${\bf -0.312}$$\,\,$& --- &
4400$^*$ &  14.038$^*$&139.5702 & $0.0$& --- &3661$^*$\cr
$\eta$ & $0$ & {\bf 10.684} & {\bf 604} & ${\bf 0.622}$$\,\,$& --- &
4400$^*$& {\bf 4.386} & 547.51 & $0.0$& --- &3661$^*$\cr
$\sigma_0$ & $0$ & {\bf 2.307} & {\bf 429} & ${\bf -15.169}$$\,\,$& --- &
{\bf 1435} & {\bf 4.486} & {\bf 478} & ${\bf -2.594}$& --- &3661$^*$\cr
$\sigma_1$ & $1$ & {\bf 0.539} & {\bf 515} & ${\bf 4.763}$$\,\,$& --- &
1435$^*$ & {\bf 0.477} & {\bf 454} & ${\bf 9.875}$& --- &3661$^*$ \cr
$\omega$ & $0$ & {\bf 3.456} & {\bf 657} & ${\bf 0.843}$$\,\,$& 
${\bf 0.048}$ &{\bf 1376} & {\bf 8.711} & 782.65 & $0.0$& $0.0$ &{\bf 1591} \cr
$\rho$ & $1$ & {\bf 0.327} & {\bf 787} & ${\bf -1.263}$\,\,$$& ${\bf 6.536}$ &
  1376$^*$ & {\bf 0.626} & 775.50 & ${\bf -2.787}$& ${\bf 5.099}$ &  1591$^*$ 
 \cr
$h_1$ & $0$ & {\bf 0.0026} & --- & ---$\,\,\,\,\,\,$& ---$\,\,$ &1376$^*$ 
& --- & &&&\cr
$a_1$ & $1$ & ${\bf -0.436}$ & --- & ---$\,\,\,\,\,\,$& ---$\,\,$ &1376$^*$ 
& --- &&&&\cr
\mr
\noalign{\smallskip}
& & \multicolumn{5}{c}{$\Lambda_N={\bf 1656};\; E_t=-8.48\, (-8.48)$}
& \multicolumn{5}{c}{$\Lambda_N={\bf 1739};\; E_t=-8.50\, (-8.48)$ }\\
\br
 \end{tabular}
 }
\end{table}

The OBE kernels are the sum of one-boson-exchange terms of the form 
\bea
V^b_{12}(p_1,k_1;P)=\epsilon_b\delta\frac{\Lambda^b_1(p_1,k_1)
\otimes \Lambda^b_2(p_2,k_2)}{m_b^2+|q^2|} f(\Lambda_b,q) \, ,
\label{OBE}
\eea
with $b=\{s, p, v, a\}$ denoting the boson type (scalar, pseudoscalar, vector, 
and axial vector), $q$ 
 the four-momentum transfer, $m_b$ the boson mass, $\epsilon_b$ a phase 
factor, $\delta=1$ for isoscalar bosons and 
$\delta={\bm \tau}_1\cdot{\bm \tau}_2=-1-2(-)^I$ for isovector bosons.  
The use of the absolute value $|q^2|$ amounts to a covariant redefinition 
of the propagators and form factors in the region $q^2>0$, done to remove all 
singularities \cite{Gro08}.    
 The boson form factors $f(\Lambda_b,q)$ are normalized to unity at $q^2=0$ 
and approach zero at large $q^2$ with a mass scale of $\Lambda_b$, treated as 
one of the OBE parameters. 
%
%
The axial vector bosons are treated as contact interactions, with the 
structure as in equation~(\ref{OBE}), but with the propagator replaced 
by a constant, $m_a^2+|q^2|\to m^2$, where the nucleon mass $m$ sets a 
convenient scale not related to a boson mass (the effective boson mass in a 
contact interaction is infinite).   The explicit forms of the numerator 
functions 
$\Lambda^b_1\otimes \Lambda^b_2$ can be inferred from table~\ref{tab:Ls}.
Note that $\lambda_p=0$ corresponds to pure pseudovector coupling, and that 
the definitions of the off-shell coupling parameters $\lambda$ or $\nu$ 
differ for each boson. 

The $bNN$ vertex functions, as written in equation~(\ref{OBE}), also include 
strong nucleon form factors
\bea
\Lambda^b(p,k)=h(p)h(k)\widetilde \Lambda^b(p,k) \, ,
\eea
where $p$ $(k)$ is the four-momentum of the incoming (outgoing) nucleon,  
$h(p)$ is a strong form factor associated with an off-shell nucleon with 
four-momentum $p$, and $\widetilde \Lambda^b$ is a reduced $bNN$ vertex 
containing no form factors.  The strong form factors $h$  provide both 
needed convergence and a phenomenological description of the short range 
structure of the off-shell nucleons.  The recent models~\cite{Gro08} use the 
form
\bea
h(p)=\left[\frac{(\Lambda_N^2-m^2)^2}
{(\Lambda^2_N-m^2)^2+(m^2-p^2)^2}\right]^2\, ,
\label{nuclff}
\eea
with $h(p)$ a function of $p^2$ only, subject to the constraint that  
$h(p)=1$ when $p^2=m^2$.

The three types of OBE models used in applications of the CST equations are 
summarized in tables~\ref{tab:cstOBEa} and~\ref{tab:cstOBE}.     
The oldest models (1992) \cite{Gro92} did not have any off-shell terms in the 
scalar meson exchanges (all $\nu=0$) and Model IIB (one of four models 
developed at that time) fit the $np$ data with a $\chi^2/N_{\rm data}=3.40$, 
not a precision fit by todays standards (a revision of this model with 
slightly altered parameters gave a better fit with 
$\chi^2/N_{\rm data}=2.53$~\cite{Gro98}).  Shortly afterward, in connection 
with the study of the three-nucleon bound state~\cite{Sta97}, it was realized 
that including the $\nu$-dependent off-shell terms in the scalar exchanges 
would greatly improve both the fit to the two-body data and the three-body 
binding energy.  Model W16  gave both the best fit to the $np$ scattering 
data (with  $\chi^2/N_{\rm data}=2.25$) and the correct three-body binding 
energy~\cite{Pin09a}.  

\begin{figure}
   \begin{tabular}{ll}
 \centering
 \includegraphics[width=0.5\textwidth]{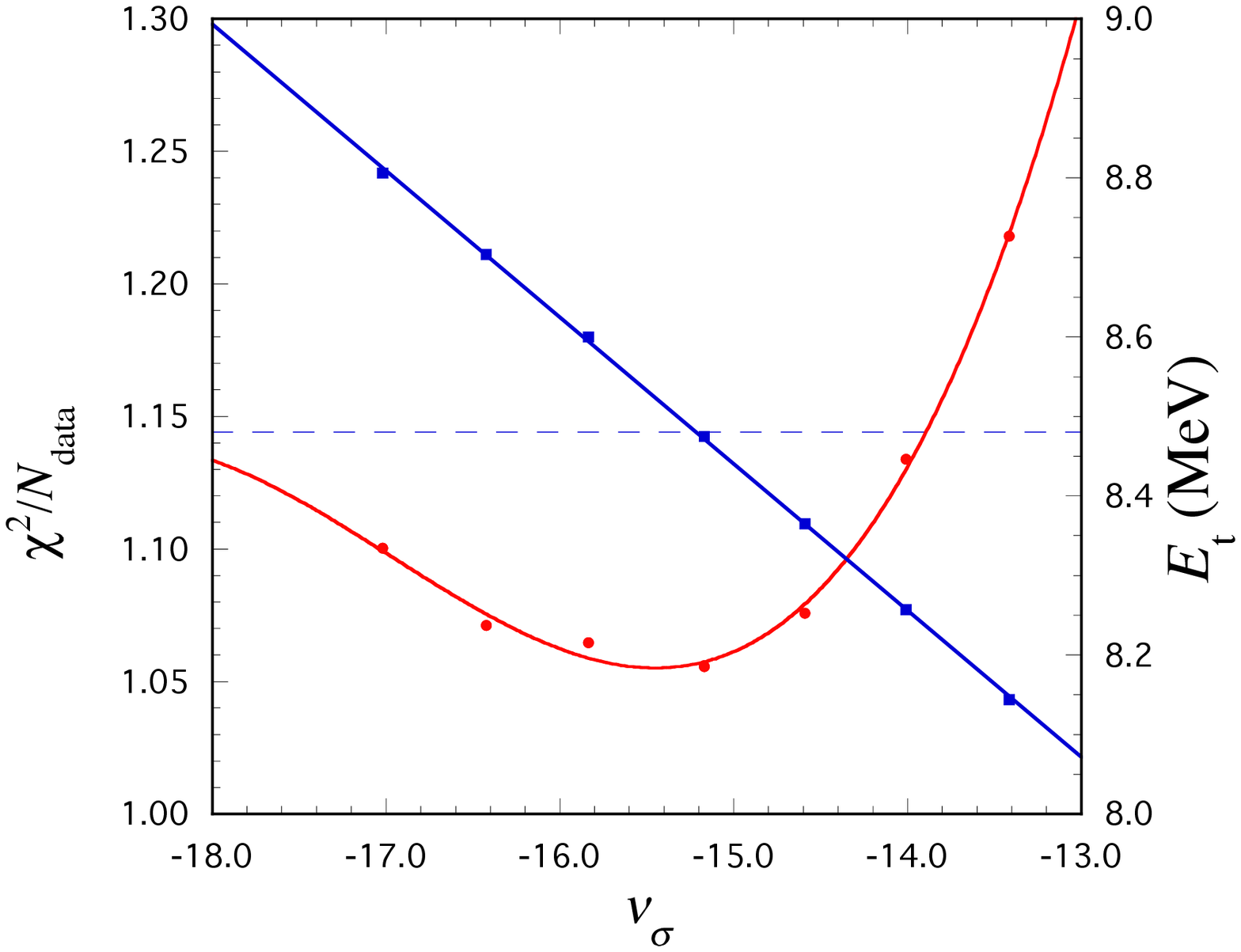}
 &
 \includegraphics[width=0.5\textwidth]{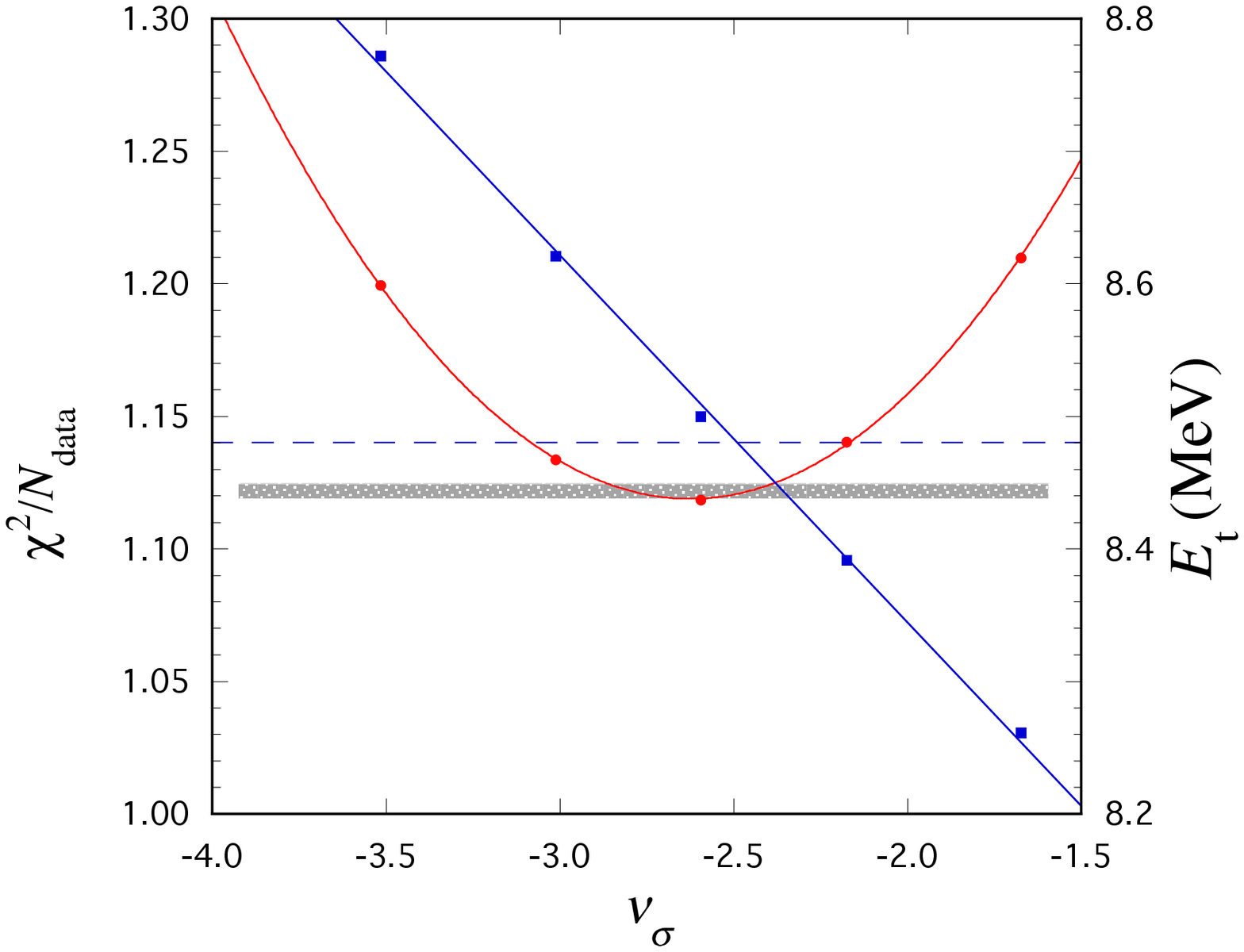} \\
 \end{tabular}
 \caption{
(Color online) Results for calculations of $\chi^2/N_\mathrm{data}$ 
(curves with a minimum and 
left scale) to a 2007 $np$ data base, and triton binding energy $E_t$ 
(linear curves and right scale) for WJC-1 family (left panel) and for 
WJC-2 family of 
models (right panel) as function of $\nu_\sigma$. 
The points with the lowest $\chi^2/N_\mathrm{data}$ are 
models WJC-1 and WJC-2, respectively.  For further details, see 
Ref.~\cite{Gro08}. 
}
 \label{fig:WJC1-WJC-2-chi2-Et}
\end{figure}

The parameters for both of these older models were determined by fits to 
the Nijmegen phase shifts~\cite{Stoks:1993uq}, and it was found that fitting 
these phase shifts did not give the best fit to the data itself.  A major 
effort was made to fit the $np$ data base (below $E_{\rm lab}=350$ MeV) 
directly, which produced the two precision fits (with  
$\chi^2/N_{\rm data}\simeq 1$)  WJC-1 and WJC-2~\cite{Gro08}.  
Model WJC-1 was constructed with the goal to obtain the best possible fit, 
while the objective of WJC-2 was to use the smallest number of parameters 
without destroying  the quality of the fit significantly.   The variation 
of the three-body binding energy and the $\chi^2/N_{\rm data}$ for both of 
these models is quite striking and is shown in 
figure~\ref{fig:WJC1-WJC-2-chi2-Et}.  As with the family of models associated 
with W16, the best fit to the two-body data also gave the correct three-body 
binding energy.  This result  is a simple demonstration that the CST OBE 
models, which generate no irreducible three-body forces when applied 
consistently to the three-nucleon sector, seem to automatically include the 
physics contained in the explicit three-body forces required by 
nonrelativistic models~\cite{Gro14}.

\subsubsection{Two-body current operator and the deuteron form factors}

\begin{figure*}
\begin{center}
\includegraphics[width=1.0\textwidth]{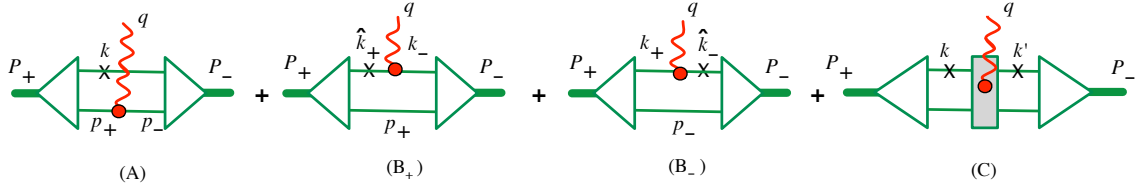}
\end{center}
\caption{
(Color online) Diagrammatic representation of the current operator of the 
CST with particle 1 on-shell.  
}
\label{Fig1}
\end{figure*}

The basic idea of the derivation of the current is to couple the photon to all
propagators and momentum dependent couplings in each of the infinite series 
of two-nucleon diagrams.  This yields, according to a general argument 
developed by Feynman, a conserved current.  If all contributions before 
and after the interaction with the photon are summed up using the two-body 
CST equation, figure~\ref{fig:Meqoffa}, it emerges that only the four 
diagrams shown in figure~\ref{Fig1} are needed to fully describe elastic 
electron scatting from  the two-body bound state~\cite{Gro87,Ada98}.  
These can be written 
%
\bea
\fl J^\mu_{\lambda\lambda'}(q)\!=\!\!\int_{kk'} 
\!\! \overline{\Psi}_{\lambda_n\alpha}^{\lambda}(k,P_+)
\Big[j^\mu_{\alpha\alpha'}(p_+,p_-)\delta_{\lambda_n\lambda'_n}
\delta({\bf k}-{\bf k}')+V^\mu_{\lambda_n\lambda_n',\alpha\alpha'}(k\,P_+; k'\,P_-)
\Big] \Psi_{\alpha'\lambda_n}^{\lambda'}(k',P_-)
\nonumber\\
 +\int_{k_+} 
 \overline{\Psi}_{\lambda_n\alpha}^{\lambda}(\hat k_+,P_+)\,
{\cal G}_{\alpha\beta}^{\lambda'}(k_-,P_-)\Big[\bar u_\gamma({\bf k}_+,\lambda_n) 
\,j^\mu_{\gamma\gamma'}(\hat k_+, k_-)\,S_{\gamma'\beta}(k_-)\Big]^T
\nonumber\\
+\int_{k_-} 
\overline{\cal G}_{\beta\alpha'}^{\lambda}(k_+,P_+)\,
{\Psi}_{\alpha'\lambda_n'}^{\lambda'}(\hat k_-,P_-)\Big[S_{\beta\gamma}(k_+) \,
j^\mu_{\gamma\gamma'}(k_+, \hat k_-)\,u_{\gamma'}({\bf k}_-,\lambda_n') 
\Big]^T\label{eq:23}
\eea
where here, and in the equations to follow, sums over repeated polarization 
indices are implied,  and  $V^\mu$ is the two-nucleon interaction current.     
%
The (B$_\pm$) diagrams 
need the vertex function ${\cal G}_{\alpha\beta}(k,P)$ for {\it both} 
particles off-shell (the CST-BS vertex function).  This vertex function 
allows $k^2\ne m^2$, but when $k^2=m^2$ it is related to the vertex function 
of equation~(\ref{eq:deuteronpole}) by 
${\cal G}^{\lambda_d}_{\alpha\lambda}(k,P)={\cal G}^{\lambda_d}_{\alpha\beta}(k,P)
u_\beta({\bf k},\lambda)$.  It can be calculated from the CST-BS 
equation~(\ref{eq:CSTBS-NN}), represented diagrammatically in 
panel (A) of figure~\ref{fig:Meqoffa}.  As suggested by the figure, 
the OBE kernels, obtained from Feynman diagrams, are already defined 
off-shell, and therefore once the CST vertex function is  known, the 
CST-BS vertex function can be calculated by quadratures.  These off-shell 
vertex functions are necessary to properly account for the interactions of 
the photon with particle 1, where energy-momentum conservation inside the 
loop does not allow the on-shell constraint to be maintained simultaneously 
before and after the interaction.  Since the $dnp$ vertex function is 
explicitly antisymmetric, it might be possible to replace these diagrams  
by (A)-type diagrams, but this would require an unconventional redefinition 
of the interaction current contribution (C) and has not been attempted.

As shown by Riska and Gross (RG)~\cite{Gro87}, the two-body current operator 
defined in figure~\ref{Fig1} (and its generalization to inelastic scattering) 
will yield a conserved current, even in the presence of phenomenological 
form factors, provided the one- and two-body nucleon currents satisfy the 
appropriate Ward-Takahashi (WT) identities.  
In order for the RG prescription to work, the strong form factors at 
the meson-baryon vertices must be factorizable into a product of individual 
form factors associated with each hadron, and the models discussed above 
all have this property. 

With this structure, the strong nucleon form factors may be moved from the 
kernels to the nucleon propagators, leading to dressed (or damped) 
propagators of the form
\bea
S_d^{-1}(p)&=&\frac{m-\slashed{p}}{h^2(p)}=\frac{S^{-1}(p)}{h^2(p)}=
\frac{2m\Theta(p)}{h^2(p)} \, ,\label{eq:Sd}
\eea
where the $h$ occurs squared because one comes from the initial and one 
from the final interactions that connect the propagator
and $\Theta(p)\equiv (m-\slashed{p}-i\epsilon)/(2m)$.  
A conserved two-nucleon current can then be constructed using a 
{\it dressed\/} single nucleon current of the 
form~\cite{Adam:2002cn,Gross:2014zra}
\bea
j^\mu(p,p')=h(p)h(p')j_R^\mu(p,p')\, . \label{eq:3.4}
\eea
The {\it reduced\/} current $j_R^\mu$  satisfies the WT identity
\bea
q_\mu \, j_R^\mu(p,p')&&=e_0\Big[S^{-1}_{d}(p')-S^{-1}_{d}(p)\Big]
\, , \label{eq:210a}
\eea
where $e_0=e(1+\tau_z)/2$ is the charge operator.

The simplest solution to equation~(\ref{eq:210a}),  for the off-shell 
{\it isoscalar\/} part of the nucleon current needed for the 
description of the deuteron form factors, can be written as
\bea
 \fl \qquad\quad j^\mu(p',p)&=&e_0\,\frac{q^\mu}{q^2}
\Big[f_0(p',p)\slashed{q}+g_0(p',p)\Theta(p')\slashed{q}\,\Theta(p)\Big]
\nonumber\\&&
+e_0\,f_0(p',p)\left\{F_1\widetilde\gamma^\mu+F_2
\frac{i\sigma^{\mu\nu}q_\nu}{2m}\right\} 
+ e_0\,g_0(p',p)F_3\Theta(p')\widetilde\gamma^\mu\Theta(p) \, ,
\label{3.1}
\eea
where now $e_0=\frac12 e$, $q=p'-p$, $F_i=F_i(q^2)$ are the isoscalar 
form factors of the nucleon, with $F_3$ a new form factor that contributes 
only when both nucleons are off-shell, and the transverse gamma matrix is
%
$\widetilde\gamma^\mu=\gamma^\mu-\slashed{q}q^\mu/q^2$.  
%
Using the shorthand notation $h=h(p)$ and $h'=h(p')$, the functions 
$f_0$ and $g_0$ are
\begin{equation}
\fl f_0(p',p)=\frac{h'}{h} \frac{(m^2-p^2)}{p'^2-p^2}+\frac{h}{h'} 
\frac{(m^2-p'^2)}{p^2-p'^2} \, ,
\qquad
g_0(p',p)=\frac{4m^2}{p'^2-p^2}\left(\frac{h}{h'}-\frac{h'}{h}\right)
\, .
\end{equation}
The apparent singularity at $q^2=0$ in the first line of 
equation~(\ref{3.1})  
is cancelled by the second line, because both $F_1$ and $F_3$ are subject 
to the constraints $F_i(0)=1$ (for $i=1,3$).  Note that 
equation~(\ref{3.1}) 
displays the very interesting property that all the physics is in the 
transverse term (second line), and that the longitudinal part (first line), 
uniquely fixed by the WT identity, does not contribute to a physical 
amplitude, because $q^\mu$ gives zero when contracted into another conserved 
current or into any of the real or virtual photon polarization vectors.

The condition that the reduced interaction current 
$\widetilde{V}^\mu(kP_+;k'P_-)$ (the current with the strong form factors 
$h$ removed) must satisfy in order that the total current, $J^\mu(q)$, 
be conserved (the two-body WT identity) can be written as
\bea
\fl q_\mu  \widetilde{V}^\mu_{\beta\beta',\alpha\alpha'}(k\,P_+;k'\,P_-)
&=&e_0\Big[ \widetilde{V}_{\beta\beta',\alpha\alpha'}(k,k';P_-)- 
\widetilde{V}_{\beta\beta',\alpha\alpha'}(k,k';P_+)
 \nonumber\\
 &&+  \widetilde{V}_{\beta\beta',\alpha\alpha'}(k-q,k';P_-)- 
 \widetilde{V}_{\beta\beta',\alpha\alpha'}(k_,k'+q;P_+)\Big] . 
 \label{eq:29}
\eea
Recently the isoscalar part of this interaction current, 
needed for calculation  of the deuteron form factors, has been uniquely 
determined~\cite{Gross:2014zra}.

\subsubsection{Three-nucleon form factors in the CST}
\label{sec:3NCST}

\begin{figure}
\centerline{
\includegraphics[width=0.9\textwidth]{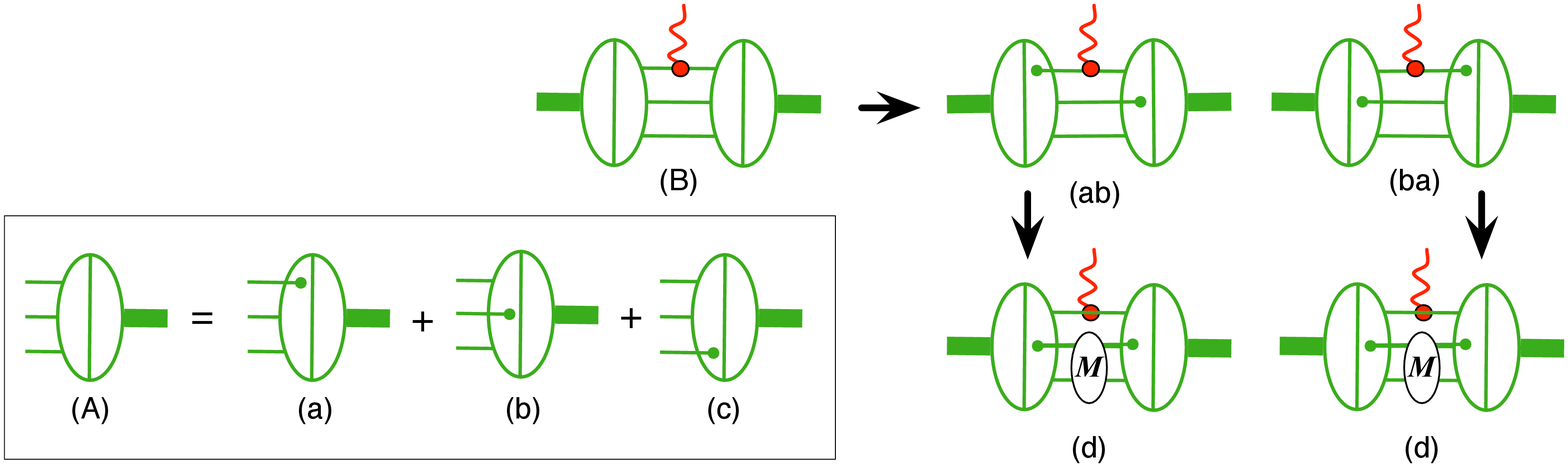}
}
\caption{
(Color online) The figures enclosed in the rectangle show the BS vertex 
function as a
sum of three subamplitudes for the three different choices of
the last interacting pair.  The rest of the figure shows that naive use
of this vertex function in the form factor gives overlap terms
(a)$\times$(b)=(ab) and (b)$\times$(a)=(ba) leading (after use of the
equation) to two terms of type (d), where only one should be
present.}
\label{fig:BSdouble}
\end{figure}  

To expose one of the central issues in the construction of
three-body currents, we begin by looking at what appears to be
the lowest order result in the BS formalism, and
show that this expected result leads to overcounting.

As discussed above, the full BS three-body vertex function is the sum of 
three subamplitudes $\Gamma^i$, as shown in panel (A) of 
figure~\ref{fig:BSdouble}.  Guided by nonrelativistic theory, we might 
expect the
impulse approximation to the current to be related to the square of the wave
function, as illustrated in panel (B).
However, if this proposed
current is expanded using the wave equations, it leads to two terms of 
type (d), 
while direct examination of the ladder sum (for example)
shows that there should be only one such term.  Unless an interaction term 
of type
(d) is explicitly subtracted from the ``impulse'' approximation, it will
 be double
counted.  The same problem does not arise in nonrelativistic theory because 
there the
diagrams represent a sequence of operators which, in general, do not commute. 
The iteration of (a)$\times$(b) gives a
different contribution from that of (b)$\times$(a), and both must be present.
(The treatment of this problem in the context of the BS theory is discussed in
Refs.~\cite{Kvi99a,Kvi99b}.)

It turns out that the spectator theory, like nonrelativistic
theory, also does not suffer from double counting.
Furthermore, the topology of the terms shown
in figure~\ref{fig:BSdouble} can be used to simplify the
spectator formalism.  A detailed demonstration of this is given in 
Refs.~\cite{Gro04,Ada05}.

\begin{figure}
\centerline{
\includegraphics[width=0.4\textwidth,angle=-90]{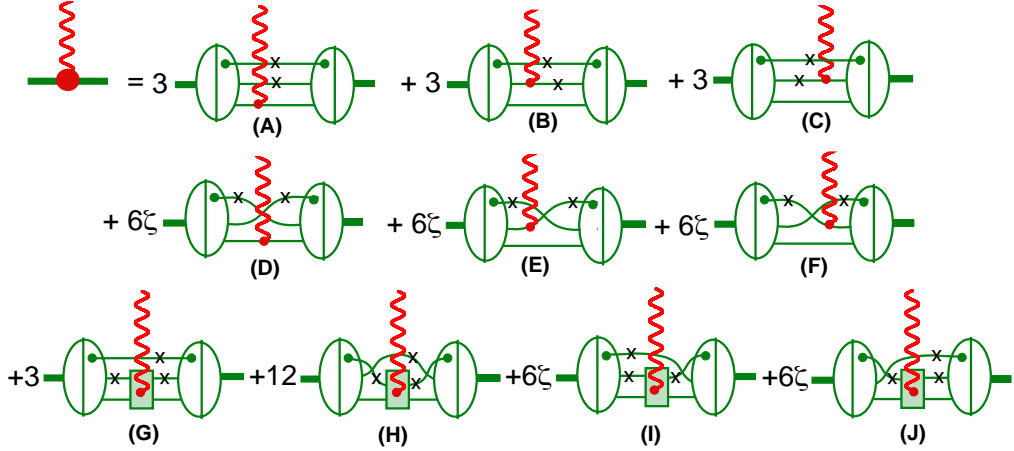}}
\caption{
(Color online) The electromagnetic three-nucleon current in CST for elastic 
electron scattering from the three-nucleon bound state.  
}
\label{fig:coreFF}
\end{figure} 


In practice, the three-nucleon current is constructed by first coupling 
the photon to internal lines and vertices,  which can be done very nicely
using the ``gauging of equations'' method of Kvinikhidze
and Blankleider~\cite{Kvi97b}. In a second step, the three-body equations 
are employed to rearrange the result into a more usable form. For example, 
the final expression of the current obtained in Ref.~\cite{Kvi97b} includes 
a contribution in
which the spectator (particle 1) is off-shell.  In Ref.~\cite{Gro04}, 
 the three-body bound-state and scattering equations 
 are used to replace this amplitude by an equivalent one in which the
spectator is on-shell, as required in the CST.  This replacement also 
leads to a nice
demonstration of how the spectator equations avoid the double-counting problem
in a natural way.
The final result for the three-nucleon current is shown in 
figure~\ref{fig:coreFF}.  The
diagrams (A)--(F) are referred to as 
the complete impulse approximation (CIA), while diagrams
(G)--(J) are contributions from the interaction currents.

In algebraic form, 
the six diagrams that make up the CIA can then be written
\begin{eqnarray}
\fl 
J^\mu_{\rm CIA} = 3e\int_{k_1k_2}
\Bigl\{
\bar\Psi_{\lambda_1\lambda_2\alpha'}(k_1,k_2,k^+_3)\,
{\cal O}_{12}
\,j_{\alpha'\alpha}^\mu(k^+_3,k_3)\,
\Psi_{\lambda_1\lambda_2\alpha}(k_1,k_2,k_3)\nonumber\\
\fl \qquad 
{}+\bar\Gamma_{\lambda_1\beta'\alpha}(k_1,k^+_2,k_3)
\,S_{\beta'\beta}(k_2^+)
\,j_{\beta\gamma}^\mu(k_2^+,k_2)\,
{\cal O}_{12}\,u_{\gamma}(k_2,\lambda_2)\,
\Psi_{\lambda_1\lambda_2\alpha}(k_1,k_2,k_3)\nonumber\\
\fl \qquad
{}+\bar\Psi_{\lambda_1\lambda_2\alpha}(k_1,k_2,k^+_3)
\,\bar u_{\gamma}(k_2,\lambda_2)\,
{\cal O}_{12}
\,j_{\gamma\beta'}^\mu(k_2,k^-_2)\,S_{\beta'\beta}(k_2^-)\,
\Gamma_{\lambda_1\beta\alpha}(k_1,k^-_2,k^+_3)
\Bigr\}\, , 
\label{CIA}
\end{eqnarray}
where 
${\cal O}_{12}=1+2\zeta {\cal P}_{12}$,
%
$j^\mu_{\alpha'\alpha}(k',k)$ is the single-nucleon current discussed above, 
$k_i^\pm=k_i\pm q$, and in every term
$k_1^2=k_2^2=m^2$ and $P=k_1+k_2+k_3$ is the
four-momentum of the incoming three-nucleon bound state.  
The interaction current diagrams give
\begin{eqnarray}
\fl J^\mu_{\rm I} = & 3e\int_{k_1k_2k'_2}
\bar\Psi_{\lambda_1\lambda'_2\alpha'}(k_1,k'_2,k'_3)
{\cal O}_{12}
V_{\lambda_2'\lambda_2,\alpha'\alpha}^\mu
(k'_2 P'_{23};k_2 P_{23}){\cal O}_{12}
\Psi_{\lambda_1\lambda_2\alpha}(k_1,k_2,k_3)\, ,
\label{IACelastic}
\end{eqnarray}
where 
$P'=P+q=k_1+k'_2+k'_3$, $k_1^2=k'^2_2=k_2^2=m^2$,  
$P^{(\prime)}_{23}=k^{(\prime)}_2+k^{(\prime)}_3$, and
$V^\mu$ is the symmetrized two-body interaction
current  satisfying the two-body WT identity (\ref{eq:29}).  

Note that this calculation  requires knowledge of the three-nucleon vertex 
function with the {\it two interacting\/} nucleons off-shell.  This vertex 
function was defined in Ref.~\cite{Gro04} and can be obtained using the 
Faddeev equation~(\ref{eq:3NFad}), generalized to the case when 
{\it both\/} of the final state interacting nucleons are off-shell.
It requires the scattering amplitude for both of the final particles 
off-shell, which is obtained by quadratures from the off-shell kernel 
and the on-shell scattering amplitude using equation~(\ref{eq:CSTBS-NN}),
illustrated in figure~\ref{fig:Meqoffa}(A).
The off-shell kernel 
is  known (in principle), and is discussed in more detail in 
Refs.~\cite{Pin09a,Gross:2014zra,Gro14b}.

When the three-nucleon 
form factors were calculated with high-precision models 
WJC-1 and WJC-2~\cite{Pin09b}, the three-nucleon 
vertex functions with two nucleons
off-shell were not available, and were replaced by vertex functions
with only one off-shell nucleon. This approximation was called ``CIA-0'', 
because it can be interpreted as the zeroth-order Taylor expansion of the 
vertex function in the momentum of one nucleon around its on-shell point. 
To test its quality, CIA-0 was compared with CIA for the case of the W16 
model and found to be an excellent approximation~\cite{Pin09b}. 

When $q\to0$, the nucleon propagators in the second and
third terms of the CIA result of equation~(\ref{CIA}) develop
singularities that cancel, leading to terms involving the
derivatives of the two-body kernel. 
Diagrams (B) and (C), as well as (E) and (F), of figure~\ref{fig:coreFF} should 
therefore always be considered together to allow  these cancellations to take 
place.  The same situation occurs in the two-body case, where the (B$_\pm$) 
diagrams of figure~\ref{Fig1} must be considered together.

\subsection{Appraisal of the different theoretical approaches}
\label{subsec:comp}

The conventional and $\chi$EFT approaches
use Hamiltonian dynamics, and treat relativity and electromagnetic gauge invariance 
perturbatively, expanding in powers of the ratio of the typical momentum of the 
process to the nucleon mass.  Using  $\chi$EFT, the potential is  expanded in a 
power series.  The advantage of these methods is that the underlying formalism is 
well-known and familiar, and the perturbative treatment of the  potential permits 
an estimate of the theoretical error.  The disadvantage is that the range of 
convergence of the $\chi$EFT series (the chiral symmetry breaking scale, 
$\Lambda_\chi\simeq1$ GeV) is low for some of the calculations of few-body form 
factors presented here, which extend to momentum transfers beyond 1 GeV.  
The method in its purest form should produce results independent of the cutoff 
scale, because all divergences can be absorbed into parameters that encode the 
short range physics that cannot be calculated dynamically~\cite{Kap98}, yet in 
the practical applications reviewed here, the requirement of cutoff independence 
limits the  cutoff to  a range of about 500 to 600 MeV, near the scale of the 
important physics described by  two-pion exchange.  Furthermore, the number of 
short-range parameters is not small; there are 24 undetermined parameters needed 
for the nuclear force at N3LO or $({\cal Q}/\Lambda_\chi)^4$, and 5 more for the  
electromagnetic current to the orders discussed in this review.  While these 
parameters can, in principle, be calculated from QCD,  today they must be fit 
to experiment.  However, this is the only approach that can currently be used 
to describe nuclei for mass numbers $A\ge 4$.

 The CST approach 
 is almost completely complementary to the $\chi$EFT approach (which, for the 
purposes of this discussion, includes the
material presented in
section~\ref{subsec:conv} 
as well as section~\ref{subsec:chieft}).  It is based on covariant field theory 
and requires the development of a less familiar formalism (using covariant 
equations with relativistic normalization conditions) needed to treat the 
nonperturbative features of hadronic interactions at high energy.  
An advantage of the formalism is that it is fully covariant, exactly gauge 
invariant, and includes a four-current consistent,  to all orders, with the 
dynamics.  Disadvantages are that there is no systematic way to construct the 
kernel ({\it i.e.}, potential) as there is in  $\chi$EFT, making it difficult 
to estimate the theoretical error, and the needed  regularization is achieved by 
using hadronic form factors with cutoff masses that depend very sensitively on 
the fit.  However, the  OBE models used with this approach can be very efficient: 
one model has only 15 parameters and yet gives a fit to the $np$ data of the same 
quality as the best conventional models.    Unfortunately,  the necessary 
equations  have not yet been extended beyond the $A=3$ sector, and fits to the 
$pp$ data have yet to be done.

These two methods ($\chi$EFT and CST)  are so completely different from start to 
finish that it is difficult to compare them. Here we look at a few examples of 
how they might be describing the same physics in completely different language.

\begin{figure}
\begin{minipage}{0.4\textwidth}
\includegraphics[width=1.2\textwidth]{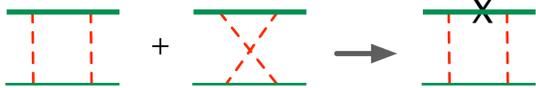}
\end{minipage}
 \begin{minipage}{0.6\textwidth}
\caption{
(Color online) Diagrammatic representation of the cancellation theorem discussed in the text.  The heavy baryon is represented by the thick line.  
}
\label{fig:cancellation}
\end{minipage}
\end{figure}

One key to this understanding may be the {\it cancellation theorem\/}, 
illustrated diagrammatically to 4th order  
in figure~\ref{fig:cancellation}~\cite{Gro93}.    
In cases where a neutral meson is exchanged between a heavy baryon with mass 
$M$ and another baryon of mass $m$, 
the sum of the Feynman ladder and crossed ladder diagrams is well approximated 
by the single box diagram with the heavy baryon on its mass shell.  
Briefly, this comes about because the evaluation of both of these diagrams reduces 
to the sum of contributions from nucleon and meson poles, and in the heavy mass 
limit  the ``large'' meson pole contributions tend to cancel. 
This cancellation works to all orders, and in the limit when $M\to\infty$ it is 
exact. Therefore, the success of the CST-OBE models may be due in part to the 
fact that 
CST sums these leading contributions to all orders 
(recall figure~\ref{fig:Meqoffa}).  If a charged meson is exchanged  the 
cancellation is not exact, but it has been shown that for charged pion exchange 
the residue is well approximated by the exchange of scalar mesons with a 
distributed mass~\cite{Gro82a,Pena:1996tf}, giving another reason why scalar 
meson exchange is so important.  In $\chi$EFT, the same physics is included 
order-by-order through (i) explicit calculations of crossed pion boxes and other 
time orderings (including contributions from so-called ``stretched boxes''), 
and (ii) non-static corrections arising from the iteration of static 
approximations to lower order irreducible potentials (recall the discussion 
in section~\ref{sec:road}).
The conclusion to be drawn from this discussion is that a simple comparison of 
the physics included in the two methods, including isolation of any differences 
that might exist,  is difficult.

\begin{figure}
\includegraphics[width=1\textwidth]{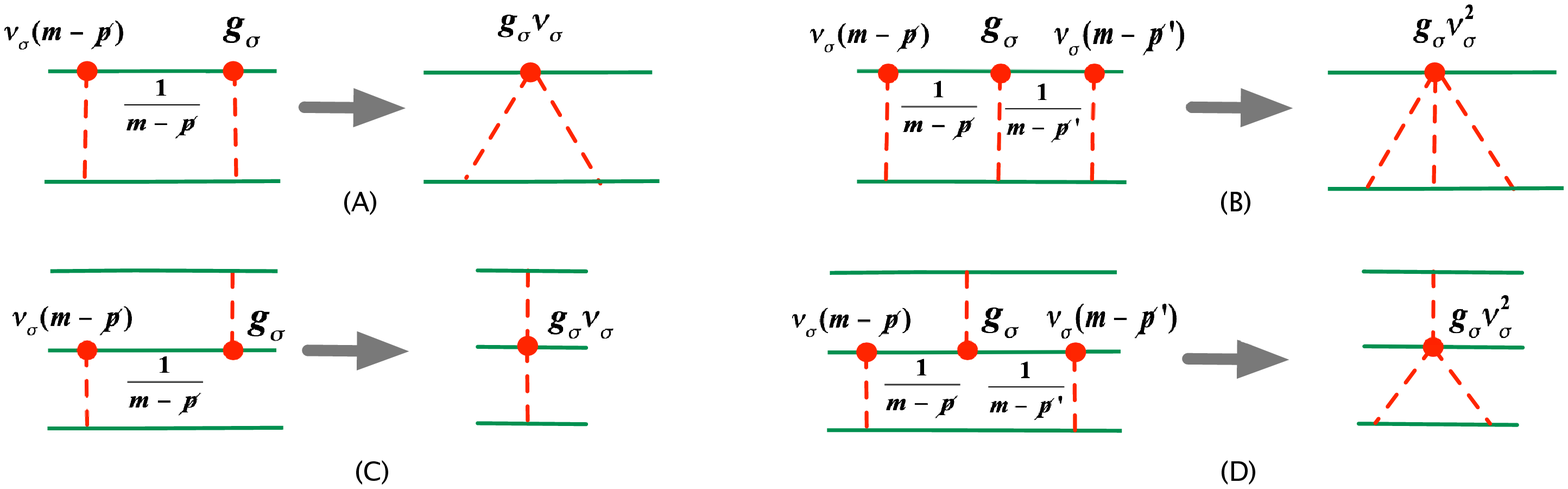}
\caption{
(Color online) Boson-nucleon vertices with off-shell couplings (those that 
depend on the operator $\Theta$) can generate effective $NN$ force diagrams 
beyond OBE, and effective thre-nucleon forces. In these examples the off-shell 
scalar $\sigma$ meson coupling generates an effective triangle contribution to 
the $NN$ interaction [panel (A)] and a double triangle contribution [panel (B)]. 
In the three-nucleon sector, off-shell exchanges between two different nucleons 
can also generate effective three-nucleon forces [panels (C) and (D)].  
Similar mechanisms arise from the pseudovector $\pi NN$ vertex.   
}
\label{fig:vertex-contact}
\end{figure}

A second key to the comparison is illustrated in figure~\ref{fig:vertex-contact}.   The CST off-shell OBE couplings  will cancel neighboring propagators, making it possible to reinterpret the iteration of off-shell contributions to OBE  exchanges as contact interactions, which can generate many non-OBE contributions in the two-body sector and effective three-body force contributions in the three-body sector.  As suggested by the figure, this can happen in an infinite number of ways, showing how some of the triangle and three-body force diagrams needed in  $\chi$EFT emerge from CST-OBE models, even though they do not appear to be included anywhere.  This insight probably explains why CST-OBE can predict the three-body binding energy (recall figure~\ref{fig:WJC1-WJC-2-chi2-Et})  without adding CST  three-body forces, while they are essential if one uses $\chi$EFT or the conventional approach.

As has been understood since the first days of the CST 
(see, for example, Ref.~\cite{Gross:1972ye}), the antiparticle contributions 
from the CST off-shell propagators also make higher order contributions similar 
to those just described.  In the rest frame of a two-body system, the propagator 
of the off-shell particle 2 may be decomposed into its positive and negative 
energy contributions
\bea
\fl G(k)=\frac1{m-\not{k}-i\epsilon}=\frac{m}{E_k}\sum_\lambda 
\Bigg\{\frac{u({\bf k},\lambda)\bar u({\bf k},\lambda)}{2E_k-W}-\frac{v(-{\bf k},
\lambda)\bar v(-{\bf k},\lambda)}{W}\Bigg\}\, ,
\eea   
where $W$ is the energy of the two-body system and $u$ ($v$) are the positive 
(negative) energy Dirac spinors of the nucleon with three-momentum ${\bf k}$.  
Using this decomposition, the deuteron bound state equation can be written in a 
coupled channel form, which in coordinate space becomes, approximately, 
%
\bea
(2E_\nabla-M_d)\psi^+(r)&=&-V^{++}(r)\psi^+(r) - V^{+-}(r)\psi^-(r)
\nonumber\\
\qquad\quad -M_d\, \psi^-(r)&=&-V^{-+}(r)\psi^+(r) - V^{--}(r)\psi^-(r)\, . 
\label{eq:Vpm}
\eea
Eliminating $\psi^-$ gives an effective potential for the positive energy sector
\bea
(2E_\nabla-M_d)\psi^+(r)&=&-\Big[V^{++}(r) + \frac{V^{+-}(r)V^{-+}(r)}{M_d-V^{--}(r)}
\Big]\psi^+(r)\, .  \label{eq:Veff}
\eea
The short-range piece proportional to $V^{+-}V^{-+}$ is positive definite 
in central channels, and when combined with a modest  $\omega$-exchange term 
gives a credible explanation of the  short-range repulsion needed for an 
explanation of the $NN$ data.  Perhaps more significantly, this simple argument 
demonstrates how  negative-energy (antiparticle) channels generate higher order 
terms that could be compared to the  $\chi$EFT expansion.  As is obvious from 
the discussion in the last two paragraphs, the comparison of the dynamics 
included in CST-OBE with that included in $\chi$EFT is a challenging problem that 
has yet to be seriously studied. 

All of these considerations also apply to the construction of the current 
(which, in the language of the $\chi$EFT or conventional approaches, is separated 
into charge and three-vector current operators).  The current should be given 
by the theory, and in this respect $\chi$EFT provides clear guidance for its 
construction. Unfortunately, both $\chi$EFT and CST depart from their underlying 
field theories by using the measured structure of the nucleon to describe the 
$\gamma N\to N$ vertices, thereby giving up the goal of describing the few body 
systems entirely from first principles.  
Furthermore, the internal couplings of the photon to the pion (and, in CST, 
other mesons)  are also not calculated from first principles.
In the isoscalar channel needed for the deuteron form factors, coupling to mesons 
in flight does not contribute to the currents, so the isoscalar   interaction 
currents are more tightly constrained.  The freedom that might have existed 
because of the momentum dependence arising from the off-shell 
($\nu$-dependent terms) in the CST has recently been 
constrained~\cite{Gross:2014zra}.  However, there are always exotic interaction 
currents, such as the famous $\rho\pi\gamma$ and $\sigma\omega\gamma$ 
contributions~\cite{Hum89}.  These effects are purely transverse, and hence cannot 
be constrained in CST by the WT identities that link the currents to the $NN$ 
interaction. The two isoscalar contact interactions that arise in the $\chi$EFT 
expansions include these unknowns, but the unknown form factors associated with 
these terms are still a problem.  Fortunately, the data show that these effects 
are small.

We conclude this discussion with the observation that  
the interactions between hadrons can be described using either confined 
quark-gluon degrees of freedom or physically observable hadronic (i.e. colorless) 
degrees of freedom. 
Which choice is most efficient  depends on the energy scale of the physics being 
studied.    In this respect $\chi$EFT  and CST are similar: CST uses nucleons and 
mesons with masses below 1 GeV, while $\chi$EFT  uses nucleons and pions (only).  
The great advantage of $\chi$EFT  is the organizational principle it provides: 
the precise definition of a perturbation series for the potential rooted in the 
approximate chiral symmetry of QCD.  Unfortunately the scale at which this series 
diverges is probably less than 1 GeV.  
The organizational principle behind the CST is similar to that in vogue for many 
years: the exchange of the lightest mesons should account for the longest range 
(and hence largest) force, as suggested by the famous Yukawa potential 
proportional to 
\bea
V_b(r)\simeq C_b\frac{e^{-m_br}}{r}.
\eea 
The longest range forces are  described by the pion, and the intermediate and 
shorter range forces described by the exchange of heavier mesons (or, in the 
language of $\chi$EFT, multiple pion exchanges). This organizational principal 
is less precise and not as well defined as the one provided by $\chi$EFT, but 
works extremely well if CST equations are used with OBE parameters adjusted to 
fit the data.  The most unifying  conclusion to be drawn is that perhaps a 
comparison of how the physics is described by these two different approaches 
may teach us more than the study of either of them alone.  This can be done by 
making a systematic nonrelativistic expansion  of the CST equations, along the 
lines outlined in equations~(\ref{eq:Vpm}) and~(\ref{eq:Veff}); it is yet to be done.

\section{Results: theory vs. experiment}
\label{sec:res}

In this section the theoretical
predictions  for the electromagnetic
form factors of the Hydrogen and Helium isotopes  corresponding to the 
conventional, $\chi$EFT,
and CST approaches described in section~\ref{sec:theory} are compared to the 
experimental data obtained in the global analysis discussed in
section~\ref{sec:exp}.  We first discuss the sensitivity of the various 
theoretical approaches to their physical content (section~\ref{sec:ms}), 
and then compare the predictions of the different approaches to data 
(section~\ref{sec:com}).    
In section~\ref{sec:com} we also present results for the 
charge and magnetic radii, and the magnetic
dipole and electric quadrupole moments.


\subsection{Model sensitivity} \label{sec:ms}

Each of the theoretical approaches makes certain assumptions about the physics 
of electron scattering, and the final result is usually the sum of several 
different contributions.  In this subsection we discuss the relative sizes of 
these contributions and the sensitivity of predictions to the values of
unknown input parameters entering the various approaches.

%
\begin{figure}[bth] 
\vspace*{1cm}
\centerline{
\includegraphics[width=14cm
]{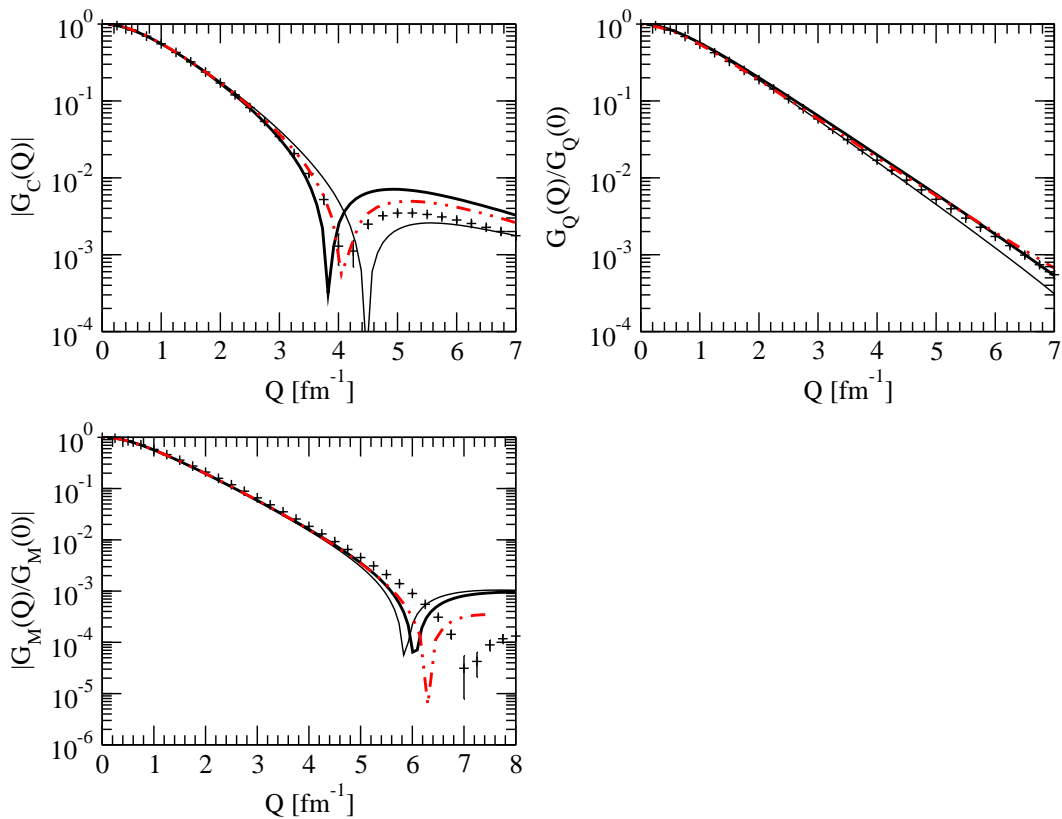}
}
\caption[]{(Color online) The deuteron 
form factors calculated in the conventional approach
are compared to data.  The thin and thick (black) solid lines
correspond, respectively, to results obtained by retaining 
in the charge and current operators only one-body (IA) or
both one- and two-body contributions.  The dash-double-dotted (red)
line corresponds to the approximately relativistic calculation
mentioned in the text, and includes contributions from one- and two-body 
electromagnetic operators.}
\label{fig:ff1}  
\end{figure}

\begin{figure}[bht] 
\centerline{\includegraphics[width=14cm
]{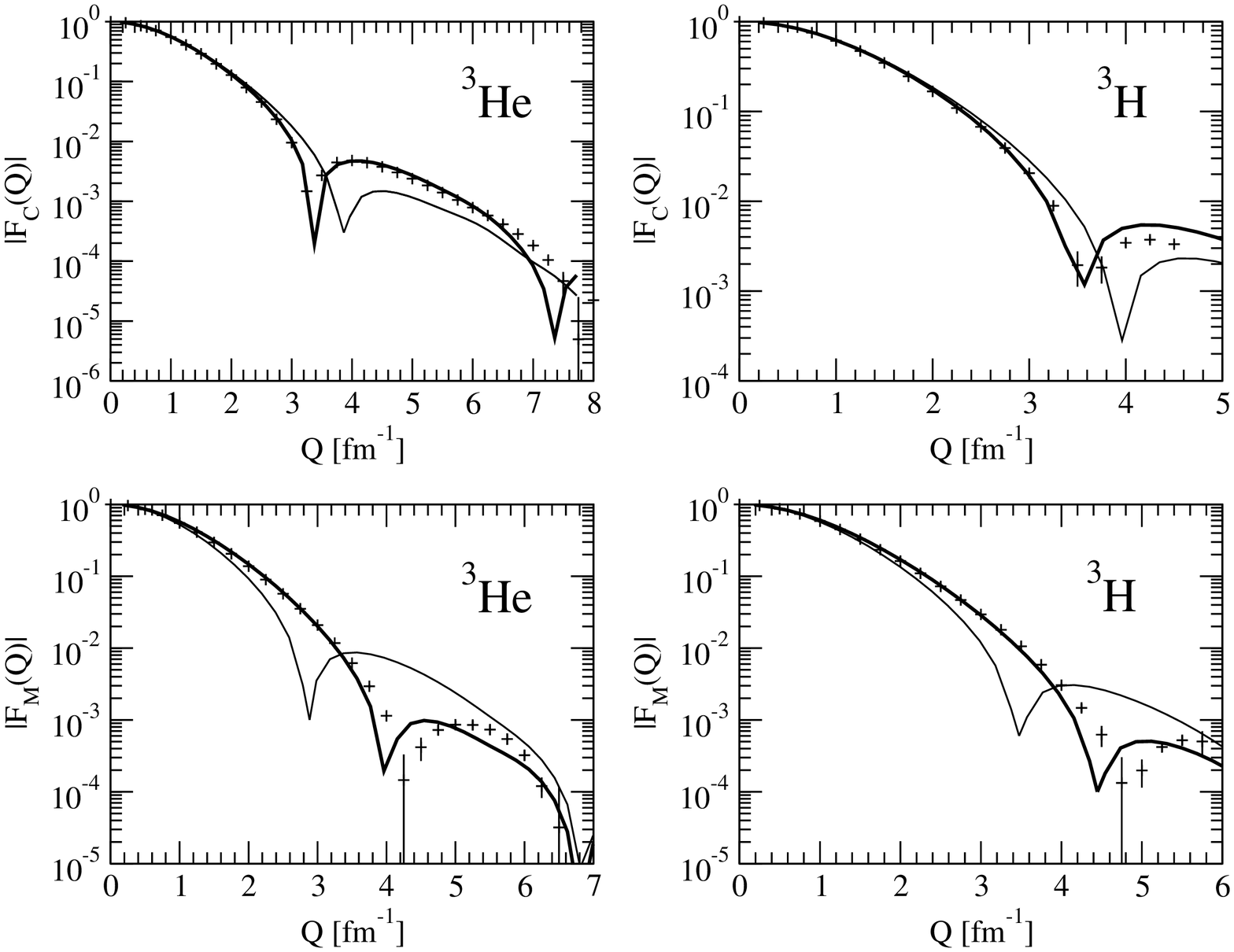}}
\caption[]{$^3$He and $^3$H charge and magnetic form factors 
as function of the momentum transfer $Q$ (in fm$^{-1}$),
calculated within the conventional approach based
on the AV18/UIX realistic potentials, are compared with
experimental data.  The thin and thick solid lines
correspond to results obtained by retaining, respectively,
one-body only or both one- and many-body contributions 
in the current and charge operators.}
\label{fig:ff5}  
\end{figure}

\subsubsection{Conventional approach}

This approach  uses the
non-relativistic Argonne $v_{18}$ (AV18) two-nucleon potential~\cite{Wir95}
for the $A=2$ system.  When applied to the $A=3$ and 4 systems,
it includes the (non-relativistic) Urbana IX (UIX) three-nucleon
potential~\cite{Pud95}, whose strength parameters are adjusted to
reproduce the $^3$H binding energy in exact Green's function Monte
Carlo calculations and the saturation density of nuclear matter
in hypernetted-chain variational calculations.  The resulting
AV18/UIX Hamiltonian then leads to $^3$He and $^4$He binding
energies in excellent agreement with the empirical values.  Leading
two- and three-body terms in the electromagnetic charge and current
operators are constructed from the AV18 and UIX potentials, as outlined in
subsection~\ref{subsec:conv} and described in detail in Ref.~\cite{Mar05}.
For the case of the deuteron only, we also present results obtained with
an approximately relativistic treatment of nuclear dynamics, based on
the relativistic Hamiltonian in equation~(\ref{eq:hhrr}) and by including
one- and two-body terms in the electromagnetic operator as well as
boost effects to order $(v/c)^2$ in the initial and final states.


Figure~\ref{fig:ff1} compares
theoretical predictions for the deuteron form factors obtained with the 
AV18 potential
using a one-body current only, and the full one plus two-body current
operators.  For this latter case 
results
obtained with the approximately relativistic Hamiltonian are also presented.  
Two-body charge contributions, predominantly
due to the pion-range operator of equation~(\ref{eq68a}), have
opposite signs in $G_C$ and $G_Q$, substantially
reducing $G_C$ while moderately increasing $G_Q$.
%

The sensitivity of the three-body form factors to many-body terms in the
electromagnetic current is displayed in figure~\ref{fig:ff5}.  
These many-body terms
make important contributions for momenta larger than about 2 fm$^{-1}$.

\subsubsection{$\chi$EFT approach}   
The $\chi$EFT calculations are based on the next-to-next-to-next-to-leading
order (N3LO) two-nucleon potentials of Refs.~\cite{Ent03,Machleidt11}
corresponding to short-range cutoffs $\Lambda=500$ and 600 MeV/c---denoted
respectively as N3LO(500) and N3LO(600)---and retain, in the case of the
$A=3$ and 4 systems, the next-to-next-to-leading order (N2LO) three-nucleon 
potential
in the local form of Ref.~\cite{Nav07}.  The low-energy constants (LEC's)
$c_D$ and $c_E$ (in standard notation) that characterize this three-nucleon 
potential 
have been constrained by reproducing, in essentially exact calculations based
on hyperspherical-harmonics techniques~\cite{Mar12}, the $^3$H/$^3$He binding
energies and the tritium Gamow-Teller matrix element for each of the $\Lambda$
values considered---one of the LEC's also enters
the nuclear weak axial current and can therefore be
determined in a weak transition.  The resulting Hamiltonians are denoted as
N3LO/N2LO(500) and N3LO/N2LO(600) below.   Up to one loop,
no unknown LEC's enter in the electromagnetic charge operator,
beyond the nucleon axial coupling constant $g_A$, pion decay
amplitude $f_\pi$, and proton and neutron magnetic moments,
the latter associated with a next-to-next-to-leading order (N2LO)
relativistic spin-orbit correction (see the N2LO panel in figure~\ref{fig:f5}).
In contrast, the electromagnetic current operator up to one loop is
characterized by an additional five LEC's.  In the $\chi$EFT
results presented in the following subsections, the values
listed in tables~\ref{tb:tds} and~\ref{tb:tdv} (set III) are considered:
the two LEC's multiplying isoscalar operator structures are
fixed by reproducing the deuteron and trinucleon isoscalar 
magnetic moments, while estimates for two of the three isovector LEC's
are obtained from $\Delta$-isobar saturation
arguments, with the remaining LEC fixed by
reproducing the trinucleon isovector magnetic moment.

\begin{figure}[bht] 
\vspace{0.8cm}
\centerline{
\includegraphics[width=14cm
]{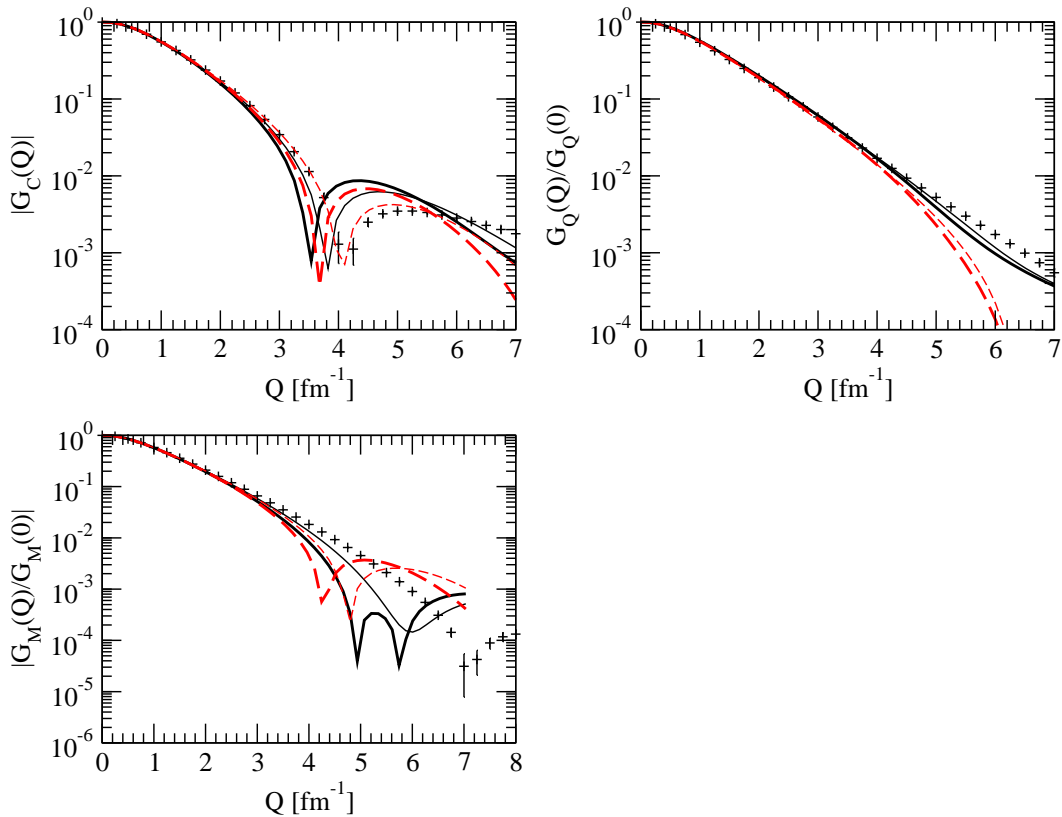}
}
\caption[]{(Color online) The deuteron 
form factors calculated in the $\chi$EFT approach are
compared to data.  The thick and thin dashed (solid) lines
are results for
the N3LO(500)  and N3LO(600) chiral potentials using
LO (up to N3LO) corrections in the
charge and current operators. 
}
\label{fig:ff2}  
\end{figure}

Figure~\ref{fig:ff2} shows predictions for the deuteron form factors based on 
the N3LO(500) and
N3LO(600) chiral potentials and including either LO electromagnetic
operators only or corrections up to N3LO to these operators (see
figures~\ref{fig:f2} and~\ref{fig:f5}).  Loop corrections
at N4LO in the charge operator do not contribute to the observables
under consideration, since they are isovector.  The full results
reproduce well the measured form factors for momentum transfers
up to 2--3 fm$^{-1}$.  For $G_Q$, the agreement between theory and
experiment extends over a considerably wider region of momentum
transfers, in fact well beyond the range that one would naively expect
to be valid for the present $\chi$EFT expansion.  The cutoff dependence
is modest for $G_Q$, but pronounced for $G_C$ and $G_M$.
Finally, we note that, as in the case of the conventional approach,
the N3LO one-pion-exchange charge operator in panel (c) of figure~\ref{fig:f5}
gives the dominant contribution beyond LO.  However, its effect on
$G_C$ and $G_Q$ is significantly smaller than obtained in the conventional
approach.   This is due to the fact that the functional forms of the cutoff 
used to regularize the
operator in these two approaches are different and, more importantly, the value 
for $\Lambda$
in the conventional AV18 calculation ($\Lambda \simeq 1200$  MeV/c)
is much larger than those adopted in the $\chi$EFT
calculations ($\Lambda$ either 500 or 600 MeV/c).

\begin{figure}
\centerline{\includegraphics[width=14cm
]{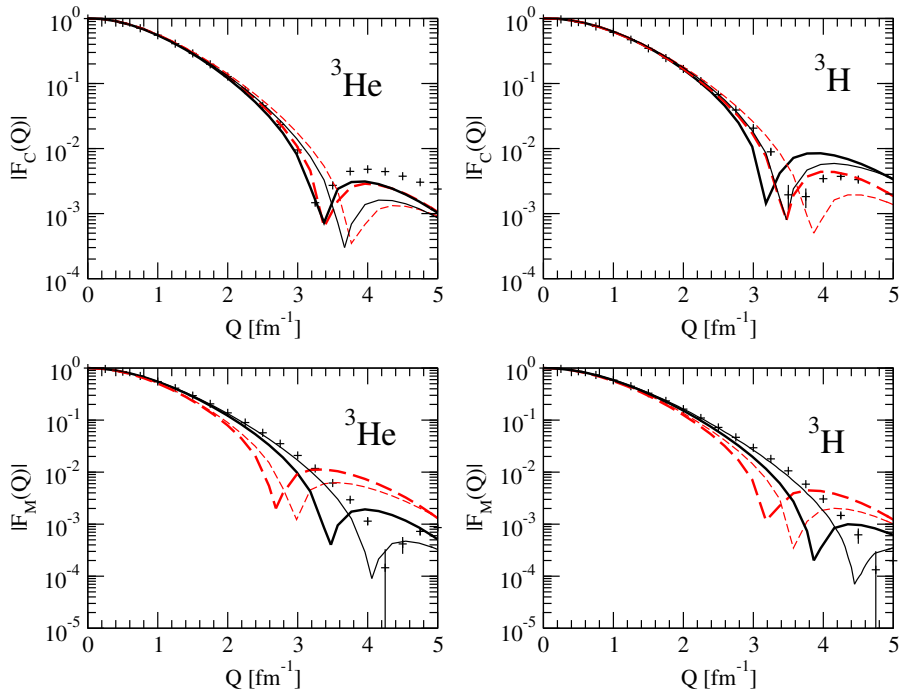}}
\caption[]{(Color online) $^3$He and $^3$H
charge and magnetic form factors 
as function of the momentum transfer $Q$ (in fm$^{-1}$),
calculated within the $\chi$EFT approach, are compared with
experimental data.  The thick and thin dashed (solid) lines
correspond to results obtained using
the N3LO/N2LO(500) and N3LO/N2LO(600) chiral potentials,
with corrections in the charge and current operators up to LO (up to N4LO).}
\label{fig:ff6}  
\end{figure}

Figure~\ref{fig:ff6} shows the sensitivity of the three-body form factors to 
the chiral cutoff and the order of the calculation.  Here, since there are 
isovector contributions to the currents, corrections to the current up to N4LO 
are retained. 

\subsubsection{CST approach}

Unfortunately, 
the most recent calculation for the deuteron form factors, reported in 
1995~\cite{VanOrden:1995ca}, uses the older Model IIB (with parameters given 
in table~\ref{tab:cstOBEa}).  This model has no off-shell scalar meson 
couplings 
(the parameter $\nu_s=0$, ensuring that the  off-shell projection operators 
$\Theta$ given in table~\ref{tab:Ls} are absent), yet these terms  are 
needed to 
give the high precision fits to the $np$ scattering data~\cite{Gro08}. 
 Since the momentum dependence in these terms generates a new class of 
isoscalar 
interaction currents, a new generation of calculations that includes these 
currents is required.  In 2014, using principles of simplicity and picture 
independence, these isoscalar interaction currents were uniquely 
determined~\cite{Gross:2014zra}, and excellent results for the deuteron 
magnetic 
moment~\cite{Gro14b} and quadrupole moment~\cite{Gross:2014tya} were obtained 
(reported below).  The new results for the form factors were not available at 
the time this review was prepared.

\begin{figure}
\vspace{0.8cm}
\centerline{
\includegraphics[width=14cm
]{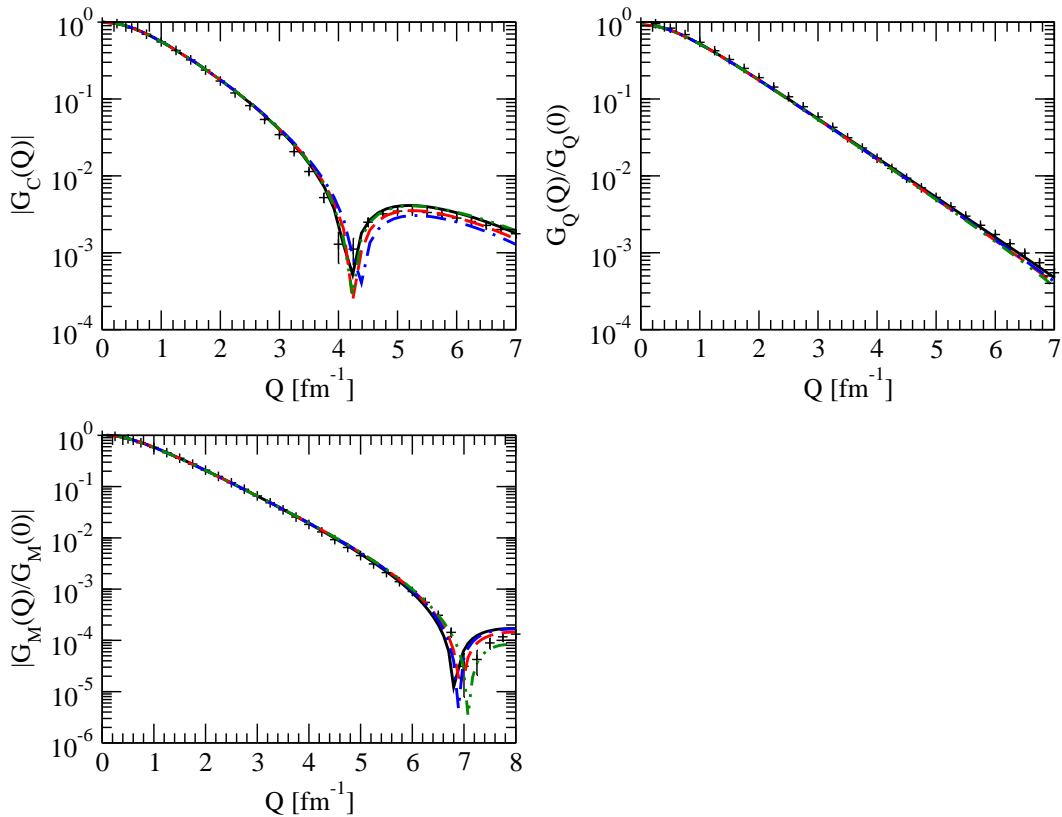}
}
\caption[]{(Color online) The deuteron form factors calculated in the CST 
approach are
compared to data.  The solid line (black) is the full calculation, the dashed  
line (red) has no $\rho\pi\gamma$ exchange current (the CIA), the dash-dotted 
line (blue) is the extreme limit $F_3=1$, and the dashed-dot-dot line (green) 
is the extreme limit $F_3=0$.) } 
\label{fig:ff3}  
\end{figure}

Model IIB requires no interaction currents, but the results depend on the new, 
unknown, off-shell nucleon form factor, $F_3(Q)$, which contributes 
only when 
the incoming and outgoing nucleons are {\it both\/} off-shell.  In has also 
been 
customary to add the contributions of a $\rho\pi\gamma$ interaction current, 
which is separately conserved and therefore not constrained by the requirements 
of current conservation.  The size of these effects are shown in 
figure~\ref{fig:ff3}, where the the $\rho\pi\gamma$ form factor and coupling 
constant were taken from Ref.~\cite{Cardarelli:1995ap} 
(with the model 2 form factor).  The figure shows the extreme limits of 
$F_3(Q)=0$ or 1 [actually, $F_3(Q)=0$ is impossible because of 
the constraint $F_3(0)=1$, and is shown only to provide a lower limit, 
and while $F_3(Q)=1$
is possible, it violates our expectations that 
$F_3\to0$ as $Q\to\infty$].  
Note that all of these effects are small except at high momentum transfers. 

\begin{figure}[bht] 
\vspace*{0.8cm}
\centerline{
\includegraphics[width=14cm,angle=0]{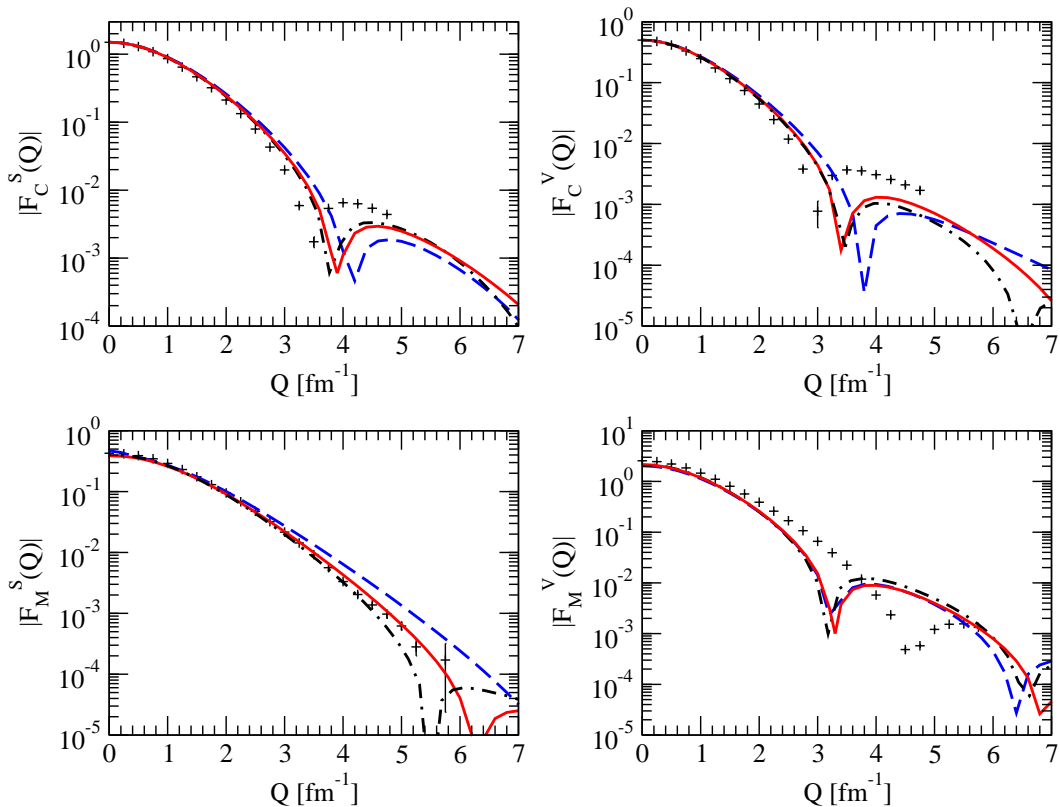}}
\caption[]{(Color online) Isoscalar (left panels) and isovector (right panels) 
combinations of the charge (upper panels) and magnetic (lower panels) $A=3$ 
form factors as function of the momentum transfer $Q$ (in fm$^{-1}$),
calculated in the CST approach, are compared with experimental data. 
The results for the WJC-2 (red solid lines) and 
WJC-1 (blue dashed lines) were calculated in CIA-0. 
The results obtained within the conventional approach based on the 
AV18/UIX potentials using the one-body operator and the same (Galster)
nucleon form factors are also shown as dash-dotted lines.}
\label{fig:fcm_cst}  
\end{figure}

The three body calculations are more recent~\cite{Pin09a,Pin09b}, and were 
done when the models WJC-1 and WJC-2 were available. Figure~\ref{fig:fcm_cst} 
shows the isoscalar and isovector combinations of the $A=3$ charge and 
magnetic form factors  
in the complete impulse approximation CIA-0 described in 
section~\ref{sec:3NCST}. A moderate model dependence can be observed for 
momentum transfer above about 3 fm$^{-1}$ (in the isovector magnetic form 
factor above 5 fm$^{-1}$). The difference can be traced back to the different 
pion-nucleon coupling of the two models: the pseudoscalar admixture in 
WJC-1 automatically generates strong Z-diagram contributions suppressed by 
the pure pseudovector coupling used in WJC-2. 

To compare the CST calculations to the experimental data, at least the 
dominant pion-exchange currents would have to be taken into account, in 
particular the $\gamma \pi NN$ contact terms induced by pseudovector 
$\pi NN$ coupling. They are part of diagrams (G) - (J) of 
figure~\ref{fig:coreFF}, which are not present in CIA.
But Z-diagrams for pseudoscalar coupling are roughly equivalent to these 
pseudovector contact terms, so they are already partially included in the 
WJC-1 calculations in CIA. However, the corresponding mixing parameter has 
the opposite sign, so---compared to WJC-2---this contribution actually moves 
the form factors away from the data. In a complete calculation this will be 
compensated by a stronger contact term coming from WJC-1's pseudovector 
$\pi NN$ component. 

A direct comparison of a CIA calculation with data makes sense only for model 
WJC-2 in the case of the isoscalar magnetic form factor, where the 
$\gamma \pi NN$ contact term is suppressed. And indeed, as 
figure~\ref{fig:fcm_cst} shows, the model describes the data very well.
Figure~\ref{fig:fcm_cst} also compares the CST models in CIA with a 
calculation in the conventional approach performed with the AV18/UIX potential 
in impulse approximation with relativistic corrections and using the same 
(Galster) nucleon form factors. The close agreement between the results 
of AV18/UIX and WJC-2 can again be attributed to the suppression of 
Z-diagrams in WJC-2.

These results alone would not be sufficient to give preference to either 
WJC-1 or WJC-2. However, recent precision calculations of the deuteron 
quadrupole moment \cite{Gross:2014tya} seem to prefer WJC-2. For this reason, 
when comparing later on to the other approaches the $A=3$ results with WJC-2 
are shown.

\subsection{Comparison of the different approaches with data} \label{sec:com}

Electromagnetic form factors characterizing the coupling
of the external field to individual hadrons enter the nuclear current and
charge operators.  Those for the proton and neutron in the dominant
one-body terms of these operators are taken from fits to elastic
electron scattering data off the proton and deuteron, specifically
the dipole parameterization (including the Galster factor for the
neutron electric form factor) in the conventional approach, the
H\"ohler parameterization~\cite{Hoh76} in the $\chi$EFT approach,
and the GKex05 parameterization~\cite{FF_GK_05_2_Lomon_02,FF_GK05_1_Lomon_06}
in the CST approach.  We note
that for momentum transfers up to $\simeq 6$--7 fm$^{-1}$, i.e., the
$Q$-range over which most of the results are presented below,
these various parameterizations do not differ significantly.
Hadronic electromagnetic form factors also enter the nuclear
many-body current and charge operators, and the specific
parameterizations adopted for these have been 
briefly discussed in
section~\ref{sec:theory} and more extensively in the original
references.

\begin{figure}[bht] 
\vspace*{0.8cm}
\centerline{
\includegraphics[width=14cm
]{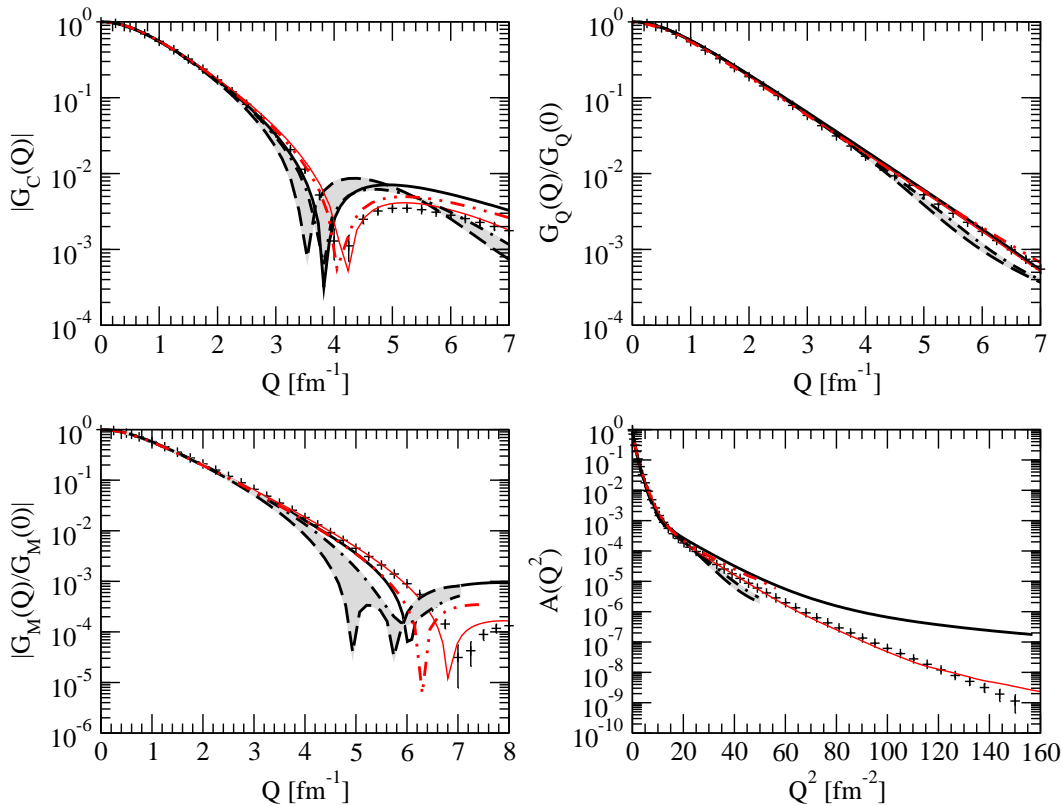}
}
\caption[]{(Color online) 
The deuteron experimental charge, quadrupole, and magnetic 
form factors compared to results obtained from (i) the conventional
approach using the AV18 potential and including one- and two-body terms in 
the charge and current operators  
(thick solid line); (ii) an approximately relativistic version of the 
conventional calculation using equation~(\ref{eq:hhrr}) with the AV18 
potential and including contributions from one- and two-body electromagnetic
operators (dash-double-dotted
line); (iii) the $\chi$EFT approach using 
rthe N3LO(500) (dashed line)
and N3LO(600) (dash-dotted line) potentials and including
corrections up to N3LO  in the charge and current operators 
(with the region between these two lines lightly shaded);
and (iv) the CST approach using model IIB (thin solid line).
Also shown are conventional, $\chi$EFT, and CST predictions for the
$A$ structure function measured out to
$Q^2 \simeq 150$ fm$^{-2}\simeq 6$ GeV$^2$. 
} 
\label{fig:ff4}  
\end{figure}

\subsubsection{$^2$H nucleus}
\label{subsec:2h}

Figure~\ref{fig:ff4} compares the experimental data for the three deuteron 
form factors 
and for the $A$ structure function (which has been measured out to
$Q^2 \simeq 150$ fm$^{-2}$), to the full predictions obtained
from the conventional, $\chi$EFT, and CST approaches.
Note that the $\chi$EFT
predictions for $A$ only extend up to  $Q^2 \simeq 45$ fm$^{-2}$, which
is already well beyond the range of applicability of this approach.

At low and moderate values of the momentum transfer ($Q \lesssim 4$
fm $^{-1}$), the three theoretical approaches reproduce the data well.
At larger values of $Q$, particularly in the diffraction region of the
magnetic form factor and for the $A$ structure function, the conventional
and $\chi$EFT results are at variance with data, in the case of
$A$ by orders of magnitude at the highest $Q$'s.  On the other hand,
the CST results provide a remarkably good reproduction of the
data over the whole $Q$-range covered by experiment---note that the
$A$ structure function drops by 9 orders of magnitude over
this range!  The quantitative success of the CST approach
clearly demonstrates the need for a fully relativistic treatment 
of nuclear dynamics at high momentum transfers ($Q \gtrsim 5$ fm$^{-1}$).
It also suggests that, even in the extreme kinematical range
covered by the $A$ measurements, the description of nuclei in terms
of protons and neutrons interacting via effective forces (and
via effective currents with external electroweak fields) is far
more robust than one would have naively expected.  Indeed,
there are no indications for the need to explicitly account for
quarks and gluons, the degrees of freedom of the
fundamental theory (QCD) governing their dynamics.  It
appears, instead, that the effects of the nucleon substructure
can be subsumed in these effective forces and currents.

The deuteron charge radius, and magnetic dipole and electric quadrupole
moments are listed in table~\ref{tb:res_mom2}.  The results in the
conventional approach based on the AV18 potential include one- and
two-body terms in the electromagnetic charge and current operators,
and under-predict the charge radius by 0.5\%, the magnetic moment
by 1.2\%, and the quadrupole moment by 2.0\%.  We note that the
isoscalar two-body current contributions from the momentum-dependent
spin-orbit, ${\bf L}^2$, and quadratic spin-orbit components of the AV18
are individually small and tend to cancel out, since they have opposite signs.
The $\rho\pi\gamma$ contribution is also found to be negligible.
Indeed, the value for $\mu(d)$ reported here, 0.847 n.m., coincides with
that obtained in IA, i.e. with one-body currents only.  The charge radius too
is unaffected by isoscalar two-body contributions in the charge operator.
However, these contributions increase the IA prediction for $Q(d)$ by
over 3\%, but do not fully resolve the discrepancy between theory and
experiment.
\begin{table}[bth]
\caption[Table]
{\label{tb:res_mom2} The deuteron charge radius $r_c(d)$ (in fm), 
magnetic dipole
moment $\mu (d)$ (in nuclear magnetons),  and electric quadrupole moment, 
$Q(d)$ (in fm$^2$), obtained with the three different theoretical approaches, 
are compared to experimental values.  The CST charge radius (marked with *) 
was calculated with the older model IIB, the magnetic and quadrupole moments 
with model WJC-2. 
The numbers
in parentheses at the side of the WJC-2 calculations are an estimate of
numerical errors.
The experimental charge radius is taken from
Ref.~\protect\cite{Sick98}.  The experimental error for $\mu(d)$ is not shown, 
since it is negligible. The numbers in parentheses at the side of the
$\chi$EFT predictions for 
$Q(d)$ and $r_c(d)$ give the cutoff dependence
of the results. The $\chi$EFT result for $\mu(d)$ is underlined, since
$\mu(d)$ is used to constrain one of the two LEC's in the isoscalar
current at N3LO.}
\begin{center}
\begin{tabular}{cccc|c}
\br
Observable & Conv. & $\chi$EFT & CST & Exp.\\
\mr
$r_c(d)$ & 2.119 & 2.126(4) & 2.085*& 2.130(10) \\
$\mu(d)$ & 0.847 & {\underbar{0.8574}} & 0.864(2) & 0.8574 \\
$Q(d)$ & 0.280 & 0.2836(16) & 0.2836(3) & 0.2859(6)\\
\br
\end{tabular}
\end{center}
\end{table}

The $\chi$EFT approach 
leads to
values for the deuteron static properties in very close agreement with
experimental data.  The value for the deuteron magnetic moment is
underlined in table~\ref{tb:res_mom2}, since this observable is used
to constrain one of the two LEC's in the isoscalar current at N3LO.
As already noted, loop corrections at N4LO in the charge operator 
(see figure~\ref{fig:f5})
are isovector and do not contribute to the charge radius and quadrupole
moment.  Among the N2LO and N3LO contributions to $Q(d)$,
the dominant one is from the two-body charge operator
associated with one-pion exchange, panel (c) of figure~\ref{fig:f5}.
Lastly, the cutoff dependence represented in table~\ref{tb:res_mom2}
by the numbers in parentheses, is modest for these observables.

The CST numbers for the magnetic and 
quadruple moments~\cite{Gro14b,Gross:2014tya} reported in the table  
are the recent results for model WJC-2.  These include the new interaction 
currents needed to conserve current  that arise from the momentum dependent 
off-shell couplings that depend on the parameter $\nu_s$.  As discussed in 
section~\ref{sec:OBE}, these off-shell couplings are needed to give a high 
precision fit to the $np$ data, and also predict the correct three-body 
binding energy.  The charge radius is from the older 1995 
calculation~\cite{VanOrden:1995ca} and will be replaced once the new results 
of models WJC-1 and WJC-2 are obtained. Note that both the magnetic and 
quadrupole moments are predicted by this calculation, and  agree with 
experiment to about 1\%.

\subsubsection{$^3$He and $^3$H nuclei}
\label{subsec:3h}

\begin{figure}
\centerline
{\includegraphics[width=16cm]
{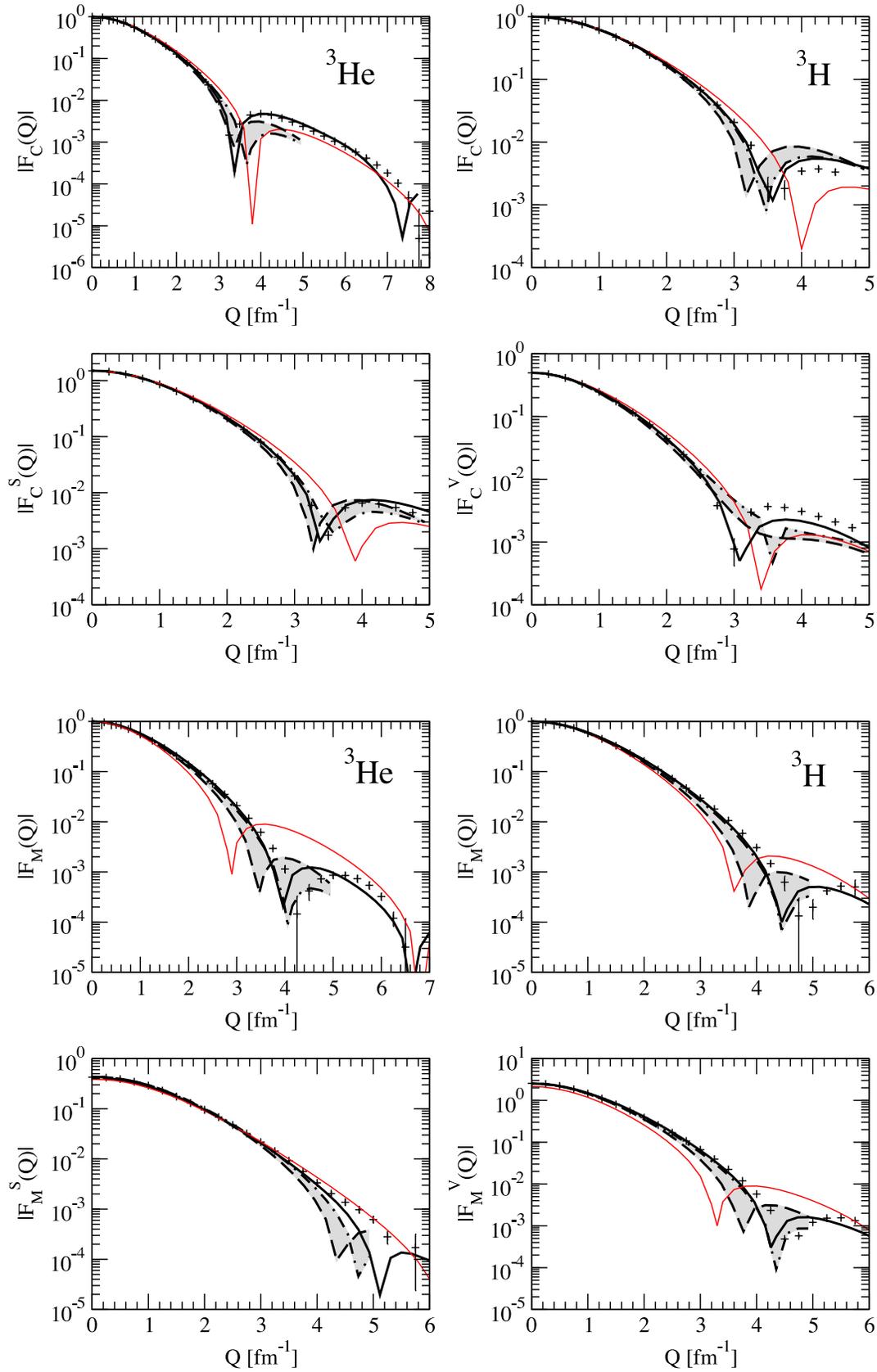}
}
\caption[]{
(Color online) 
The predictions for the form factor from the conventional approach 
(thick solid line), the $\chi$EFT approach for cutoffs of 500 
(thick dashed line) and 600 MeV/c
(thick dot-dashed line)
and the CST approach in 
CIA-0 approximation (thin solid line) are compared to data.}
\label{fig:ff8}  
\end{figure}

The charge and magnetic form factors obtained using the
three theoretical approaches are compared to data in figure~\ref{fig:ff8}.  
The full result for the conventional and $\chi$EFT approaches successfully 
reproduce
the measured form factors up to $Q\simeq 3$ fm$^{-1}$; as a matter
of fact, 
the conventional calculation 
agrees well 
with experiment 
up to 
the first diffraction
region 
and beyond.  On the other hand, even making allowance for the
significant cutoff dependence, the $\chi$EFT results tend to underestimate
the data beyond $Q \gtrsim 3$ fm$^{-1}$; in particular, they predict the
zeros in both $F_C$ and $F_M$ at significantly lower values of $Q$
than observed.  As discussed above, the CST calculation is limited to the 
CIA-0 approximation, which omits interaction current contributions and does 
not fully treat the off-shell behavior  of the vertex functions.  Therefore, 
only the isoscalar magnetic contribution for model WJC-2 should be compared 
to data, and for this one observable (bottom left panel of 
figure~\ref{fig:ff8})  the agreement is good.   


The charge and magnetic radii of $^3$He, the magnetic dipole
moments of $^3$He and $^3$H, and their isoscalar and isovector
combinations, are listed in table~\ref{tb:res_mom3}.  The determination
of the $^3$H radii is affected by the unavailability of accurate data at
low momentum transfer, and values for these radii are not given here.

The conventional ($\chi$EFT) approach uses trinucleon wave
functions obtained with the hyperspherical-harmonics method from the 
AV18/UIX [N3LO/N2LO(500) and N3LO/N2LO(600)]
Hamiltonian.  
``Conventional'' results include one-body, two-body, and three-body
terms in the electromagnetic operators, while the $\chi$EFT results
retain, beyond the LO terms, corrections up to N3LO in the current
and N4LO in the charge operator.  The $\chi$EFT results for the magnetic
moments are underlined, since these observables have been used
to fix the LEC's in the current.  Both the conventional
and $\chi$EFT approaches lead to values
quite close to the empirical ones.  As is well known, IA (one-body or
LO) predictions for the magnetic moments typically underestimate the data 
by $\simeq 15$\%.  Finally, as previously noted,  the CST results given here 
were obtained only in the complete impulse approximation (CIA-0); a more 
reliable prediction must await
a calculation that includes isovector two-body currents induced by pion 
exchange.

\begin{table}
\caption[Table]
{\label{tb:res_mom3} The $^3$He charge and magnetic radii, respectively
$r_c(^3{\rm He})$ and $r_m(^3{\rm He})$ (both in fm), and trinucleon magnetic
dipole moments $\mu$ (in
nuclear magnetons), obtained with the three different theoretical approaches,
are compared to experimental values. The isoscalar $(S)$ and isovector $(V)$ 
combinations of the magnetic dipole moments are also
listed.   The CST results for $A=3$ were obtained in CIA-0.
The experimental $^3$He charge radius is taken from
Ref.~\protect\cite{Sick15}, while
experimental errors  on the $\mu$'s are negligible.
The numbers in parentheses in the $\chi$EFT results give the cutoff
dependence, while the $\chi$EFT results which are underlined correspond to 
those observables which have been used
to fix the LEC's in the current.}
\begin{center}
\begin{tabular}{cccc|c}
\br
Observable & Conv. & $\chi$EFT & CST & Exp.\\
\mr
$r_c({^3{\rm He}})$ & 1.928 & 1.962(4) & 1.879 & 1.973(14)\\
$r_m({^3{\rm He}})$ & 1.909 & 1.920(7) & 2.035 & 1.976(47)\\
\mr
$\mu({^3{\rm H}})$ & 2.953 & {\underbar{2.979}}& 2.441 & 2.979 \\
$\mu({^3{\rm He}})$ & --2.125 & {\underbar{--2.128}}& --1.648 & --2.128\\
$\mu^S$ & 0.414 & {\underbar{0.426}}& 0.396 & 0.426\\
$\mu^V$ & --2.539 & {\underbar{--2.553}}& --2.044 & --2.553\\
\br
\end{tabular}
\end{center}
\end{table}
\subsection{$^4$He nucleus}
\label{subsec:4he}

\begin{figure}[bht] 
\vspace*{0.8cm}
\centerline{\includegraphics[width=10cm]{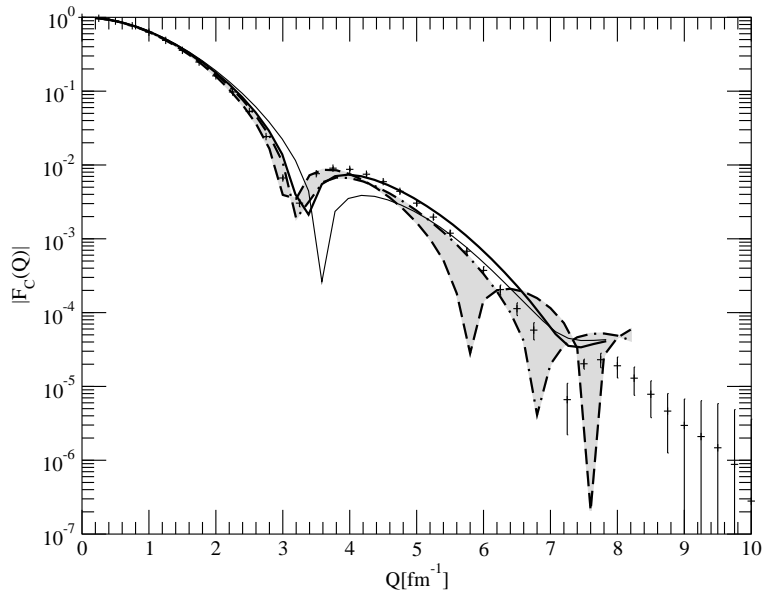}}
\caption[]{The $^4$He charge form factor 
as function of the momentum transfer $Q$ (in fm$^{-1}$),
calculated within the conventional (AV18/UIX) and $\chi$EFT (N3LO/N2LO(500)
or N3LO/N2LO(600)) approaches
is compared with experimental data.  The thin and thick solid lines
correspond to AV18/UIX results obtained by retaining, respectively,
one-body only or both one- and two-body contributions 
in the charge operator; the dashed and dot-dashed lines bounding
the lightly shaded area correspond, respectively, to
the N3LO/N2LO(500) and N3LO/N2LO(600) results, and both include up to N3LO corrections in the charge
operator.}
\label{fig:ff10}  
\end{figure}

Only results obtained in the conventional and $\chi$EFT
approaches are available for $^4$He.  The ``conventional'' ($\chi$EFT)
results for the charge radius obtained with the AV18/UIX 
[N3LO/N2LO(500) and N3LO/N2LO(600)] are listed
in table~\ref{tb:res_4he}. Corrections
beyond IA (or LO) in the charge operator increase IA (or LO) predictions
by about 1\%.   The cutoff sensitivity, shown in parentheses, between
N3LO/N2LO(500) and N3LO/N2LO(600) is at the \% level
for this observable.
Conventional (AV18/UIX) and  $\chi$EFT 
[N3LO/N2LO(500) and N3LO/N2LO(600)] predictions
for the form factor are compared to data in figure~\ref{fig:ff10}.
In the range up to $Q\simeq 8$ fm$^{-1}$ over which calculations
have been carried out,  there is satisfactory agreement between
the conventional calculation  
and experiment.
\begin{table}[bth]
\caption[Table]
{\label{tb:res_4he} The $^4$He charge radius (in fm) obtained
in the conventional and $\chi$EFT approaches. The experimental
value is taken from Ref.~\protect\cite{Sick03c}.}
\begin{center}
\begin{tabular}{ccc|c}
\br
Radius & Conv. & $\chi$EFT &Exp.\\
\mr
$r_c({^4{\rm He}})$ & 1.639 & 1.663(11) & 1.681(4)\\
\br
\end{tabular}
\end{center}
\end{table}

\section{Conclusions}
\label{sec:conc}
Since the first measurements at Stanford, data on form factors of few-nucleon
systems obtained from elastic electron scattering experiments have provided 
crucial benchmarks for testing our understanding of nuclear dynamics, 
{\it i.e.}, of nuclear interactions
and associated nuclear electromagnetic currents.  The high $Q$ measurements 
at SLAC and JLab
have now pushed the experimental knowledge of these 
form factors (or of combinations of them) in the deuteron out to 
$Q \sim 12$ fm$^{-1}$
and in $^3$He to
$Q \sim 9$ fm$^{-1}$.  
Precision measurements at low $Q$
from JLab
and Mainz have also become recently available.   The difficult 
(but successful) efforts
to measure the deuteron tensor polarization, $T_{20}$ has made it possible 
to separate
the three deuteron form factors out to 
$Q\sim 9$ fm$^{-1}$.  
The data set is 
now sufficiently
robust to permit the three deuteron and four three-body form factors to be 
extracted using
the global analyses reported in this review.

This precision data presents a challenge for nuclear theory.  The three 
approaches discussed
in this review assume that the nucleus consists of interacting protons and 
neutrons and that
the effects of their internal substructure can be accounted for by effective 
forces and currents.
The two-body forces are described, at long distance, by one-pion exchange and, 
at short
distance, either by phenomenological terms of two-pion and shorter range in 
the conventional
approach, or by two-pion (and three-pion) exchange and contact terms 
consistent with the
symmetries of the strong interaction in $\chi$EFT, or by exchange of effective 
mesons in CST.
These two-body forces are constrained to reproduce the large database of 
elastic $NN$
differential and total cross sections and polarization 
observables
up to energies close to the pion production threshold, and do so with a 
$\chi^2$/datum $\simeq 1$.
Three-body forces, predominantly induced by two-pion exchange but supplemented
by short-range or contact terms, are needed in the conventional and $\chi$EFT 
approaches in
order to reproduce the three- and four-nucleon binding energies.  
In the CST calculations, irreducible three-body forces are not explicitly 
included. However, certain types of physical processes generated by the 
equations can be reinterpreted as three-body forces if one wants to establish 
connections to other approaches.  A surprising result is that pure 
one-boson-exchange models (which generate {\it no\/} three-body forces) will 
give a precision fit to the $NN$ data and correctly reproduce the $^3$H 
empirical binding energy, {\it provided these models include off-shell 
couplings\/} at the boson exchange vertices.  Since these off-shell couplings 
have no nonrelativistic analogue, a better way to compare the CST with the 
other approaches is to transform the CST dynamics so that  the off-shell 
couplings are replaced by an infinite tower of three-body forces with a unique 
structure guaranteed to give the same physics.   These predicted three-body 
forces can then be compared to those required by the other approaches.

Effective currents consist of a one-body component and interaction, or 
many-body currents.  In all of the approaches, the one-body current is 
parametrized in terms
of the measured nucleon electromagnetic form factors,  but in order to ensure 
current conservation in the presence of off-shell nucleons, the CST requires 
an additional, unknown nucleon form factor, $F_3$.  In the conventional and 
$\chi$EFT approaches, the  many-body
components are induced primarily by the exchange of pions, but also the
exchange of effective (and heavier) mesons, and by the excitation of
intermediate low-energy resonances of the nucleon (like the $\Delta$ isobar)
or due to transition mechanisms (like the $\rho\pi\gamma$ current).  These
many-body contributions are necessary for bringing theory into close agreement
with experiment, especially in the case of the charge form factors of $A=$2--4
nuclei and the isovector combination of the trinucleon magnetic form factors
(and magnetic moments).  In the CST, the one-boson-exchange mechanisms will 
generate interaction currents only, but for the reasons discussed above, 
transformations of these currents that remove their off-shell couplings should 
produce predicted many-body currents that can be compared to those found in 
the other approaches.  These studies are still in their infancy.  

One would have expected that probing few-body systems at very high $Q$
might have revealed a new role for quark degrees-of-freedom at short distances.
Except for their implicit role in determining the effective forces and currents
(as well as nucleon electromagnetic form factors), the models discussed in
this review do not include any such effects.  However, the high 
$Q$ ($Q \gtrsim 5$ fm$^{-1}$)
data have unequivocally shown the need for
a relativistic treatment of nuclear dynamics in this regime.  The CST approach,
which includes relativistic corrections to all orders, provides an excellent
description of the deuteron data up to 
$Q \simeq 12$ fm$^{-1}$.
While this
conclusion was also evident from the 2001 and 2002 reviews on the deuteron
form factors~\cite{Garcon:2001sz,Gilman:2001yh}, it is strengthened by the new
data and broader perspective of the present review.

While the high $Q$ 
CST calculations for the deuteron form factors seem
to be a success (to be confirmed by the next generation calculation in 
progress),
the three-body calculations are still incomplete, and $^4$He calculations have
yet to be attempted.  The fully relativistic CST is probably too cumbersome to
be extended much beyond its current scope, and  further progress with this
method 
would be aided by the development of some new approximation schemes that
would permit it to be extended to more complicated systems.  Pending such a
development, the other approaches are the only way to study systems for
$A>3$ (and even to describe the $A=3$ system fully).

Perhaps the most significant new developments reviewed here are
predictions for few-body form factors that are obtained from  the new 
higher-order
$\chi$EFT calculations of the two-body currents. Moreover,
in the present review, we have reported the first $\chi$EFT calculation
of the $^4$He form factor.  
The ability of 
these
calculations to successfully predict the low $Q$ 
form factors shows that our
understanding can be traced directly to QCD, but the sensitivity of these 
calculations
to the cutoffs, which becomes evident at high $Q$,  
shows that their 
predictive
power is limited to low $Q$ 
phenomena.   This is really no surprise, since
$\chi$EFT employs a perturbative expansion that explicitly breaks down at high
$Q$.  
Perhaps this very promising technique can be extended to higher $Q$ 
by employing some form of Hamiltonian dynamics~\cite{Keister:1991sb}, but this
is yet to be investigated.

\section*{Acknowledgements}
The work of  F.G., R.S., and J.W.V.O. is partially supported by the by 
Jefferson Science
Associates, LLC, under U.S.\ DOE Contract No.\ DE-AC05-06OR23177. 
A.S.\ and M.T.P.\ received partial financial support by 
Funda\c c\~ao para a Ci\^encia e a 
Tecnologia (FCT) under grant Nos.~PTDC/FIS/113940/2009, 
CFTP-FCT (PEst-OE/FIS/U/0777/2013), and 
by the European Union under the HadronPhysics3 Grant No.~283286.
The calculations were made possible by grants of computing time
from the National Energy Research Supercomputer Center.

\end{document}